\documentclass[a4paper,12pt]{article}
\usepackage[english]{babel}
\usepackage{jheppub2}
\usepackage{subfig}
\usepackage[T1]{fontenc} 
\usepackage{graphicx}
\usepackage{amsmath}
\usepackage{slashed}
\usepackage{amssymb}
\usepackage{float}
\usepackage{placeins}
\usepackage{braket}
\usepackage{simplewick}
\allowdisplaybreaks[4]

\title{Conserved vector current in QCD-like theories and the gradient flow}

\author[a]{Marco Boers}
\author[a,b]{and Elisabetta Pallante}

\affiliation[a]{Van Swinderen Institute for Particle Physics and Gravity, University of Groningen, 9747 AG, The Netherlands}
\affiliation[b]{Nikhef, Science Park, Amsterdam, The Netherlands}

\emailAdd{m.r.boers@rug.nl}
\emailAdd{e.pallante@rug.nl}

\abstract{We present analytical results for the 
Euclidean 2-point correlator  of the  flavor-singlet vector current evolved by the gradient flow at next-to-leading order ($\order{g^2}$) in perturbatively massless QCD-like theories. We show that 
 the evolved 2-point correlator requires multiplicative renormalization, in contrast to the nonevolved case,
 and confirm, in agreement with other results in the literature, that such renormalization ought to be identified with a universal renormalization of the evolved elementary fermion field in all evolved fermion-bilinear currents, whereas the gauge coupling renormalizes as usual. We explicitly derive the asymptotic solution of the Callan-Symanzik equation for the connected 2-point correlators of these evolved currents in the limit of small gradient-flow time $\sqrt{t}$, at fixed separation $|x-y|$. Incidentally, this computation determines the leading coefficient of the operator-product expansion (OPE) in the small $t$ limit for the evolved currents in terms of their local nonevolved counterpart. Our computation also implies that, in the evolved case, conservation of the vector current, hence transversality of the corresponding 2-point correlator, is no longer related to the nonrenormalization, 
in contrast to the nonevolved case. Indeed, for small flow time the evolved vector current is conserved up to $\order{t}$ softly violating effects, despite its $t$-dependent nonvanishing anomalous dimension. 
}

\newcommand{\be}{\begin{equation}}
\newcommand{\ee}{\end{equation}}
\newcommand{\nn}{\nonumber}
\newcommand{\bea}{\begin{eqnarray}}
\newcommand{\eea}{\end{eqnarray}}
\newcommand{\bfig}{\begin{figure}}
\newcommand{\efig}{\end{figure}}
\newcommand{\bc}{\begin{center}}
\newcommand{\ec}{\end{center}}

\newcommand{\f}[2]{\frac{#1}{#2}}

\newcommand{\eps}{{\varepsilon}}
\newcommand{\Lap}{\Delta}

\newcommand{\dD}{\delta^{\scriptscriptstyle{(d)}}}
\newcommand{\dsx}{ \slashed {\partial}_x}
\newcommand{\dsy}{ \slashed {\partial}_y}
\newcommand{\dsum}{ \overrightarrow{\slashed {\partial}}+    \overleftarrow{\slashed {\partial}}  }
\newcommand{\dsumxy}{ \overrightarrow{\slashed {\partial}}_y+ \overleftarrow{\slashed {\partial}}_x  }
\newcommand{\dsumxyi}{ \overrightarrow{\slashed {\partial}}_x+ \overleftarrow{\slashed {\partial}}_y  }

\newcommand{\etsx}{e^{(t-s)\Lap_x}}
\newcommand{\xt}{\frac{x^2}{8t}}
\newcommand{\order}[1]{\mathcal{O}(#1)}
\newcommand{\del}{\partial}
\newcommand{\psibar}{\bar{\psi}}
\newcommand{\half}{\frac{1}{2}}
\newcommand{\aslash}{\slashed{A}}
\newcommand{\dslash}{\slashed{\partial}}
\newcommand{\mn}{{\mu\nu}}
\newcommand{\pslash}{\slashed{p}}
\newcommand{\qslash}{\slashed{q}}
\newcommand{\kslash}{\slashed{k}}

\newcommand{\chibar}{\bar{\chi}}
\newcommand{\xbar}{\bar{x}}
\newcommand{\ybar}{\bar{y}}
\newcommand{\zbar}{\bar{z}}
\newcommand{\gmu}{\gamma_\mu}
\newcommand{\gnu}{\gamma_\nu}
\newcommand{\grho}{\gamma_\rho}

\begin{document}
\maketitle

\section{Introduction and main results} \label{sec:intro}
The gradient flow for classical Yang-Mills theory was first formulated in \cite{AtiyahBott1982,Donaldson1985}, see also \cite{Neuberger2011} for a nice review of the subject. The gradient flow equation is a specific map of the elementary gauge field $A_\mu(x)$ to its gradient-flow evolved (smeared) version $B_\mu(t,x)$, for a given initial condition at $t=0$. The parameter $\sqrt{t}\geq 0$, with the dimension of a time, or equivalently a length, can be interpreted as the gradient-flow time, or equivalently the smearing radius. 

The last decade has seen a revived interest in the gradient flow, whose properties and uses are being further investigated in the context  of quantum Yang-Mills theory coupled to matter fields, more specifically, QCD-like theories formulated in the  continuum or on a Euclidean lattice. In \cite{Neuberger2006} the authors provided a lattice formulation of the Yang-Mills gradient flow to study large-$N$ properties of Yang-Mills theory. 

Later on, the one-loop renormalization  of the gradient-flow evolved Yang-Mills Lagrangian density in the Wilsonian normalization\footnote{See, e.g., \cite{Becchetti} for further details on Wilsonian versus canonical normalization.}, i.e.,  $g^2B_{\mn}^aB_{\mn}^a$ with $B_{\mn}^a$ the canonically normalized evolved gauge-field strength, was derived in \cite{Luscher1}. A systematic analysis of the all-order renormalization properties of gradient-flow evolved elementary gauge fields, and implications for the renormalization of pure-glue local composite operators was then provided in \cite{Luscher2}, see also \cite{J}. The renormalization of 
gradient-flow evolved elementary fermion fields was first investigated in \cite{LuscherFerm}.

Interestingly, gradient-flow type equations -- that can be seen as generalized diffusion equations -- are employed in many other contexts, ranging from physics to engineering, often to smear microscopic effects in mechanical systems. In quantum field theory, we note that stochastic quantization -- see, e.g., \cite{DamgaardHuffel1987} for a review -- involves analogous techniques, though fundamentally different in purpose. 

In fact, the 
gradient-flow equation for gauge fields used here coincides with the Langevin equation for the stochastic quantization of a Yang-Mills theory with the noise term removed. Yet, the gradient flow in the present paper only acts on the operators in the correlators and it never involves the Lagrangian that occurs in the definition of the vacuum expectation value.

In this work we further explore the properties of gradient-flow evolved composite operators in the fermion sector of QCD-like theories. In particular, we will compute  to $\order{g^2}$ the Euclidean 2-point correlator $\Pi_\mn^V(t,x-y)=\braket{J_\mu^V(t,x)J_\nu^V(t,y)}$ of 
the  flavor-singlet vector current $ J_\mu^V(t,x)=\bar{\chi}(t,x)\gamma_\mu\chi(t,x)$ evolved to a gradient-flow time $\sqrt{t}$. Our main result is derived in Sec. \ref{sec:isolatingdivs}:
\be\label{eq:intro}
\begin{split}
\Pi^V_{R,\mn}(t,x-y, \mu,g(\mu))&= Z_{J_t}^2(g(\mu),\eps)\, \Pi_{\mn}^V(t,x-y, \eps,g)\\
&\hspace{-4.cm}=\Big(1+\frac{g^2(\mu)}{(4\pi)^2}\frac{6}{\eps}C_2(R)+\ldots\Big) 
\Big(1-\frac{g^2(\mu)}{(4\pi)^2} C_2(R) \Big( \frac{6}{\eps}+ 6\log{ (t\mu^2)} +\text{finite}  \Big)\Big)\\
&\hspace{-3.5cm}\times  \Pi^V_{\mn,0}(t,x-y)+\ldots \\
&\hspace{-4.cm}=\Big( 1- \frac{g^2(\mu)}{(4\pi)^2}
6 C_2(R)\log{ (t\mu^2)} +\text{finite}
\Big)\Pi^V_{\mn,0}(t,x-y)+\ldots
\end{split}
\ee
where $\Pi_{\mn}^V$ is the bare correlator in dimensional regularization, 
with $g$ the bare coupling, and $\Pi^V_{\mn,0}(t,x-y)$ is the leading order evolved correlator in Eq.~\eqref{eq:evolved2pointLO}:
\begin{equation}
\begin{split}
\Pi_{\mn,0}^V(t,x-y)
&=-\frac{2\,d(R)}{\pi^4}\frac{\gamma\big(2,\frac{(x-y)^2}{8t}\big)^2}{\big((x-y)^2\big)^3}\Big(\frac{\delta_\mn}{2} -\frac{(x-y)_\mu (x-y)_\nu}{(x-y)^2} \Big)
\end{split}
\end{equation}
Equation \eqref{eq:intro} shows that the evolved 2-point correlator requires a multiplicative renormalization. 
 
Hence,  the evolved vector current acquires a $t$-dependent anomalous dimension, in contrast to the nonevolved case, where it has no anomalous dimension. The result in Eq.~\eqref{eq:intro} is consistent with  results for 1-point correlators of evolved fermion bilinears \cite{LuscherFerm,MakinoSuzuki2014,Harlander3}\footnote{We note that a sign in the renormalization factor of the evolved fermion field, crucial for this discussion, has been corrected in the Erratum of \cite{Harlander3}.}. Importantly, it confirms that such renormalization ought to be identified with a universal renormalization of the evolved elementary fermion field \cite{LuscherFerm}, whereas the gauge coupling renormalizes as usual.

Our result is thus consistent with studies so far \cite{Luscher2,LuscherFerm}, which suggest that the only renormalization of evolved fermionic composite operators is the one induced by the 
renormalization of the evolved elementary fermion fields and the gauge coupling, independently of their tensor structure. Equation \eqref{eq:intro} thus yields the leading $\order{g^2}$ contribution to the anomalous dimension of all evolved fermion-bilinear currents. 
 
Moreover, we will make contact with the nonevolved case by deriving the leading contribution to the OPE of the evolved fermion-bilinear currents from the corresponding connected 2-point correlators in the limit  of small gradient-flow time $\sqrt{t}$, at fixed separation $|x-y|$. The universal UV asymptotics of the leading coefficient $c(t)$ in the OPE \cite{Luscher2,LuscherFerm}:
\be
 J_{R}(t,x)=\braket{J(t,x)}_{R}+c(t)J_R(x)+\order{t}
 \ee 
 of a generic renormalized evolved fermion-bilinear current $J_R(t,x)$ -- with $J(t,x)$ and $J(x)$ multiplicatively renormalizable -- as the renormalization-group invariant coupling $g(\sqrt{t})\to 0$ reads:
\be
c(\sqrt{t}\mu, g(\mu))\sim \Big(\f{g(\sqrt{t})}{g(\mu)} \Big)^{\f{\gamma_{J_t}^{(0)}-\gamma_{J}^{(0)}}{\beta_0}}e^{\order{g^2(\mu)}}
\ee
from Eq.~\eqref{eq:UVct}, with $\gamma_{J_t}^{(0)}$ and $\gamma_{J}^{(0)}$ the leading-order coefficients of the  anomalous dimension of the evolved and nonevolved current, respectively. 
 
We will also clarify how, in the evolved case, current conservation up to terms that vanish for small $t$ no longer entails nonrenormalization.

At various intermediate steps we will be employing coordinate space techniques, instead of the more conventional momentum space approach. They are especially useful to illustrate the behavior of two limits, vanishing flow time at fixed separation and vanishing separation at fixed flow time. 

The paper is organized as follows. The gradient flow in QCD-like theories is reviewed in Sec. \ref{sec:GF}, where the notation and useful expressions in coordinate space are introduced. 

Section \ref{sec:1point} discusses the vanishing of the 1-point correlator of the divergence of the evolved vector current $\braket{\del_\mu J_\mu^V(t,x)}$, and it is mainly used to introduce some coordinate space  techniques. 

In Sec. \ref{sec:CM} we review the nonevolved flavor singlet 2-point vector correlator 
 to $\order{g^2}$, generalizing the original calculation \cite{CM} to $SU(N)$. For later use, we also discuss the all-order solution of the Callan-Symanzik equation for the connected 2-point correlators of multiplicatively renormalizable fermion-bilinear currents.

Section \ref{sec:2point} is devoted to the calculation of the  evolved 2-point  vector correlator.  
Sections \ref{sec:GF2pointLO} and \ref{sec:NLO} contain the analytic calculation to $\order{g^2}$.
Renormalization at $\order{g^2}$ is treated in Sec. \ref{sec:isolatingdivs}.
In Sec. \ref{sec:RGsolev} we determine the leading contribution to the OPE of evolved fermion-bilinear currents for small $t$ from the corresponding connected 2-point correlators. 

In Sec. \ref{sec:nonrenorm}
we discuss the conservation of the evolved vector current up to terms of $\order{t}$. We conclude in Sec. \ref{sec:conc}. 
The diagrammatic notation, useful intermediate calculations and
expressions in terms of momentum integrals are in appendices \ref{sec:diagramnotation} to \ref{sec:eps-exp}. Appendix \ref{sec:classic} shows the classical nonconservation of the evolved vector current to leading order in the gauge coupling. 
\section{The gradient flow in QCD-like theories} \label{sec:GF}
We consider perturbatively massless QCD-like theories with gauge group $SU(N)$ and $N_f$ flavors of Dirac fermions in the representation $R$. We work in Euclidean metric throughout this paper, with Hermitian gamma matrices $\gamma_\mu$, and we employ anti-hermitian generators for $SU(N)$, hence   $[T^a,T^b]=f^{abc}T^c$.

The gradient-flow equation for the gauge field reads \cite{Luscher1}:
\be
\label{eq:GF}
\dot{B}_\mu=D_\nu G_{\nu\mu}+\alpha_0 D_\mu\del_\nu B_\nu
\ee
with:
\be
G_{\mu\nu}(t,x)=\partial_\mu B_\nu - \partial_\nu B_\mu  +[B_\mu,B_\nu]
\ee
in the Wilsonian normalization of the gauge field. The gauge field $B_\mu(t,x)$ is the solution of Eq.~\eqref{eq:GF} with initial condition $B_\mu(t,x)|_{t=0}\!=\!A_\mu(x)$, where $A_\mu(x)=A_\mu^a(x)T^a$ is the bare gauge field and $T^a, a=1,\ldots N^2-1$ are the generators of $SU(N)$.  The dot in Eq.~\eqref{eq:GF} stands for the derivative with respect to $t$ and $D_\mu\!=\!\partial_\mu\!+\![B_\mu, \cdot\,]$. 
The flow time, $\sqrt{t}$, acts as the smearing radius for the gauge field $B_\mu(t,x)$. The parameter $\alpha_0$ can be seen as a gauge fixing parameter in Eq.~\eqref{eq:GF}, and we will be working with $\alpha_0=1$, along with the Feynman gauge chosen in the Lagrangian. 

The generalization of Eq.~\eqref{eq:GF} to fermion fields can be formulated as in \cite{LuscherFerm}, and reads:
\begin{equation}\label{eq:gfeqnfermions}
\begin{split}
\dot{\chi}&=(\Delta-\alpha_0\del_\mu B_\mu)\,\chi,\hspace{1cm}\chi(t,x)|_{t=0}=\psi(x)\\
\dot{\chibar}&=\chibar(\overleftarrow{\Delta}+\alpha_0\del_\mu B_\mu),\hspace{1cm}\chibar(t,x)|_{t=0}=\psibar(x)
\end{split}
\end{equation}
where
\begin{align}
\Delta&=D^2,\hspace{1cm}D_\mu=\del_\mu+B_\mu\\
\overleftarrow{\Delta}&=\overleftarrow{D}^2,\hspace{1cm}\overleftarrow{D}_\mu=\overleftarrow{\del}_\mu-B_\mu
\end{align}
and here it is implicit that the $T^a$ in the $B_\mu$ field are in the representation $R$ of the fermions.
\subsection{Solutions of the gradient-flow equations}\label{sec:GFsolutions}
We start by considering the well-known integral form of Eq.~\eqref{eq:GF} that is solved iteratively for the gauge field:
\begin{equation}\label{eq:GFgaugesolutionnobox}
B_\mu(t,x)=\int_y K_{t,\mu\nu}(x-y) A_\nu(y)+\int_y\int_0^tds\,K_{t-s,\mu\nu}(x-y) R_\nu(s,y)
\end{equation}
with kernel in Feynman gauge $K_{t,\mu\nu}(x-y)=\delta_{\mu\nu} K_t(x-y)$ , where $K_{t}(x-y) $ is the scalar kernel\footnote{Throughout this paper we use the notation $\int_x\equiv\int d^dx$ and $\int_p\equiv\int\frac{d^dp}{(2\pi)^d}$.}:
\be\label{eq:kernel}
K_t(x-y) =\int_p e^{ip(x-y)}e^{-tp^2} = \f{1}{(4\pi t)^{d/2}} e^{-\f{(x-y)^2}{4t}      }
\ee
and:
\begin{equation}
R_\mu=2[B_\nu,\del_\nu B_\mu]-[B_\nu,\del_\mu B_\nu]+[B_\nu,[B_\nu,B_\mu]]
\end{equation}
 The scalar kernel can be conveniently rewritten in operator notation as:
\be\label{eq:K}
K_t(x-y)=e^{t\Lap_x}\;\delta^{(d)}(x-y)
\ee
with $\Lap_x=\partial_\mu^x \partial_\mu^x$, i.e., it is a nonlocal Gaussian regulator of a Dirac delta distribution -- often referred to in this context as a smearing -- whose Fourier transform (FT) is an entire analytic function of order two.
By means of the ``exponential-of-Laplacian'' notation in Eq.~\eqref{eq:K}, Eq.~\eqref{eq:GFgaugesolutionnobox} reads:
\be
B_\mu(t,x)=e^{t\Lap_x}A_\mu(x) + \int_0^t ds\, e^{(t-s)\Lap_x} R_\mu(s,x)
\ee
Analogously, the integral form of Eq.~\eqref{eq:gfeqnfermions} that is solved iteratively for the fermion fields reads: 
\begin{equation}\label{eq:F}
\begin{split}
\chi(t,x)&=\int_y K_t(x-y)\psi(y) + \int_y\int_0^tds\,K_{t-s}(x-y)\Delta'\chi(s,y) \\
\chibar(t,x)&=\int_y\psibar(y) K_t(x-y)+\int_y\int_0^tds\,\chibar(s,y)\overleftarrow{\Delta}' K_{t-s}(x-y)
\end{split}
\end{equation}
where the quantities $\Delta'$ and $\overleftarrow{\Delta}'$ contain the evolved gauge field and read:
\begin{equation}
\begin{split}
\Delta'&=(1-\alpha_0)\del_\mu B_\mu+2B_\mu\del_\mu+B_\mu B_\mu\\
\overleftarrow{\Delta}'&=-(1-\alpha_0)\del_\mu B_\mu-2\overleftarrow{\del}_\mu B_\mu +B_\mu B_\mu
\end{split}
\end{equation}
By means of Eq.~\eqref{eq:K}, Eq.~\eqref{eq:F} also reads:
\begin{equation}\label{eq:Fus}
\begin{split}
\chi(t,x)&=e^{t\Lap_x}\psi(x) + \int_0^t ds\, e^{(t-s)\Lap_x}\Big\{
\Delta'(s,x)\chi(s,x)\Big\}\\
\bar{\chi}(t,x)&= e^{t\Lap_x}\bar\psi(x) + \int_0^t ds\, e^{(t-s)\Lap_x}\Big\{\bar\chi(s,x)
\overleftarrow{\Delta}'(s,x)\Big\}
\end{split}
\end{equation}
where the exponential-of-Laplacian acts on the expression inside brackets $\{\ldots\}$.
\subsection{Free propagators}\label{sec:prop}
We briefly review the expressions for the free propagators, both nonevolved and evolved, in coordinate space together with their limiting behaviors as $t\rightarrow0$ at fixed nonzero separation $(x-y)$, or $(x-y)\to 0$ at fixed positive $t$.

The nonevolved gauge field propagator in Feynman gauge is:
\begin{equation}
\braket{A_\mu^a(x)A_\nu^b(y)}=D_\mn^{ab}(x-y)=\delta^{ab}\delta_\mn D(x-y)
\end{equation}
where $D(x-y)$ is the scalar propagator in $d$ Euclidean dimensions:
\begin{equation}\label{eq:scalarpropexpl}
\begin{split}
D(x-y)&=\int_pe^{ip(x-y)}\frac{1}{p^2}\\
&=\frac{\Gamma(\frac{d}{2}-1)}{4\,\pi^{d/2}}
\f{1\hspace{.6cm}}{((x-y)^2)^{d/2-1}}
\end{split}
\end{equation}
which satisfies:
\begin{equation}
\Lap_xD(x-y)=\Lap_yD(x-y)=-\dD(x-y)
\end{equation}
The nonevolved free fermion propagator is:
\begin{equation}
\braket{\psi_i(x)\psibar_j(y)}=\delta_{ij}S(x-y)
\end{equation}
with $SU(N)$ indices (which we will mostly keep implicit) $i,j=1,...,d(R)$, where $d(R)$ is the dimension of the representation $R$, and:
\begin{equation}\label{eq:fermpropexpl}
\begin{split}
S(x-y)&=\int_pe^{ip(x-y)}\frac{-i\pslash}{p^2}\\
&=\frac{\Gamma(\frac{d}{2})}{2\,\pi^{d/2}}\frac{\slashed{x}-\slashed{y}\hspace{.6cm}\,}{((x-y)^2)^{d/2}}
\end{split}
\end{equation}
which satisfies:
\begin{equation}\label{eq:derS}
\dslash_xS(x-y)=-\dslash_yS(x-y)=\dD(x-y)
\end{equation}
We also recall the relation between the fermion and scalar propagators:
\begin{equation}
\dslash_xD(x-y)=S(y-x)=-S(x-y)
\end{equation}

The above formulas are readily generalized to the evolved case, where the free propagators do not receive contributions from the second flow-time integral terms in Eqs.~\eqref{eq:GFgaugesolutionnobox} and \eqref{eq:F}. Hence:
\begin{equation}\label{eq:propb-gf}
\begin{split}
\braket{B_\mu^a(t,x)B_\nu^b(s,y)}&=\braket{e^{t\Lap_x}A_\mu^a(x)e^{s\Lap_y}A_\nu^b(y)}\\
&=\delta^{ab}\delta_\mn e^{t\Lap_x+s\Lap_y} D(x-y)\\
&\equiv \delta^{ab}\delta_{\mu\nu} D(\xbar_t-\ybar_s)
\end{split}
\end{equation}
and:
\begin{equation}\label{eq:prop-gf}
\begin{split}
\braket{\chi(t,x)\chibar(s,y)}&=\braket{e^{t\Lap_x}\psi(x)e^{s\Lap_y}\psibar(y)}\\
&=e^{t\Lap_x+s\Lap_y}S(x-y)\\
&\equiv S(\xbar_t-\ybar_s)
\end{split}
\end{equation}
where we have introduced a convenient notation: the bar over the coordinate in the last line of Eqs.~\eqref{eq:propb-gf} and \eqref{eq:prop-gf} represents the exponential-of-Laplacian, and the subscript the associated flow time. In this notation the scalar kernel in Eq.~\eqref{eq:K} reads $K_t(x-y)=\dD(\xbar_t-y)=\dD(x-\ybar_t)$. 

The evolved gauge and fermion propagators now satisfy, respectively:
\begin{equation}
\Lap_x D(\xbar_t-\ybar_s)=\Lap_y D(\xbar_t-\ybar_s)=-K_{t+s}(x-y)
\end{equation} 
and
\begin{equation}\label{eq:dSGF}
\dslash_x S(\xbar_t-\ybar_s)=-\dslash_y S(\xbar_t-\ybar_s)=K_{t+s}(x-y)
\end{equation}
where the evolved scalar propagator is:
\begin{equation}\label{eq:propDGFevolved}
\begin{split}
D(\xbar_t-\ybar_s)&=\int_pe^{ip(x-y)}e^{-(t+s)p^2}\frac{1}{p^2}\\
&=\frac{\gamma(\frac{d}{2}-1,\frac{(x-y)^2}{4(t+s)})}{4\,\pi^{d/2}}\f{1\hspace{.6cm}}{((x-y)^2)^{d/2-1}}
\end{split}
\end{equation}
and the evolved fermion propagator is:
\begin{equation}\label{eq:propf}
\begin{split}
S(\xbar_t-\ybar_s)&=\int_pe^{ip(x-y)}e^{-(t+s)p^2}\frac{-i\pslash}{p^2}\\
&=\frac{\gamma(\frac{d}{2},\frac{(x-y)^2}{4(t+s)})}{2\,\pi^{d/2}}\frac{\slashed{x}-\slashed{y}\hspace{.6cm}\,}{((x-y)^2)^{d/2}}
\end{split}
\end{equation}
with $\gamma(a,z)$ the lower incomplete gamma function. 
The gradient-flow evolution thus amounts to replacing the gamma functions in Eqs.~\eqref{eq:scalarpropexpl} and \eqref{eq:fermpropexpl} with their lower incomplete counterpart, whose integral representation is:
\begin{equation}\label{eq:lower-gamma}
\gamma(a,z)=\int_0^z dt\,  t^{a-1}e^{-t}
\end{equation}
for $\text{Re}(a)>0$\footnote{When analytically continuing to $\text{Re}(a)<0$, $\gamma(a,z)$ has the same poles and residues at $a=0,-1,...$ as $\Gamma(a)$ has at $a=0,-1,...$, provided $z\neq 0$.}.
For $z\ll 1$, the expansion:
\begin{equation}\label{eq:incomplgammaexpansion}
\gamma(a,z)=\frac{z^a}{a}+\order{z^{a+1}}
\end{equation}
applied to Eqs.~\eqref{eq:propDGFevolved} and \eqref{eq:propf} for vanishing separation at fixed positive flow time (hence, $(x-y)^2/t\to 0$) illustrates how the gradient flow regulates the short-distance singularities in $x$-space of the free propagators at fixed  flow time. Indeed, in this limit one has:
\begin{equation}\label{eq:Dlim}
\begin{split}
D(\xbar_t-\ybar_t)&=\frac{4}{(d-2)(8\pi)^{d/2}}\,\f{1}{t^{d/2-1}} \Big(1+O\Big( \f{(x-y)^2}{t} \Big)\Big)
\end{split}
\end{equation}
and 
\begin{equation}\label{eq:Slim}
\begin{split}
S(\xbar_t-\ybar_t)&=\frac{1}{d(8\pi)^{d/2}}\,\f{\slashed {x} -\slashed {y}}{t^{d/2}} \Big(1+O\Big( \f{(x-y)^2}{t} \Big)\Big)
\end{split}
\end{equation}
Note that the evolved fermion propagator in Eq.~\eqref{eq:Slim} actually vanishes in this limit due to its Lorentz structure\footnote{For Euclidean distances $(x-y)^2=0$ implies $x_\mu=y_\mu$.}, differently from the scalar propagator in Eq.~\eqref{eq:Dlim}.

Viceversa, Eqs.~\eqref{eq:propDGFevolved} and \eqref{eq:propf} recover the nonevolved result in the limit  $t\to 0$ at fixed nonzero separation, since $\gamma(a,(x-y)^2/(8t))\to \Gamma(a)$. 

The result of the combined limits $t\to 0$ and $(x-y)\to 0$ thus depends on the order in which the two limits are taken in the following sense. For the scalar propagator one always produces a singularity, which is in $x$-space -- the one of the nonevolved case -- or in $t$-space, i.e., in the flow coordinate, when taking first $t\to 0$, or $(x-y)\to 0$, respectively. 
The fermion propagator, instead, vanishes when taking first $(x-y)\to 0$, whereas it recovers the original $x$-space singularity when taking first $t\to 0$. This makes clear that we need to consider the latter limit, i.e., $t\to 0$ at nonzero separations in order to make contact with the nonevolved correlators of the original quantum field theory. 
\subsection{The vector current evolved by the gradient flow}
\label{sec:JV}
We introduce the  evolved vector current:
\be\label{eq:JVGF}
J_\mu^V(t,x)=\bar\chi(t,x)\gamma_\mu\chi(t,x)
\ee
written in terms of the evolved fermion fields $\chi$ and $\bar\chi$.

For the perturbative calculation of the 1- and 2-point correlators in Secs. \ref{sec:1point} and \ref{sec:2point}, respectively, we go from the Wilsonian to the canonical normalization by rescaling the bare gauge field everywhere $A_\mu(x)\rightarrow g A_\mu(x)$, with $g$ the bare gauge coupling. After rescaling, we conveniently rewrite all fields as expansions in powers of $g$:
\bea\label{eq:series}
\chi(t,x) &=& \sum_{n=0}^\infty\,g^n \chi_n(t,x) \nn\\
\bar\chi(t,x) &=& \sum_{n=0}^\infty\,g^n \bar\chi_n(t,x) \nn\\
B_\mu(t,x) &=& \sum_{n=1}^\infty\,g^n B_{\mu,n}(t,x) 
\eea
and for later use we write explicitly the expressions for $\chi_n, \bar\chi_n$ and $B_{\mu, n}$ for $n\leq 2$:
\bea\label{eq:chiandBexpansions}
\chi_0(t,x)&=& e^{t\Lap_x}\psi(x)~~~~~~\bar\chi_0(t,x)= e^{t\Lap_x}\bar\psi(x) \nn\\
\chi_1(t,x)&=& 2\int_0^t ds\, e^{(t-s)\Lap_x} \left\{B_{\mu, 1}\partial_\mu\chi_0   \right\} \nn\\
\bar\chi_1(t,x)&=& -2\int_0^t ds\, e^{(t-s)\Lap_x} \left\{ \partial_\mu\bar\chi_0 B_{\mu, 1}  \right\} \nn\\
B_{\mu, 1}(t,x)&=&e^{t\Lap_x}A_\mu(x) \nn\\
B_{\mu, 2}(t,x)&=&\int_0^t ds\,  e^{(t-s)\Lap_x} \left\{ [B_{\nu, 1}, 2\partial_\nu B_{\mu, 1} - \partial_\mu B_{\nu, 1}]\right\}\nn\\
\chi_2(t,x)&=& \int_0^t ds\, e^{(t-s)\Lap_x} \left\{B_{\mu, 1} B_{\mu, 1} \chi_0 + 2B_{\mu, 2} \partial_\mu\chi_0+2B_{\mu, 1} \partial_\mu\chi_1      \right\}\nn\\
\bar\chi_2(t,x)&=& \int_0^t ds\, e^{(t-s)\Lap_x} \left\{
\bar\chi_0 B_{\mu, 1} B_{\mu, 1}  
- 2 \partial_\mu\bar\chi_0 B_{\mu, 2} -2 \partial_\mu\bar\chi_1 B_{\mu, 1}      \right\}
\eea
All fields in the flow integral in Eq.~\eqref{eq:chiandBexpansions}  are functions of $(s,x)$, derivatives are always with respect to $x$ and the exponential-of-Laplacian acts on the  expressions inside brackets $\{\ldots\}$.

Analogously, we expand the evolved vector current in Eq. \eqref{eq:JVGF} in powers of the coupling:
\begin{equation}\label{eq:Jseries}
J^V_\mu(t,x)=\sum_{n=0}^\infty g^nJ^V_{\mu,n}(t,x)
\end{equation}
where the currents $J^V_{\mu,n}(t,x)$ for $n=0,1,2$ read:
\begin{equation}\label{eq:cc}
\begin{split}
J^V_{\mu,0}(t,x)&=\chibar_0(t,x)\gamma_\mu\chi_0(t,x)\\
J^V_{\mu,1}(t,x)&=\chibar_1(t,x)\gamma_\mu\chi_0(t,x)+
\chibar_0(t,x)\gamma_\mu\chi_1(t,x)\\
J^V_{\mu,2}(t,x)&=\chibar_2(t,x)\gamma_\mu\chi_0(t,x)
+\chibar_0(t,x)\gamma_\mu\chi_2(t,x)+\chibar_1(t,x)\gamma_\mu\chi_1(t,x)
\end{split}
\end{equation}
with $\chi_n$ and $\bar\chi_n$, for $n=0,1,2$ in Eq. \eqref{eq:chiandBexpansions}.
\section{The 1-point correlator $\langle \partial_\mu J_\mu^V(t,x)\rangle$ evolved by the gradient flow }
\label{sec:1point}
Given the operators $O_\pm(t,x)$:
\be
O_\pm(t,x) = \bar\chi(t,x)\slashed {D}_\pm\chi(t,x)
\ee
with 
\be
\slashed {D}_\pm = \overrightarrow{\slashed {D}} \pm \overleftarrow{\slashed {D}}
\ee
and ($D_\mu=\partial_\mu+ B_\mu$):
\bea
\slashed {D}_+ &=& \dsum \nonumber\\
\slashed {D}_- &=& \overrightarrow{\slashed {\partial}} - \overleftarrow{\slashed {\partial}}+2 \slashed {B}
\eea
we recognize that $O_+$ is the divergence of the vector current $O_+=\del_\mu J^V_\mu$, whereas  $O_-$ is the operator that enters the fermion equation of motion. The 1-point correlator $\braket{O_-(t,x)}$ has been studied to $\order{g^2}$ in \cite{MakinoSuzuki2014}, where,  differently from its nonevolved counterpart, it was found to be nonzero and to renormalize with a new counterterm induced by the gradient flow. 

In this section we consider the 1-point correlator $\braket{O_+(t,x)}=\braket{\del_\mu J^V_\mu(t,x)}$. Its nonevolved version vanishes (before and after subtraction of divergences) independently of whether $J_\mu^V$ is conserved or not. In fact, $\braket{(\del_\mu J^V_\mu(x))_R}=\del_\mu\braket{ J^V_{\mu\,R}(x) }=0 $ holds for the simple reason that the momentum at the vertex for $J_\mu^V$ vanishes, but, in the vector case, also because $\braket{J^V_{\mu\,R}(x) }$  itself vanishes due to Lorentz invariance and  the fact that $J_\mu^V$ is odd under charge conjugation\footnote{$\braket{J_\mu^V(x)}=\braket{C^\dagger C J_\mu^V(x) C^\dagger C}=-\braket{J_\mu^V(x)}$.}. 

For the evolved correlator, the same chain of identities:
\begin{equation}\label{eq:vanishing1point}
\braket{O_{+\,R}(t,x)}=\braket{(\del_\mu J^V_\mu(t,x))_R}=\del_\mu\braket{ J^V_{\mu\,R}(t,x) }=0
\end{equation}
holds provided one can write the evolved current in terms of its Fourier transform as $J^V_\mu(t,x)=\int_p e^{ipx}\tilde{J}^V_\mu(t,p)$. This is true for the bare current, with $\chi$ and $\bar\chi$ in Eq.~\eqref{eq:F}, and it cannot be spoiled by renormalization, analogously to the nonevolved case. Again, the last equality in Eq.~\eqref{eq:vanishing1point} is also implied by Lorentz invariance and the fact that $J_\mu^V(t,x)$ is odd under charge conjugation. 

\begin{figure}[tb]
	\centering
	\begin{minipage}{.33\textwidth}
		\centering
		\includegraphics[width=.55\linewidth]{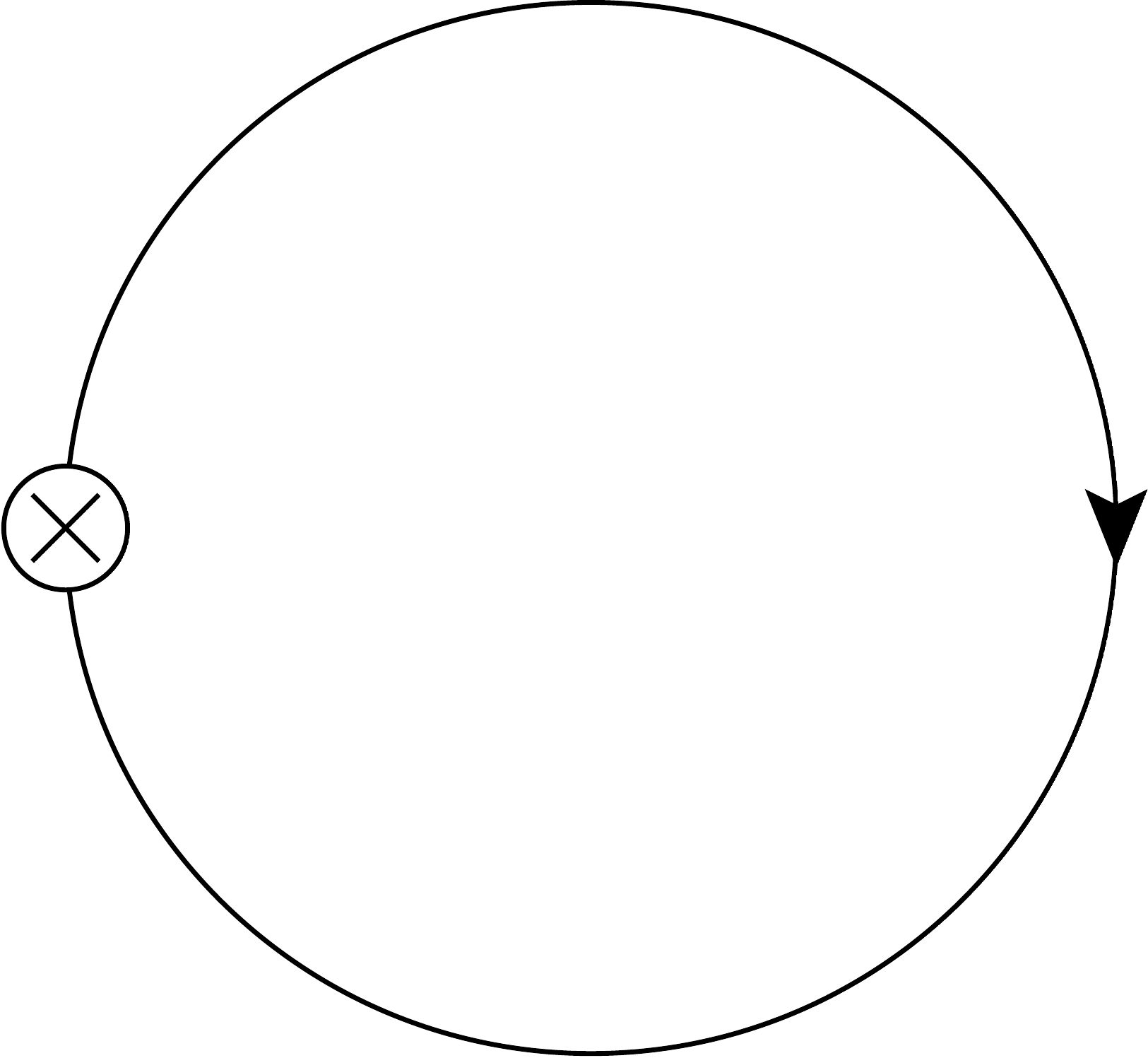}
		\caption*{D01}
		\label{fig:D01}
	\end{minipage}%
	\begin{minipage}{.33\textwidth}
		\centering
		\includegraphics[width=.55\linewidth]{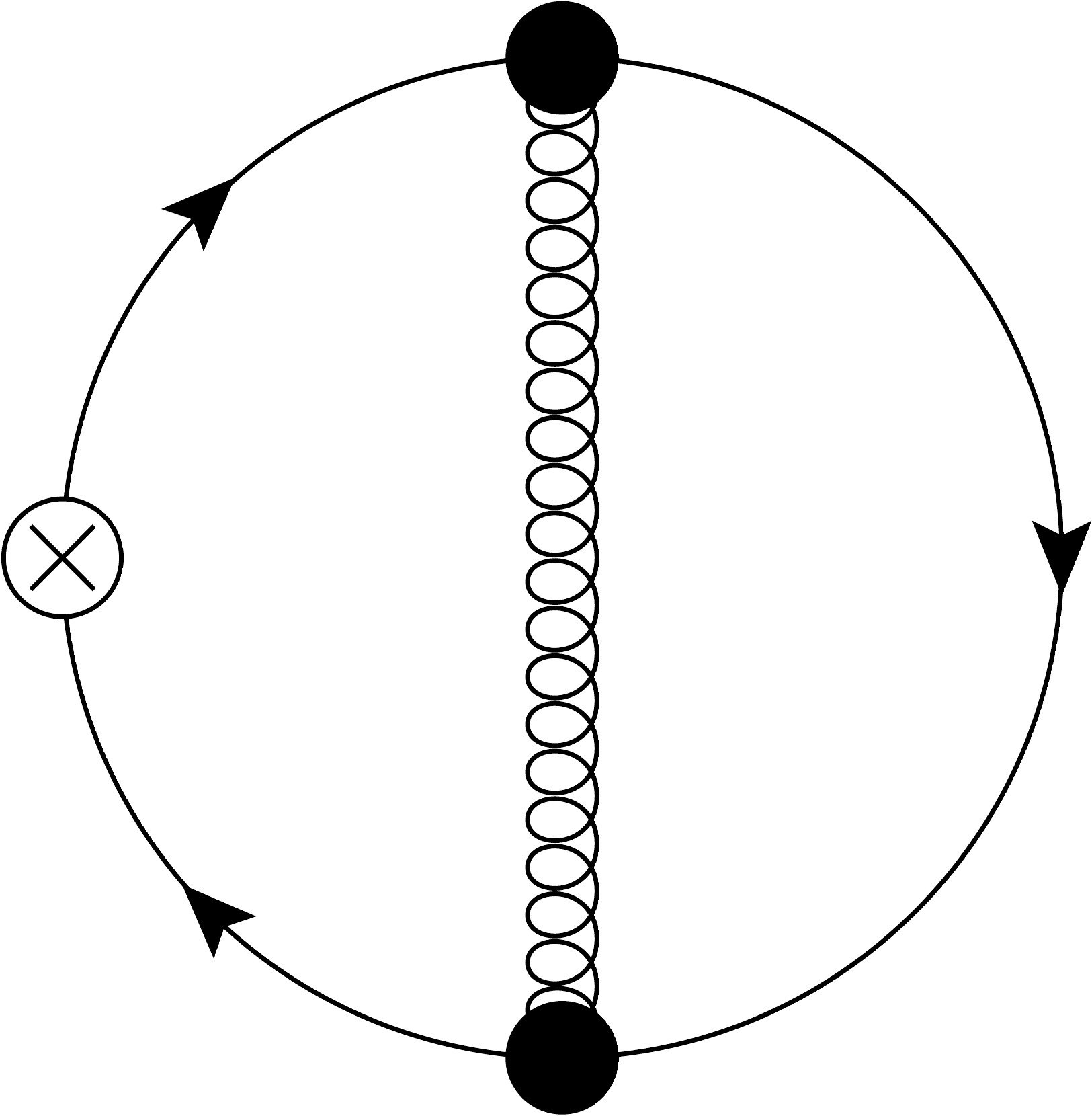}
		\caption*{D02}
		\label{fig:D02}
	\end{minipage}
	\begin{minipage}{.33\textwidth}
		\centering
		\includegraphics[width=.55\linewidth]{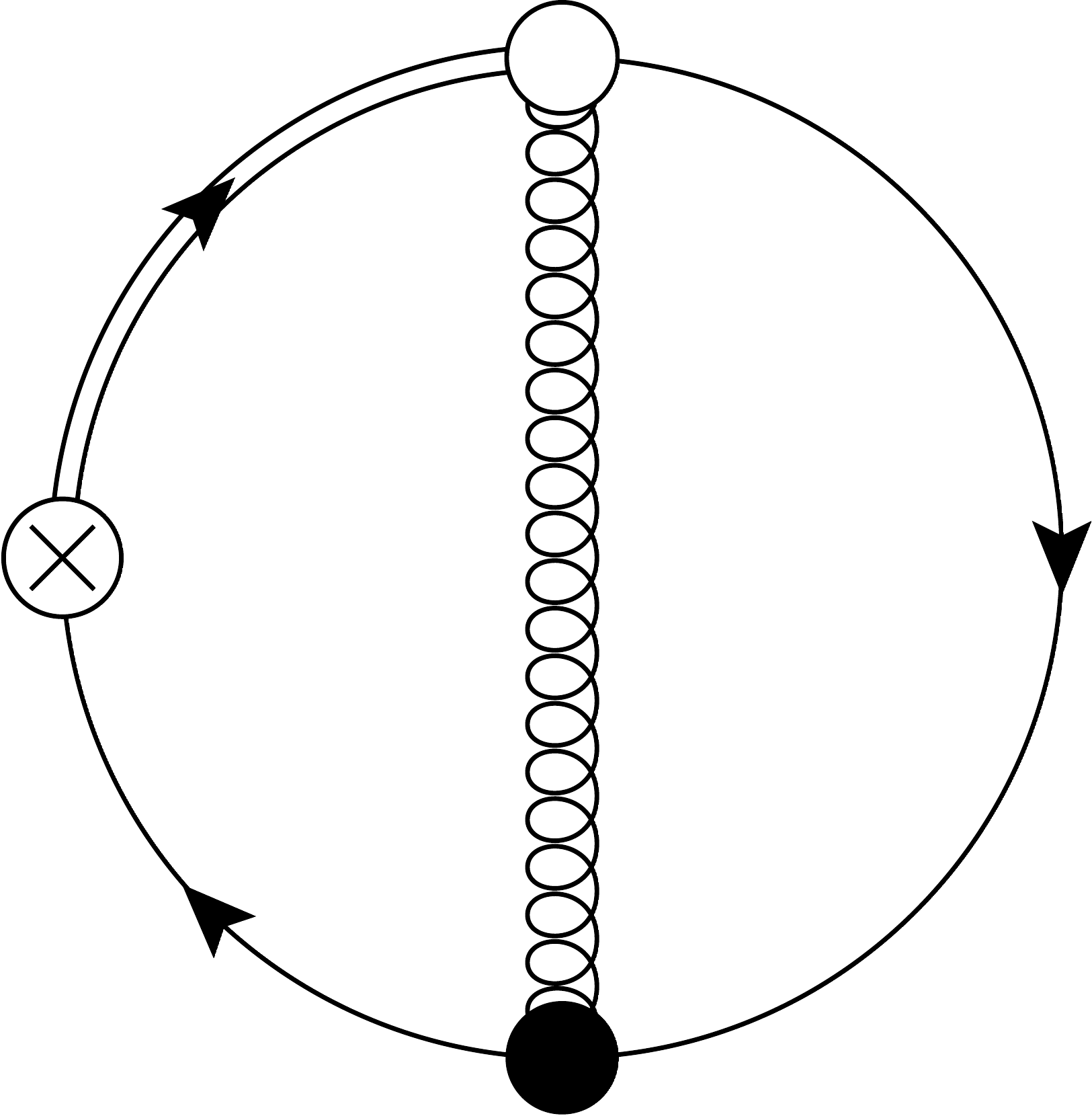}
		\caption*{D03}
		\label{fig:D03}
	\end{minipage}
	\begin{minipage}{.33\textwidth}
		\centering
		\includegraphics[width=.55\linewidth]{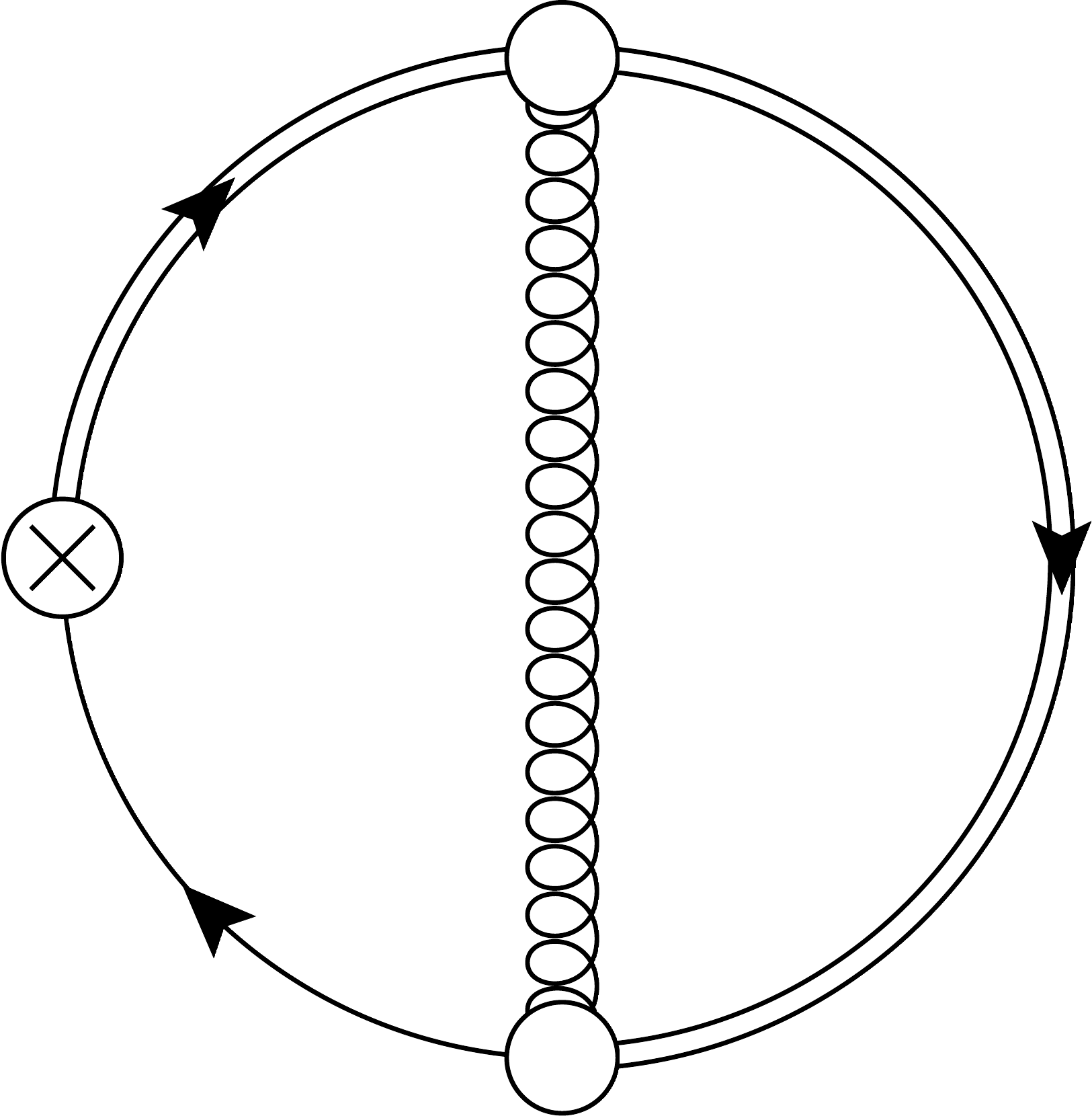}
		\caption*{D04}
		\label{fig:D04}
	\end{minipage}%
	\begin{minipage}{.33\textwidth}
		\centering
		\includegraphics[width=.55\linewidth]{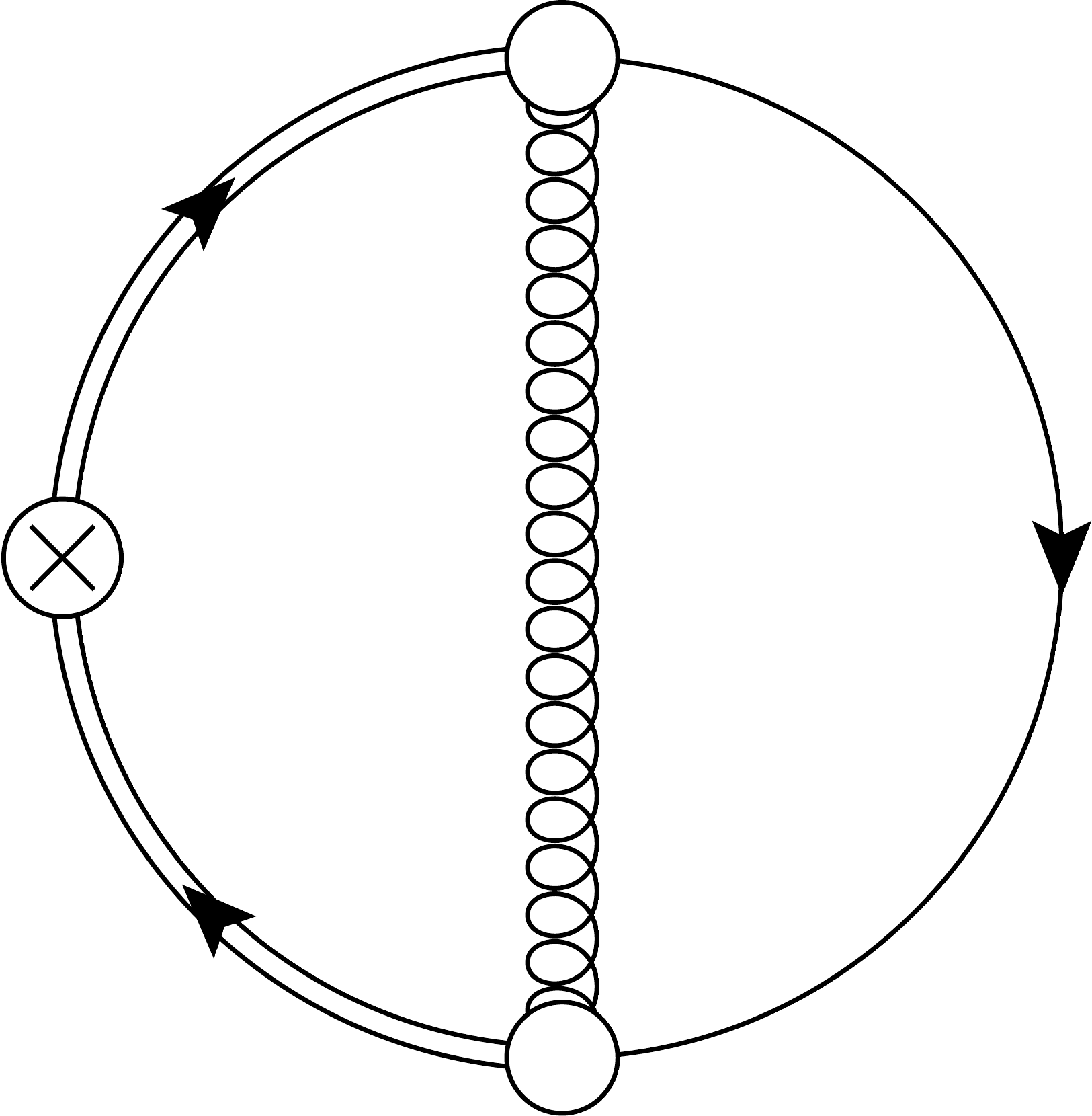}
		\caption*{D05}
		\label{fig:D05}
	\end{minipage}
	\begin{minipage}{.33\textwidth}
		\centering
		\includegraphics[width=.55\linewidth]{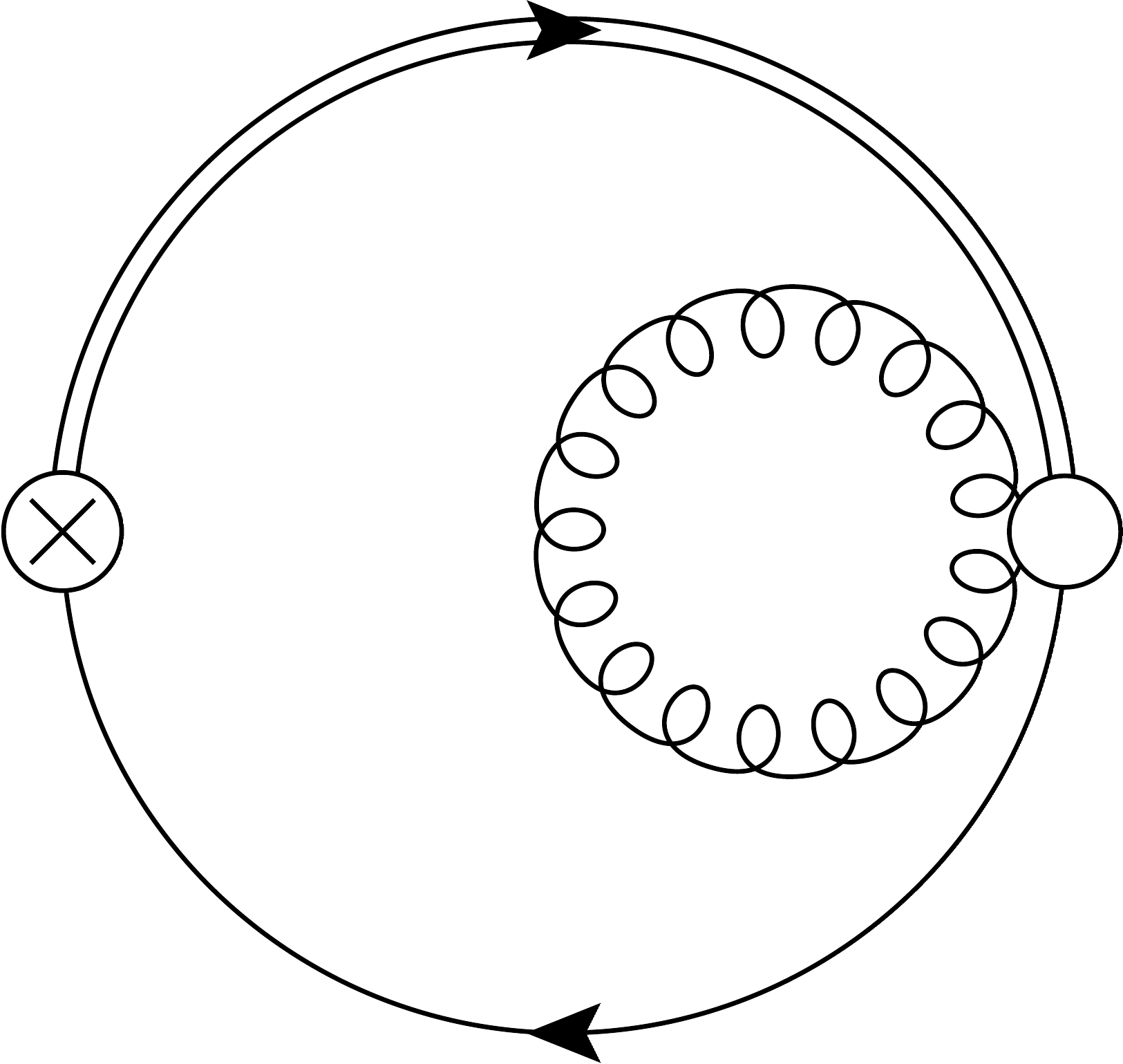}
		\caption*{D06}
		\label{fig:D06}
	\end{minipage}
\caption{Diagrams that contribute to the  1-point correlator $\braket{\del_\mu J^V_{\mu}(t,x)}$ at leading order (D01) and next-to-leading order (D02-D06). Notation is in App. \ref{sec:diagramnotation}.}
\label{fig:D01-D06}
\end{figure}

We verify Eq.~\eqref{eq:vanishing1point} diagrammatically up to $\order{g^2}$ to illustrate the coordinate space approach in the context of evolved correlators. The diagrams that contribute to $\braket{\del_\mu J^V_{\mu}(t,x)}$ to order ${g^2}$ are shown in Fig.~\ref{fig:D01-D06}\footnote{The naming of the diagrams corresponds to the one in \cite{MakinoSuzuki2014}.}, following the notation explained in App. \ref{sec:diagramnotation}. 
\subsection{Leading order, $\order{g^0}$}
The leading order contribution is given by diagram D01 in Fig.~\ref{fig:D01-D06} and reads:
\be
\begin{split}
\braket{\del_\mu J^V_{\mu,0}(t,x)}_0
&=
\int_y\, \dD(x-y) 
\langle \bar\chi_0(t,x)(\dsumxy)\chi_0(t,y)\rangle_0  \\
&=-\int_y\,\dD (x-y) \, \text{tr} [(\dsy +\dsx) S(\ybar_t-\xbar_t)] \\
&=-4\,d(R)\int_y\,\dD (x-y) \, \left(\dD(\xbar_t-\ybar_t) -\dD(\xbar_t-\ybar_t)\right)\\
&=0
\end{split}
\ee
where $d(R)$ is the dimension of the fermion representation $R$, and $\langle\ldots\rangle_0$ stands for the connected contribution to the path-integral average over the Euclidean free-theory measure. In the third line we used Eq.~\eqref{eq:derS}, which implies that the correlator vanishes at leading order. We work with dimensional regularization in $d=4-2\eps$ dimensions. 
\subsection{Next-to-leading order, $\order{g^2}$}
The next-to-leading order contribution is given by diagrams D02 through D06 in Fig.~\ref{fig:D01-D06}. Diagram D02 contains the insertion of two vertices from the  QCD action -- QCD vertices in short --, and is the only diagram present at $\order{g^2}$ in the nonevolved case, where it vanishes.
Diagram D03 contains the insertion of one QCD vertex, while D04-D06 do not contain QCD vertices. We show that each one of these five contributions vanishes separately.
The contribution from D02 reads:
\be
\begin{split}
D02&=\f{1}{2}\braket{\del_\mu J^V_{\mu,0}(t,x)  
\int_{z_1}(-g\bar\psi{\slashed A} \psi)\int_{z_2}(-g\bar\psi{\slashed A} \psi)
}{}_0\\
&= -\text{tr}(T^aT^a)  g^2\int_{z_1,z_2}
D(z_1-z_2)
\text{tr} \Big[ S(z_2-\xbar_t)     (\dsum) S(\xbar_t-z_1) \\
&\hspace{0.4 cm}\times
\gamma_\rho S(z_1-z_2)\gamma_\rho  \Big] \\
&=\text{tr}(T^aT^a)  (2-d)g^2
\int_{z_1,z_2}D(z_1-z_2)\dD(\xbar_t-z_1)\text{tr}\Big[ S(\xbar_t-z_2)   \\
& \hspace{0.4 cm}\times \left( S(z_1-z_2) +S(z_2-z_1)\right)  \Big] \\
&=0
\end{split}
\ee
where we employed $\gamma_\rho\gamma_\alpha\gamma_\rho=(2-d)\gamma_\alpha$,  the relation:
\be
S(z_2-x)(\dsum)S(x-z_1)=-\dD(x-z_1)S(x-z_2)-\dD(x-z_2)S(x-z_1)
\ee
and the symmetry relations $D(z_1-z_2)=D(z_2-z_1)$ and $S(z_1-z_2)=-S(z_2-z_1)$. 
The contribution from $D03$ reads:
\be
\begin{split}
D03&=g\braket{
\del_\mu J^V_{\mu,1}(t,x)  \int_z
(-g\bar\psi{\slashed A} \psi)}{}_0\\
&\hspace{-0.8 cm}=2g^2\text{tr}(T^aT^a)
\int_z\int_0^t ds\,
\text{tr} \Big[ \dsx\Big\{
\etsx \Big\{  D(\xbar_s-z)  S(\xbar_s-z)\overleftarrow{\slashed {\partial}}_x\Big\} S(z-\xbar_t)\Big\}\Big] \\
& \hspace{-0.4 cm}    -2g^2\text{tr}(T^aT^a)
\int_z\int_0^t ds
\,\text{tr} \Big[ \dsx\Big\{
\etsx \Big\{  D(\xbar_s-z)  \\
& \hspace{-0.4 cm}  \times \overrightarrow{\slashed {\partial}}_x S(z-\xbar_s)\Big\} S(\xbar_t-z)\Big\}\Big]\\
&\hspace{-0.8 cm}=2g^2\text{tr}(T^aT^a)
\int_z\int_0^t ds
\,\text{tr} \Big[ \dsx\Big\{
\etsx \Big\{  D(\xbar_s-z)  
 \Big(S(\xbar_s-z)\overleftarrow{\slashed {\partial}}_x
 \\
&\hspace{-0.4 cm}   +\overrightarrow{\slashed {\partial}}_x S(z-\xbar_s) \Big)
\Big\} S(z-\xbar_t)\Big\}\Big] \\
&\hspace{-0.8 cm}=0
\end{split}
\ee
where we employed Eq.~\eqref{eq:derS}. 
The final contribution is given by:
\be
D04+D05+D06=g^2\braket{\del_\mu J^V_{\mu,2}(t,x)}_0
\ee
By Eq.~\eqref{eq:cc}, one obtains:
\be
\begin{split}
D05&=g^2\braket{\bar\chi_1(t,x)(\dsum   )\chi_1(t,x)}_0\\
&=g^2\int_y \dD(x-y) 
\braket{\bar\chi_1(t,x)(\dsumxy)\chi_1(t,y)}_0  \\
&=4g^2
\int_y\dD (x-y) \int_0^t ds_1 \int_0^t ds_2\,
e^{(t-s_1)\Lap_x} e^{(t-s_2)\Lap_y}\Big\{
\text{tr}\Big[(\dsx+\dsy) \\
& \hspace{0.4 cm}  \times \Big\{ 
\partial_\mu^x\partial_\mu^y 
S(\ybar_{s_2}-\xbar_{s_1}) 
D(\xbar_{s_1}-\ybar_{s_2})    \Big\}\Big]\Big\}\\
&=4g^2
\int_y\dD (x-y) \int_0^t ds_1 \int_0^t ds_2\,
e^{(t-s_1)\Lap_x} e^{(t-s_2)\Lap_y}\Big\{ 
\text{tr}\Big[\partial_\mu^x\partial_\mu^y \\ 
& \hspace{0.4 cm}   \times (\dsx+\dsy) S(\ybar_{s_2}-\xbar_{s_1})\Big] D(\xbar_{s_1}-\ybar_{s_2})
\\ 
&\hspace{0.4 cm}  +\text{tr}\Big[(\dsx+\dsy)D(\xbar_{s_1}-\ybar_{s_2})
\partial_\mu^x\partial_\mu^y
S(\ybar_{s_2}-\xbar_{s_1})\Big] \Big\}\\
&=0
\end{split}
\ee
where 
in the last line we used $(\dsx+\dsy)S(\ybar_{s_2}-\xbar_{s_1})=0$ and $(\dsx+\dsy)D(\xbar_{s_1}-\ybar_{s_2})=-S(\xbar_{s_1}-\ybar_{s_2})+S(\xbar_{s_1}-\ybar_{s_2})=0$.
Finally\footnote{We use the notation $S(x-y)=\gamma_\mu S_\mu(x-y)$.}:
\be\label{eq:D04}
\begin{split}
D04+D06&=g^2\braket{\bar\chi_2(t,x)(\dsum)\chi_0(t,x)
+\bar\chi_0(t,x)(\dsum)\chi_2(t,x)}_0\\
& \hspace{-2.0 cm}=g^2
\int_y \dD(x-y) 
\braket{ \bar\chi_2(t,y)(\dsumxyi)\chi_0(t,x)+ 
\bar\chi_0(t,x)(\dsumxy)\chi_2(t,y)}_0  \\
&\hspace{-2.0 cm}= -4d\text{tr}(T^aT^a) g^2
\int_y\dD (x-y) \int_0^t ds\,e^{(t-s)\Lap_y} (\partial_\lambda^x+\partial_\lambda^y)
 \\
& \hspace{-1.6 cm}\times \Big\{ 
\int_z\dD (y-z) D(\ybar_s-\zbar_s)\left(S_\lambda(\ybar_s-\xbar_t)
+S_\lambda(\xbar_t-\ybar_s)\right)\Big\}\\
& \hspace{-1.6 cm}-16\text{tr}(T^aT^a)g^2
\int_y\dD (x-y) \int_0^t ds\,e^{(t-s)\Lap_y} (\partial_\lambda^x+\partial_\lambda^y)
 \\
&   \hspace{-1.6 cm}\times \Big\{ 
\int_z\dD (y-z)   \int_0^s du\,e^{(s-u)\Lap_y} \partial_\mu^y\Big\{ 
D(\ybar_u-\zbar_s)\partial_\mu^y (S_\lambda(\ybar_u-\xbar_t)\\
&   \hspace{-1.6 cm}+S_\lambda(\xbar_t-\ybar_u))\Big\}\Big\}\\
&\hspace{-1.6 cm}-16(2-d)g^2
\int_y\dD (x-y) \int_0^t ds\,e^{(t-s)\Lap_y} (\partial_\lambda^x+\partial_\lambda^y)\Big\{ 
\partial_\mu^yS_\lambda(y-x) \\
&  \hspace{-1.6 cm}\times \int_0^s du\,   e^{(s-u)\Lap_y}\Big\{\int_z\,\dD (y-z)\partial_\mu^zD(\ybar_u-\zbar_u)\Big\}\Big\}
\delta^{ab}\delta_{ij}\left[T^a,T^b\right]_{ij}\\
&\hspace{-2.0 cm}=0
\end{split}
\ee
where we employed the evolved fields in Eq.~\eqref{eq:chiandBexpansions}. 
 In the third equality of  Eq.~\eqref{eq:D04}, the first term 
comes from $D06$ and it vanishes because $S_\lambda(y-x)+S_\lambda(x-y)=0$, the second term is from D04 and vanishes for the same reason, the last term is also from D04 and vanishes since $\delta^{ab}\delta_{ij}\left[T^a,T^b\right]_{ij}=\text{tr}\big\{[T^a,T^a]\big\}=0$.

Hence, we have shown diagrammatically how the 1-point correlator of the divergence of the  evolved vector current vanishes to $\order{g^2}$, as anticipated in Eq.~\eqref{eq:vanishing1point} and in full analogy with the nonevolved case.
\section{The 2-point vector correlator in massless QCD-like theories}
\label{sec:CM}
We  introduce the  massless bare nonevolved 2-point vector correlator in Euclidean coordinate space:
\be\label{eq:PiV}
\Pi_{\mu\nu}^V(x-y)=\braket{ J_\mu^V(x)J_\nu^V(y)}
\ee
where $\langle\ldots\rangle$ defines the path-integral average in the Euclidean theory
 and  $J_\mu^V=\bar\psi\gamma_\mu\psi$ is the bare flavor singlet vector current.  
 For nonzero separations, contact terms do not occur in Eq.~\eqref{eq:PiV} and the correlator is multiplicatively renormalizable. 
\subsection{Solution of the Callan-Symanzik equation}
\label{sec:CS-NEgen}
 For later comparison with the gradient-flow evolved case, we recall the renor\-mal\-iza\-tion-group equation and its solution for the connected 2-point correlator of a general gauge invariant and multiplicatively renormalizable current of canonical energy dimension $D$. For simplicity, we consider the scalar case $\Pi =\braket{J(x)J(y)}_{conn}$ with, for example, 
 $J=\bar\psi\psi$. For correlators of pure-glue operators and further details of this construction see \cite{MBM,BPto}.
 
 Multiplicative renormalization in coordinate space at nonzero separation implies:
\be\label{eq:RvsB}
\Pi_R(x-y,\mu , g(\mu))=Z_J^2\left(\f{\Lambda}{\mu},g(\Lambda)  \right)\, \Pi(x-y,\Lambda , g(\Lambda))
\ee
where 
$\Pi_R$ and $\Pi$ are the renormalized and bare correlators, respectively, $Z_J$ is the multiplicative renormalization factor, $\Lambda$ is the ultraviolet cutoff in some regularization, $\mu$ is the renormalization scale and $g$ is the Yang-Mills running coupling. 
Based on dimensional considerations, we can rewrite $\Pi_R$ in the massless theory in terms of the dimensionless 2-point correlator $\overline{\Pi}_R$ as\footnote{For higher spin multiplicatively renormalizable currents Eq.~\eqref{eq:PiR} acquires a nontrivial Lorentz structure.}:
\be\label{eq:PiR}
\Pi_R(x-y,\mu , g(\mu))=\f{1}{|x-y|^{2D}}\overline{\Pi}_R(|x-y|\mu, g(\mu))
\ee
Then the Callan-Symanzik equation, i.e., the statement that the bare correlator is $\mu$ independent:
\be
\f{d}{d\log\mu} \Pi\vert_{\Lambda,g(\Lambda)} =0
\ee
translates into the following equation for $\overline{\Pi}_R$:
\be\label{eq:CS}
\Big(\f{\partial}{\partial\log\mu}+\beta(g)\f{\partial}{\partial g} +2\gamma_J(g)     \Big) \overline{\Pi}_R(|x-y|\mu, g(\mu))=0
\ee
where $\gamma_J$ is the anomalous dimension of $J$, and $\beta(g)$ is the QCD beta function:
\be\label{eq:gammaV}
\gamma_J(g)=-\f{d\log Z_J}{d\log\mu}\vert_{\Lambda,g(\Lambda)}~~~~~
\beta(g)=\f{d g}{d\log\mu}\vert_{\Lambda,g(\Lambda)}
\ee
Since $\overline{\Pi}_R$  depends on $(x-y)$ only through the dimensionless product $(|x-y|\mu)$, Eq.~\eqref{eq:CS} also reads:
\be\label{eq:CSx}
\Big(\f{\partial}{\partial\log |x-y|}+\beta(g)\f{\partial}{\partial g} +2\gamma_J(g)     \Big) \overline{\Pi}_R(|x-y|\mu, g(\mu))=0
\ee
The structure of Eq.~\eqref{eq:CSx} now implies that the dimensionful correlator in Eq.~\eqref{eq:PiR} factorizes as follows:
\be\label{eq:CSsol}
\Pi_R(x-y,\mu , g(\mu))=\f{1}{\left|x-y\right|^{2D}} F(g(|x-y|)) Z_J^2(g(|x-y|), g(\mu))
\ee
where $F$ is a dimensionless renormalization-group invariant (RGI) function of the RGI running coupling $g(|x-y|)$ -- therefore not determined by Eq.~\eqref{eq:CSx} --  and $Z_J^2$ is determined by integrating  Eq.~\eqref{eq:CSx} between a reference scale $\mu^{-1}$ and $|x-y|$:
\be\label{eq:Z}
\begin{split}
Z_J^2(g(|x-y|), g(\mu))&= \exp\int_{g(\mu)}^{g(|x-y|)}\,2\f{\gamma_J(g)}{\beta(g)} dg\\
&= \Big(\f{ g^2(|x-y|) }{g^2(\mu) }\Big)^{\f{\gamma_{J}^{(0)}}{\beta_0}} \,e^{\f{\beta_0\gamma_{J}^{(1)}  -\beta_1\gamma_{J}^{(0)}  }{\beta_0^{2}} (g^2(|x-y|)-g^2(\mu) ) +\ldots   }\\
&=(1+g^2(\mu)2\beta_0\log{(|x-y|\mu)}+\ldots)^{\f{\gamma_{J}^{(0)}     }{\beta_0}}\,e^{\order{g^2(|x-y|)-g^2(\mu)}}\\
&=(1+g^2(\mu) 2\gamma_{J}^{(0)} \log{(|x-y|\mu)}+\ldots)\,
e^{\order{g^2(|x-y|)-g^2(\mu)}}
\end{split} 
\ee
where we introduced the perturbative expansions:
\be
\begin{split}
\gamma_{J}(g)&=-\gamma_{J}^{(0)}g^2-\gamma_{J}^{(1)}g^4 +\order{g^6}\\
\beta(g)&=-\beta_0 g^3-\beta_1g^5+\order{g^7}
\end{split}
\ee
with $\beta_0$ and $\beta_1$ the universal, i.e., renormalization-scheme independent one- and two-loop coefficients of the QCD beta function, respectively. 

Importantly, the second line of Eq.~\eqref{eq:Z} determines the universal UV asymptotics of $Z_J^2$ as $g(|x-y|)\to 0$:
\be\label{eq:ZUV}
Z_J^2(g(|x-y|), g(\mu))\sim \Big(\f{ g^2(|x-y|) }{g^2(\mu) }\Big)^{\f{\gamma_{J}^{(0)}}{\beta_0}}
\,e^{\order{g^2(\mu)} }
\ee
Furthermore, the third and fourth line of Eq.~\eqref{eq:Z} contain the perturbative expansion of $Z_J^2$ by means of the perturbative expansion of $g^2(z)$ \cite{MBM}:
\be\label{eq:RGIg}
g^2(z)=g^2(\mu) \left(1+g^2(\mu) 2\beta_0\log(z\mu)+g^4(\mu)(2\beta_1\log(z\mu)+4\beta_0^2\log^2(z\mu))+\order{g^6}   \right)
\ee
with $z=|x-y|$, in terms of $g^2(\mu)$ to order $g^6(\mu)$, and 
valid for scales $\mu^{-1}$ and $z$ close to zero and $\order{1}$ logarithms $\log(z\mu)$. For a proof of Eq.~\eqref{eq:RGIg} see \cite{BPto}.

 For massless fermions, the validity of the above analysis extends with appropriate account of the Lorentz structure to the connected 2-point correlators of all flavor singlet and non-singlet fermion bilinear currents -- scalar, pseudoscalar, vector, axial and tensor\footnote{A detailed analysis of the multiplicative renormalization properties of flavor singlet and non-singlet axial and pseudoscalar currents and the role of the axial anomaly can be found in \cite{Larin} (see also refs. therein).}.

In the case of the 2-point vector correlator in Eq.~\eqref{eq:PiV},  the well known all-order result $Z_J=1$, hence $\gamma_J=0$ through Eq. \eqref{eq:gammaV}, is a consequence of the conservation of the flavor singlet vector current. 
It entails the renormalization-group invariance of the correlator $\Pi_{R,\mu\nu}^V =\Pi_{\mu\nu}^V$ and that of the vector current $J_{R,\mu}^V=J_\mu^V$.

An important observation is that, in order to fulfill these properties and at difference with the scalar case, the Lorentz structure of the vector correlator starts to change at $\order{g^4(\mu)}$ in perturbation theory. This was explicitly shown in \cite{CM}. Therefore, in this case we replace the all-order solution in  Eq.~\eqref{eq:CSsol} with:
\be\label{eq:LS}
\Pi^V_{R,\mn}(x-y, \mu, g(\mu)) =\f{1}{(x-y)^{2D}}\sum_{n=0}^\infty\, g^{2n}(|x-y|)\Big(A_{n}\f{\delta_\mn}{2}-B_n\f{(x-y)_\mu (x-y)_\nu}{(x-y)^2}\Big)
\ee
in terms of dimensionless coefficients $A_n$ and $B_n$. The leading ($n=0$) and next-to-leading ($n=1$) orders have $A_n=B_n$. 
\subsection{ Leading order, $\order{g^0}$ and next-to-leading order, $\order{g^2}$}
We now proceed to review the perturbative expression for the correlator $\Pi_{R,\mu\nu}^V$  up to next-to-leading order $\order{g^2}$. The calculation up to $\order{g^8}$ for $N=3$ and fermions in the fundamental representation can be found in \cite{CM}.

As also for the evolved case in Sec. \ref{sec:2point}, we present our results in the most general case, i.e., for $N$ colors and $N_f$ Dirac flavors in a representation $R$. With anti-hermitian $SU(N)$ generators $T^a$ in the representation $R$ one has:
\be 
\textrm{tr}(T^aT^b)=-T(R)\delta^{ab},~~~~~~\textrm{tr}(T^aT^a)=-C_2(R)d(R)=-T(R)d(G)
\ee
with $T(R)$, $C_2(R)$ and $d(R)$ the index, Casimir and dimension of the representation $R$, respectively, and $d(G)=N^2-1$ is the dimension of the adjoint representation $G$ of $su(N)$.
 
Our results will be expressed in terms of the dimension $d(R)$ and Casimir $C_2(R)$ of the fermion representation $R$. We recall that in the fundamental representation, $d(\textrm{Fund})=N$, $C_2(\textrm{Fund})=(N^2-1)/(2N)$ and $T(\textrm{Fund})=1/2$. 

\begin{figure}[tb]
	\centering
	\begin{minipage}{.33\textwidth}
		\centering
		\includegraphics[width=.6\linewidth]{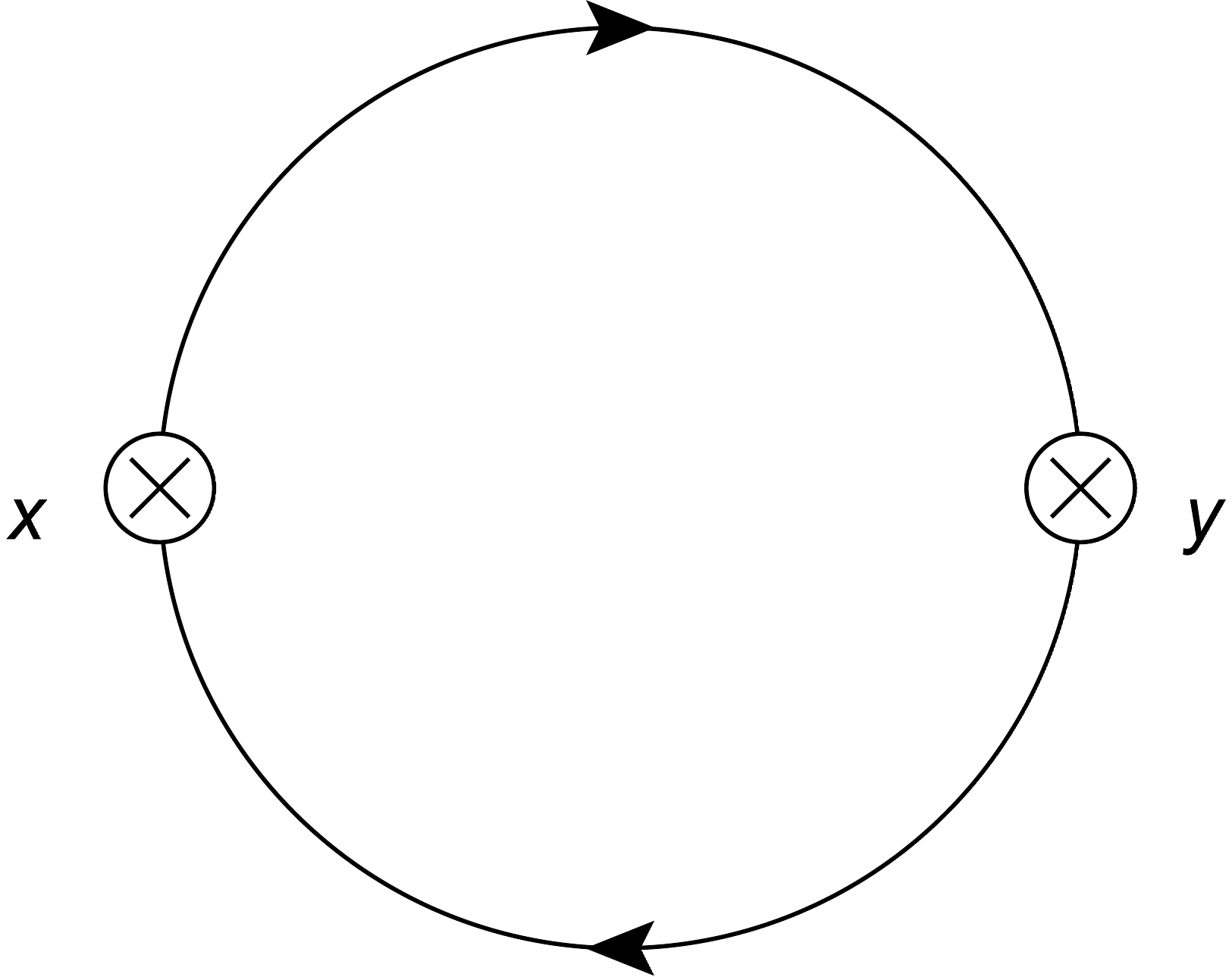}
		\label{fig:LOnonevolved}
	\end{minipage}
	\begin{minipage}{.33\textwidth}
		\centering
		\includegraphics[width=.6\linewidth]{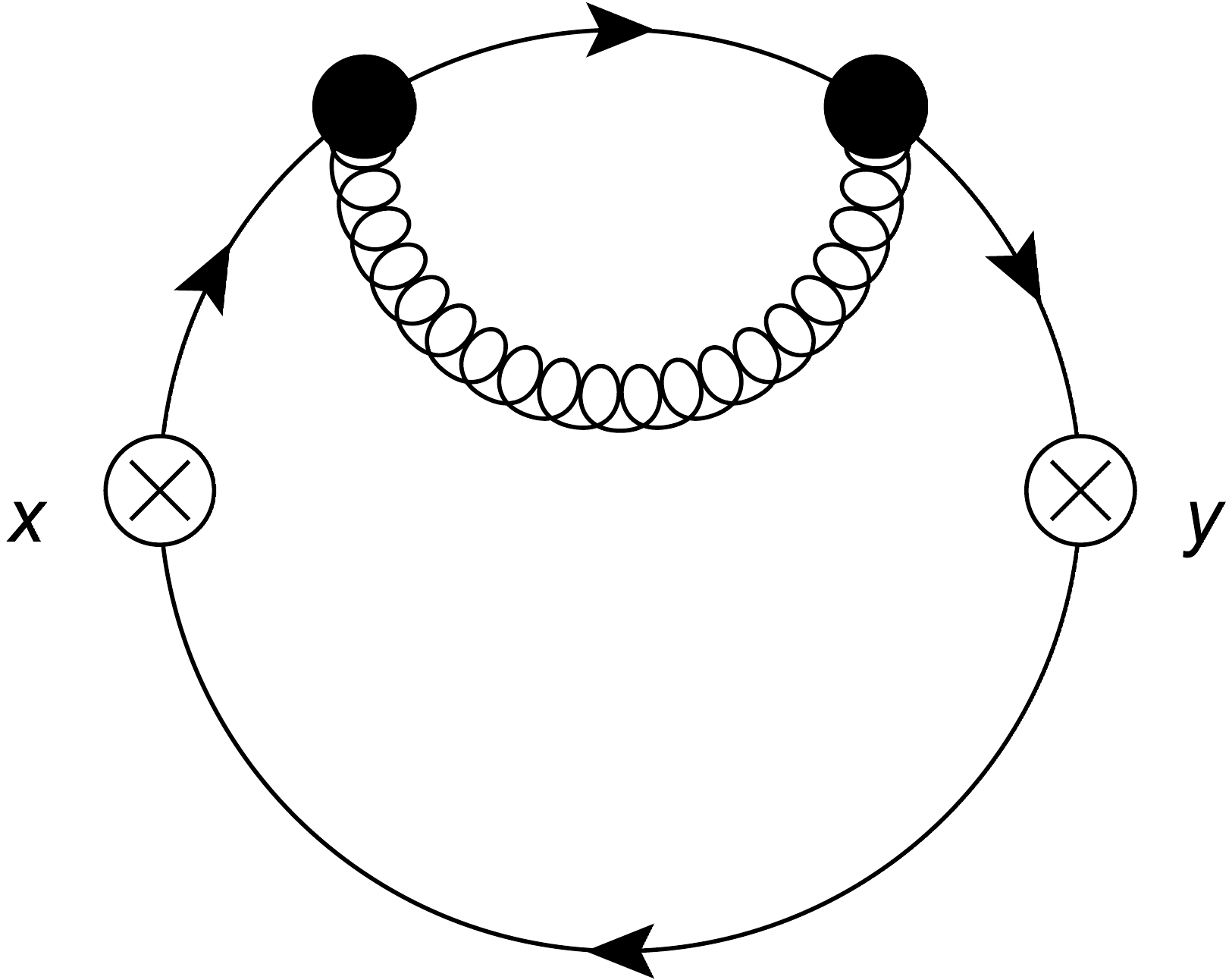}
		\label{fig:NLO1nonevolved}
	\end{minipage}%
	\begin{minipage}{.33\textwidth}
		\centering
		\includegraphics[width=.6\linewidth]{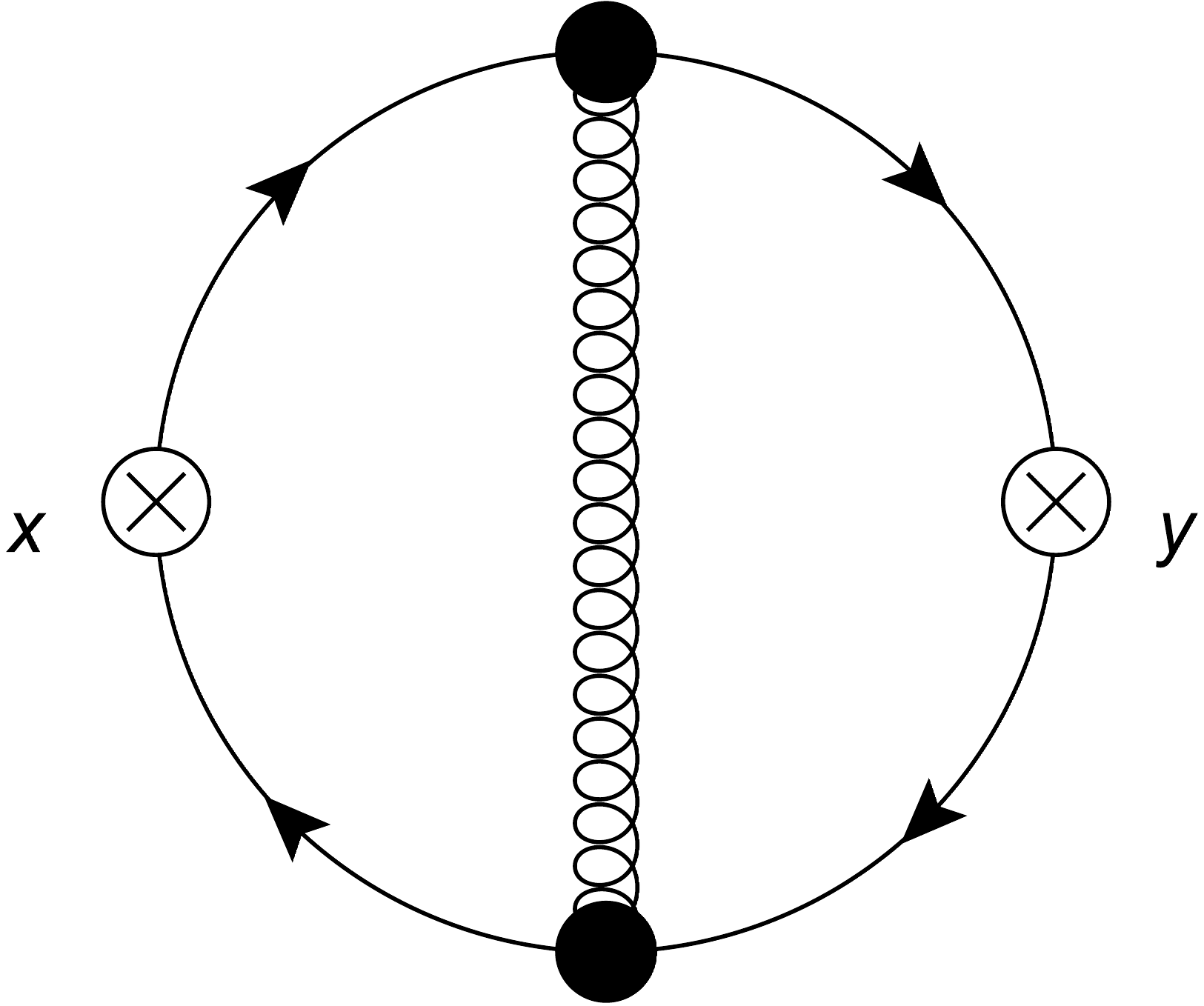}
		\label{fig:NLO2nonevolved}
	\end{minipage}
\caption{Contributions to $\Pi_{\mu\nu}^V(x-y)$ at $\order{g^0}$ (left) and $\order{g^2}$ (centre, right).}
\label{fig:2pointnonevolved}
\end{figure}

The leading order result, from the left diagram in Fig.~\ref{fig:2pointnonevolved}, is:
\begin{equation}\label{eq:nonevolved2pointLO}
\begin{split}
\Pi_{\mu\nu}^V(x-y)&=-d(R)\,\text{tr}\left[\gmu S(x-y)\gnu S(y-x)\right]+\order{g^2}\\
&=-\frac{2\,d(R)}{\pi^4}\frac{1}{\left((x-y)^2\right)^3}\left(\frac{\delta_\mn}{2} -\frac{(x-y)_\mu (x-y)_\nu}{(x-y)^2} \right)+\order{g^2}
\end{split}
\end{equation}
The next-to-leading order contributions are associated to the last two diagrams in Fig.~\ref{fig:2pointnonevolved}. 
 The UV divergences of both diagrams exactly cancel each other\footnote{One can verify the exact cancellation of the short-distance divergences in coordinate space following a known method nicely explained in \cite{Collins}, chapter 11.}, so that the next-to-leading order result is  renormalization-group invariant.
Indeed, we have verified that the results reported in \cite{CM} and ancillary files\footnote{\url{http://www-ttp.particle.uni-karlsruhe.de/Progdata/ttp10/ttp10-42/}} for the vector correlator in coordinate space with $N=3$ and fermions in the fundamental, up to and including $\order{g^6(\mu)}$, can be rewritten in terms of $g(|x-y|)$ in Eq.~\eqref{eq:RGIg} only. 

For comparison with the evolved case, our generalized result to order $g^2(\mu)$ reads:
\be\label{eq:nnlo}
\begin{split}
\Pi_{\mu\nu}^V(x-y)&=-\frac{2\,d(R)}{\pi^4}\frac{1}{((x-y)^2)^3}\Big\{ \Big(\frac{\delta_\mn}{2} -\frac{(x-y)_\mu (x-y)_\nu}{(x-y)^2}\Big) \\
&\times \Big( 1+3C_2(R)\f{g^2(\mu)}{(4\pi)^2}\Big)+\order{g^4(\mu)}
 \Big\}
\end{split}
\ee
It manifestly satisfies the transversality condition: 
\begin{equation}\label{eq:trans2}
\del_\mu^x \Pi_{\mu\nu}^V(x-y)=
\del_\mu^y \Pi_{\mu\nu}^V(x-y)
=0
\end{equation}
to order $g^2(\mu)$. 
Equation~\eqref{eq:trans2} is equivalent to the statement that the corresponding current is conserved and not renormalized. 

At order $g^4(\mu)$ and higher, transversality forces the Lorentz structure of the vector correlator to change. We can understand it as follows. 
The 2-point correlator of the conserved vector current does not renormalize, thus only the RGI coupling $g(|x-y|)$
induces a coordinate dependence beyond the one of the leading order correlator. This dependence starts at order $g^4(\mu)$, as implied by Eq.~\eqref{eq:RGIg}. Hence, the Lorentz structure of the vector correlator must change at order $g^4(\mu)$ in order to still guarantee transversality. This applies iteratively to higher orders. 

For the 2-point correlators of possibly higher spin nonconserved currents, where operator mixing may or may not occur,  a change of the Lorentz structure at higher orders is in general allowed. Interestingly, the 2-point correlator of the multiplicatively renormalizable nonconserved tensor current $\bar\psi\sigma_\mn\psi$ does not change its Lorentz structure to the orders computed in 
\cite{CM}, i.e., to $\order{g^6}$.
\section{The 2-point vector correlator evolved by the gradient flow }
\label{sec:2point}
We now consider the Euclidean 2-point correlator of flavor singlet vector currents 
 evolved to the same flow time $\sqrt{t}$.  Analogously to Eq. \eqref{eq:Jseries}, we write the bare correlator as follows: 
\begin{equation}
\Pi_{\mn}^V(t,x-y)=\braket{J^V_\mu(t,x)J^V_\nu(t,y)}=\sum_{n=0}^\infty g^{2n}\Pi_{\mn,2n}^V(t,x-y)
\end{equation}
The leading order contribution reads:
\begin{equation}
\Pi_{\mn,0}^V(t,x-y)=\braket{J^V_{\mu,0}(t,x)J^V_{\nu,0}(t,y)}_0
\end{equation}
with $J^V_{\mu,0}$ in Eq. \eqref{eq:cc}. $\Pi_{\mn,0}^V$
is derived in Sec. \ref{sec:GF2pointLO}. 

The next-to-leading order contribution is derived in Sec. \ref{sec:NLO} and can be conveniently divided into four terms:
\begin{equation}\label{eq:NLO_tot}
\Pi^V_{\mn,2}(t,x-y)=\text{I}+\text{II}+\text{III}+\text{IV}
\end{equation}
where
\begin{equation}\label{eq:types}
\begin{split}
\text{I}&= \f{1}{2}\braket{J^V_{\mu,0}(t,x) J^V_{\nu,0}(t,y)  
\int_{z_1}
(-\bar\psi{\slashed A} \psi)
\int_{z_2} (-\bar\psi{\slashed A} \psi)
}{}_0
\\
\text{II}&=\braket{J^V_{\mu,1}(t,x)J^V_{\nu,0}(t,y)
\int_z
( -\bar\psi{\slashed A} \psi )
}{}_0
+(x\leftrightarrow y;\mu\leftrightarrow\nu)\\
\text{III}&=\braket{J^V_{\mu,2}(t,x)J^V_{\nu,0}(t,y)}_0+(x\leftrightarrow y;\mu\leftrightarrow\nu)\\
\text{IV}&=\braket{J^V_{\mu,1}(t,x)J^V_{\nu,1}(t,y)}_0
\end{split}
\end{equation}
with the currents given in Eq.~\eqref{eq:cc}. 
The first two classes of diagrams, I and II, contain respectively the insertion of two QCD vertices and one QCD vertex. 

Our aim is now to establish the renormalization properties of the evolved $\Pi_\mn^V$, and how it differs from the nonevolved case where it does not renormalize.  Therefore, when deriving the next-to-leading order contribution in Sec. \ref{sec:NLO}, we will concentrate on the divergent parts only. 
The diagrammatic notation for Secs. \ref{sec:GF2pointLO} and \ref{sec:NLO} is explained in App. \ref{sec:diagramnotation}. 
\subsection{Leading order, $\order{g^0}$}\label{sec:GF2pointLO}
\begin{figure}[tb]
\centering
\includegraphics[width=.25\textwidth]{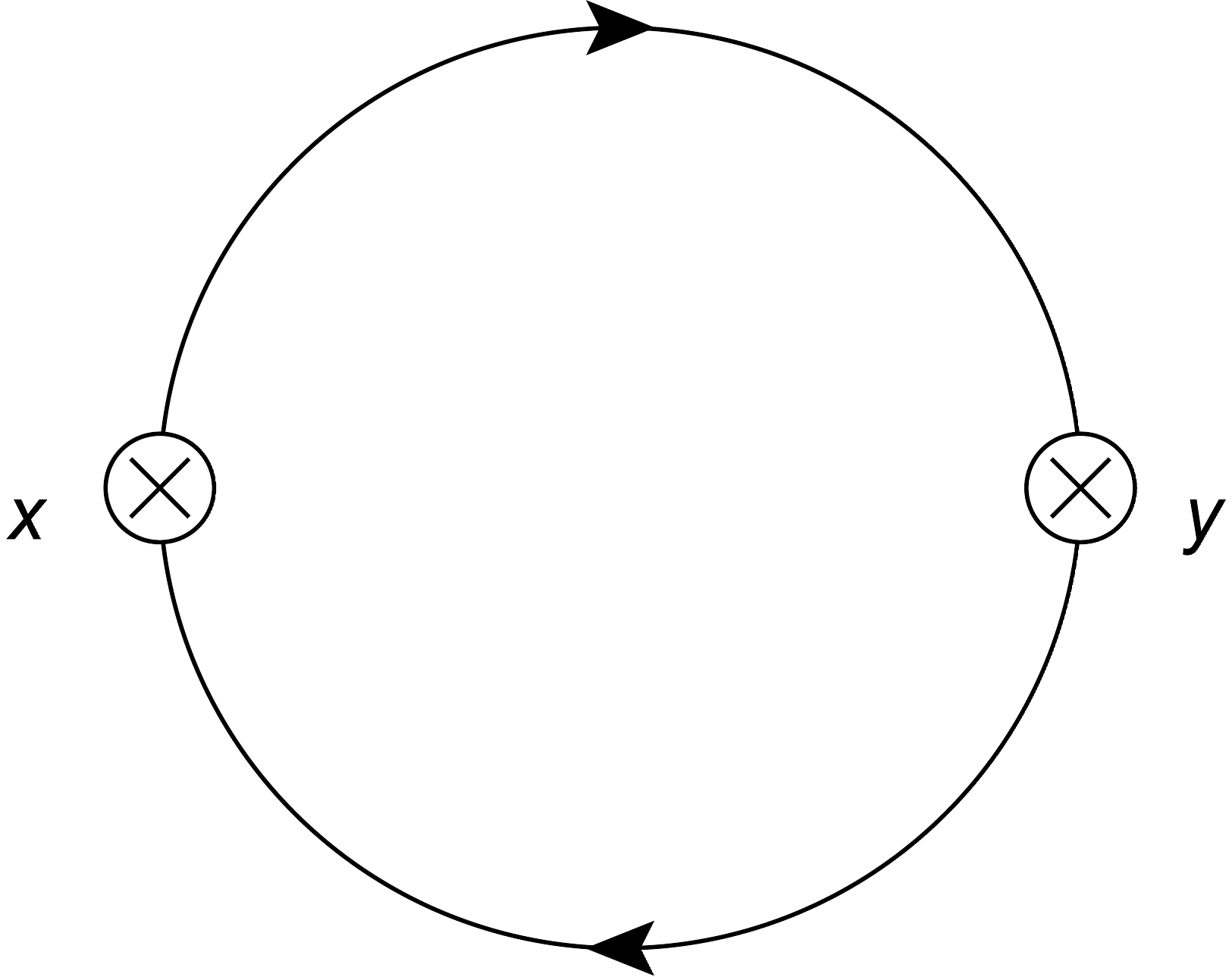}
\caption{Leading order diagram for the gradient-flow evolved 2-point vector correlator. Lines are gradient-flow evolved propagators -- see notation in App. \ref{sec:diagramnotation}. }
\label{fig:2ptLO}
\end{figure}

The leading order contribution is given by: 
\begin{equation}\label{eq:evolved2pointLO}
\begin{split}
\Pi_{\mn,0}^V(t,x-y)
&=-d(R)\,\text{tr}\big[\gmu S(\xbar_t-\ybar_t)\gnu S(\ybar_t-\xbar_t)\big]\\
&=-\frac{2\,d(R)}{\pi^4}\frac{\gamma\big(2,\frac{(x-y)^2}{8t}\big)^2}{\big((x-y)^2\big)^3}\Big(\frac{\delta_\mn}{2} -\frac{(x-y)_\mu (x-y)_\nu}{(x-y)^2} \Big)
\end{split}
\end{equation}
from the diagram in Fig.~\ref{fig:2ptLO}, and  $S(\xbar_t-\ybar_t)$ is the evolved fermion propagator in Eq. \eqref{eq:propf}.
The comparison of Eq.~\eqref{eq:evolved2pointLO} with the nonevolved result in Eq.~\eqref{eq:nonevolved2pointLO} shows that  the gradient-flow evolution replaces $\Gamma(\frac{d}{2})$ in $d$ spacetime dimensions with the lower incomplete gamma function $\gamma(\frac{d}{2},\frac{(x-y)^2}{8t})$ introduced in Sec. \ref{sec:prop}. This has the following implications. 

As $t\to 0$ at fixed nonzero separation, one recovers the nonevolved result in Eq.~\eqref{eq:nonevolved2pointLO} up to exponentially suppressed contributions:
\begin{equation}\label{eq:PVlimtsmall}
\Pi^V_{\mn,0}(t,x-y)\Big|_{t\sim 0}\hspace{-0.3 cm}= \Pi^V_{\mn,0}(x-y)\Big(1+\mathcal{O}\Big(\f{(x-y)^2}{t} e^{-\f{(x-y)^2}{t} }    \Big) \Big)
\end{equation}
where $\Pi^V_{\mn,0}(x-y)$ is the nonevolved leading order contribution in Eq.~\eqref{eq:nonevolved2pointLO}, and we employed $\gamma(2,z)=1-e^{-z}(1+z)\sim 1-ze^{-z}$ for $z$ large\footnote{$\gamma(n+1,z)=n!(1-e^{-z}e_n(z))$, with $e_n(z)=\sum_{k=0}^n z^k/k!$, for $n=0,1,2,\ldots$.}.

Thus, the spacetime derivative of the small $t$ expansion in Eq.~\eqref{eq:PVlimtsmall} yields:
\begin{equation}\label{eq:notrans-smallt}
\del_\mu^x \, \Big(\Pi^V_{\mn,0}(t,x-y)\Big|_{t\sim 0}\Big)= \Pi^V_{\mn,0}(x-y) \f{x_\mu-y_\mu}{t}\mathcal{O}\Big(\f{(x-y)^2}{t} e^{-\f{(x-y)^2}{t} }  \Big) 
\end{equation}
showing that the small $t$ evolution induces an exponentially soft violation of transversality.

We recall that in the nonevolved case the transversality condition  is respected up to contact terms. In fact, in the evolved case, the spacetime derivative of  Eq.~\eqref{eq:evolved2pointLO} reads:
\begin{equation}\label{eq:notrans}
\begin{split}
\del_\mu^x \, \Pi^V_{\mn,0}(t,x-y)
&=-d(R)\, \del_\mu^x\Big\{ \text{tr}\big[\gmu S(\xbar_t-\ybar_t)\gnu S(\ybar_t-\xbar_t)\big]\Big\}\\ 
&\hspace{-1.0cm}=2\,d(R)\, K_{2t}(x-y)\,\text{tr}\big[\gnu S(\xbar_t-\ybar_t)\big]\\ 
&\hspace{-1.0cm}=\f{4\,d(R)}{\pi^2}\, K_{2t}(x-y)\,\frac{({x}_\nu-{y}_\nu)}{((x-y)^2)^{2}} \gamma\Big(2,\frac{(x-y)^2}{8t}\Big)\\
&\hspace{-1.0cm}=-\f{16\,d(R)}{\pi^2}\, \partial_\nu^x K_{2t}(x-y)
\,\frac{t}{((x-y)^2)^{2}} \gamma\Big(2,\frac{(x-y)^2}{8t}\Big)\\
&\hspace{-1.0cm}=-\f{16\,d(R)}{\pi^2}\,
\partial_\nu^x \Big( e^{2t\Lap_x} \delta^{(4)}(x-y)\Big)
\,\frac{t}{((x-y)^2)^{2}} \gamma\Big(2,\frac{(x-y)^2}{8t}\Big)
\end{split}
\end{equation}
showing that the violation of transversality at a generic fixed $t$ is proportional to the derivative of a smeared Dirac delta distribution, i.e., a contact term smeared by the gradient flow. 
For each equality in Eq.~\eqref{eq:notrans} we employed in order Eq.~\eqref{eq:evolved2pointLO}, Eq.~\eqref{eq:dSGF}, 
Eq.~\eqref{eq:propf} for the evolved fermion propagator, 
and finally Eqs.~\eqref{eq:kernel} and \eqref{eq:K} for the scalar kernel $K_{2t}$. 

We may also consider the alternative limit of vanishing separation $(x-y)\to 0$ at fixed flow time. 
In this case, the expansion in Eq.~\eqref{eq:incomplgammaexpansion} inserted in Eq.~\eqref{eq:evolved2pointLO} yields for the 2-point correlator:
\begin{equation}\label{eq:PVlimxy}
\Pi^V_{\mn,0}(t,x-y)\Big|_{x\sim y}\hspace{-0.3 cm}= -\frac{d(R)}{2\pi^4}\frac{(x-y)^2}{(8t)^4}\Big(\frac{\delta_\mn}{2} -\frac{(x-y)_\mu (x-y)_\nu}{(x-y)^2} \Big)\Big(1+\mathcal{O}\Big(\f{(x-y)^2}{t}\Big)\Big)
\end{equation}
showing that it vanishes at zero spacetime separation, i.e., 
the short-distance singularity has been removed by the gradient flow. 

This exercise makes manifest  the noncommutativity of the two limits $t\to 0$ and $(x-y)\to 0$ at the level of 2-point correlators\footnote{Of course, the evolved correlator vanishes asymptotically for spacetime separations much larger than the smearing radius $\sqrt{t}$, hence, transversality is trivially recovered asymptotically in this limit.}.
\subsection{Next-to-leading order, $\order{g^2}$}\label{sec:NLO}
We treat separately the four contributions in Eq.~\eqref{eq:types}.
\begin{figure}[tb]
\centering
\begin{minipage}{.35\textwidth}
	\centering
	\includegraphics[width=.6\linewidth]{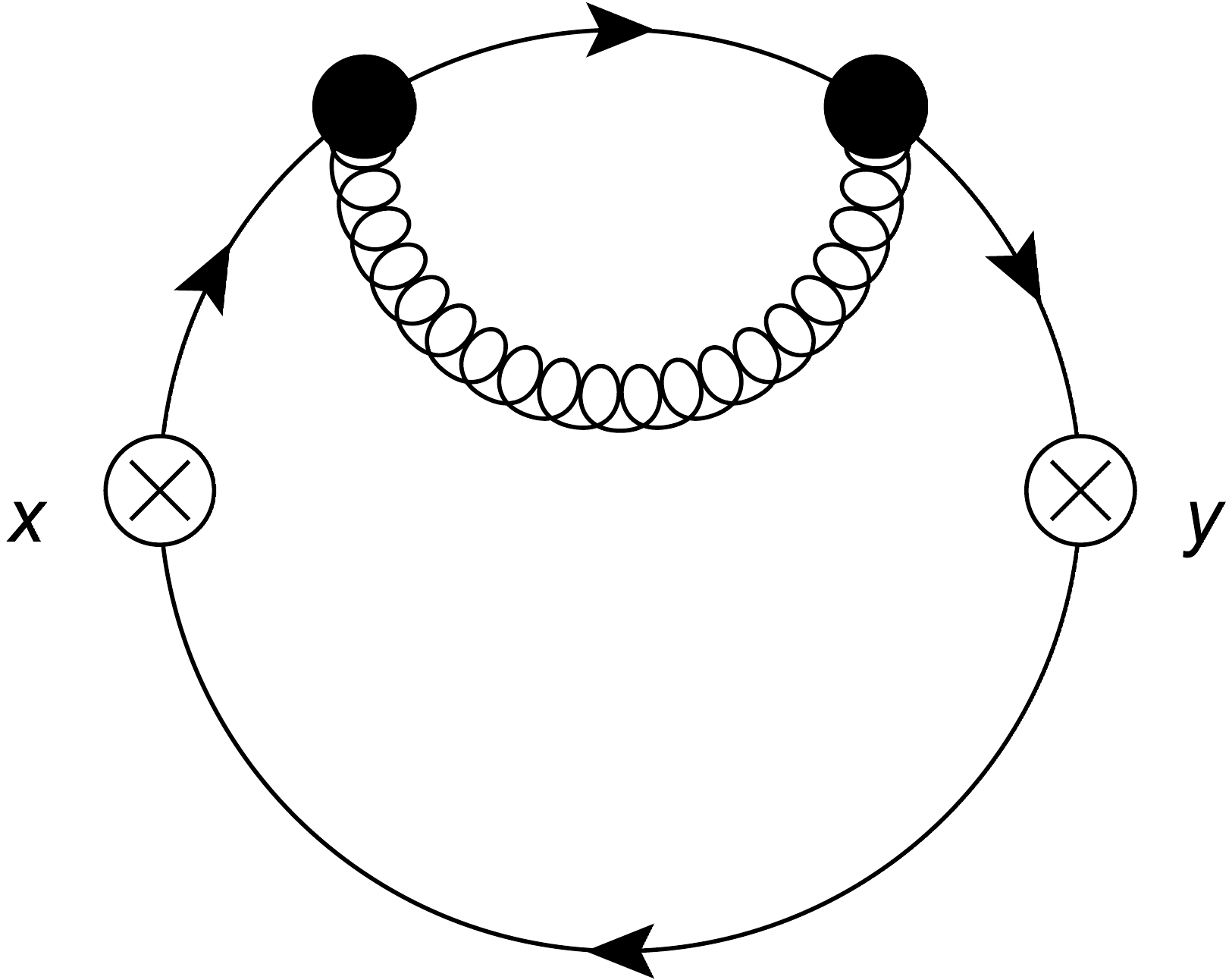}
	\caption*{(I.1)}
\end{minipage}%
\begin{minipage}{.35\textwidth}
	\centering
	\includegraphics[width=.6\linewidth]{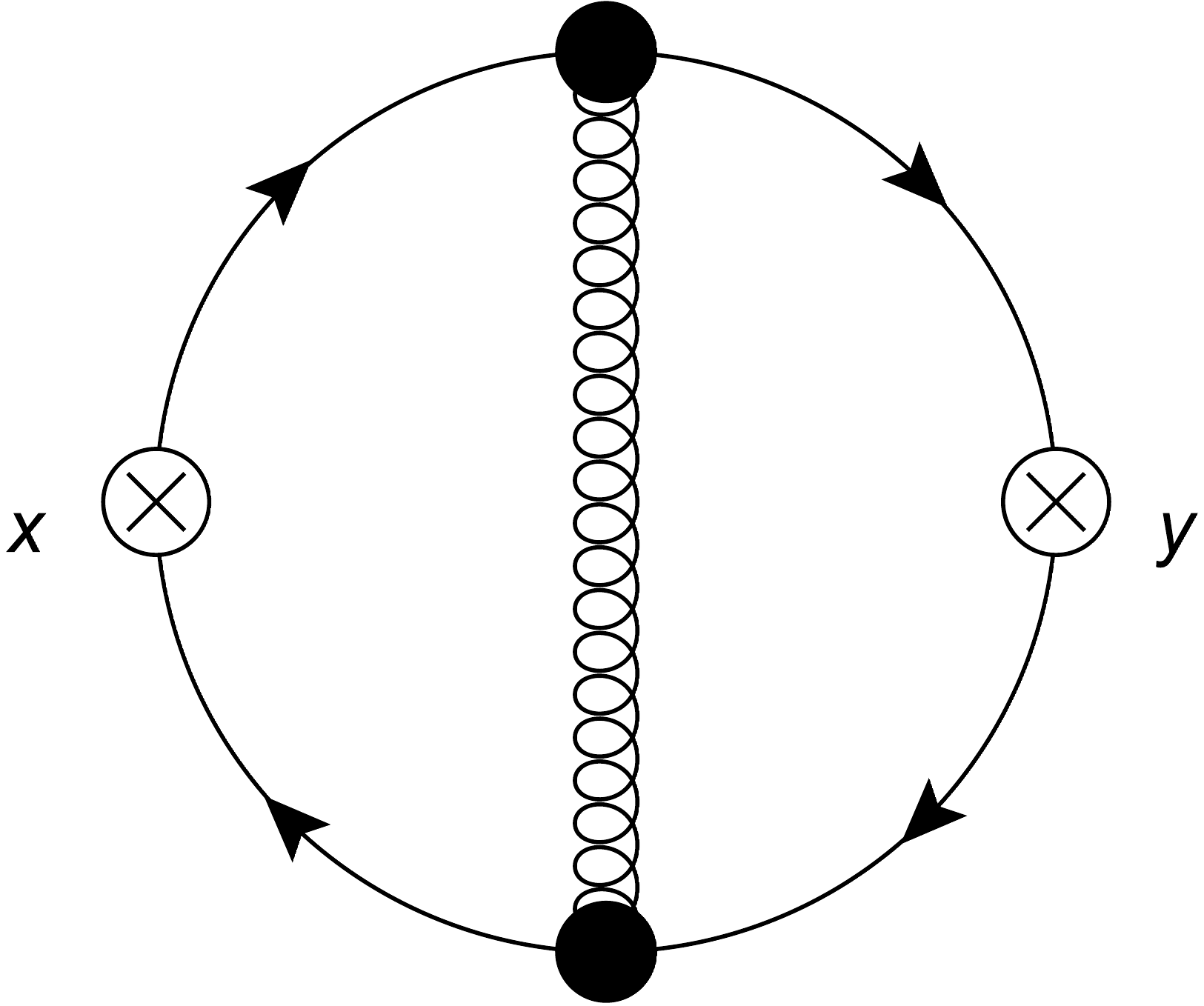}
	\caption*{(I.2)}
\end{minipage}
\caption{Type I diagrams for the evolved 2-point vector correlator at $\order{g^2}$.}
\label{fig:classI}
\end{figure}
\subsubsection{Type I contribution}
\label{sec:NLOI}
The first contribution in Eq.~\eqref{eq:types} is associated to the diagrams in Fig.~\ref{fig:classI} and it is given by:
\begin{equation}\label{eq:I1}
\begin{split}
\text{(I.1)}&=
-2\text{tr}(T^aT^a)\int_{z_1,z_2}\text{tr}\Big[\gmu S(\xbar_t-z_1)\grho S(z_1-z_2)\grho S(z_2-\ybar_t)
 \\
&\hspace{3cm}\times  \gnu S(\ybar_t-\xbar_t)\Big]  D(z_1-z_2)
\end{split}
\end{equation}
and:
\begin{equation}\label{eq:I2}
\begin{split}
\text{(I.2)}&=-\text{tr}(T^aT^a)\int_{z_1,z_2}\text{tr}\Big[\gmu S(\xbar_t-z_1)\grho S(z_1-\ybar_t)\gnu S(\ybar_t-z_2) \\
&\hspace{3cm}\times  \grho S(z_2-\xbar_t)\Big]  D(z_1-z_2)
\end{split}
\end{equation}
Type I diagrams are the direct generalization of the two diagrams that contribute at $\order{g^2}$ to the nonevolved correlator, see Sec. \ref{sec:CM}. In the latter case, the UV divergence of I.1 cancels the one of I.2 rendering the correlator finite. In particular, the divergence, i.e., the non integrable short-distance singularity of the self-energy contribution I.1 arises at $z_1=z_2$ -- the coordinates of the internal vertices in Fig. \ref{fig:classI} --, while the divergence of I.2 is at $z_1=z_2=x$ and $z_1=z_2=y$. 

In the evolved case, we note that  the propagators that contribute to the divergence of I.1 are not modified by the flow. Hence, I.1 is UV divergent and it generates the same divergence as in the nonevolved case. In I.2, the two fermion propagators that potentially contribute to the divergence are now modified by the flow, thus their short-distance behavior is altered as explained in Sec. \ref{sec:prop}. This is enough to render I.2 finite, in contrast to the nonevolved case. 
In App. \ref{app:momentumreps} we rewrite all contributions in terms of the momentum space representations of the (evolved) propagators and (evolved) Dirac delta's. Then, the fact that I.2 is finite is again manifest due to the exponential flow factors $e^{-tk^2}$, with $t$ the external flow time and $k$ the internal loop momentum  associated to the propagators that generate the UV divergence in the nonevolved case. 
\subsubsection{Type II contribution}
\label{sec:NLOII}
The second contribution in Eq.~\eqref{eq:types} is associated to the diagrams in Fig.~\ref{fig:classII} and it is given by:
\begin{figure}[tb]
	\centering
	\begin{minipage}{.35\textwidth}
		\centering
		\includegraphics[width=.6\linewidth]{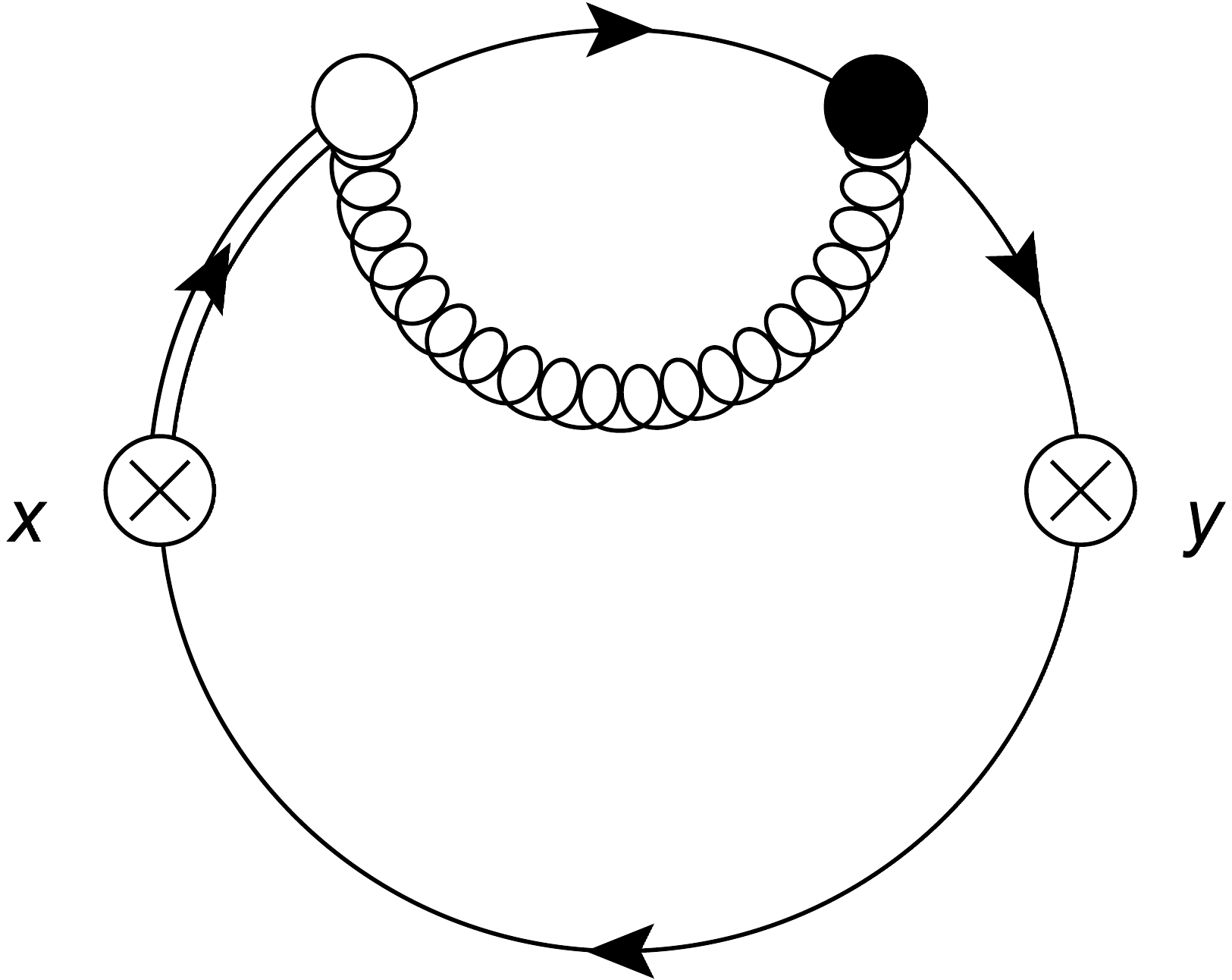}
		\caption*{(II.1)}
	\end{minipage}%
	\begin{minipage}{.35\textwidth}
		\centering
		\includegraphics[width=.6\linewidth]{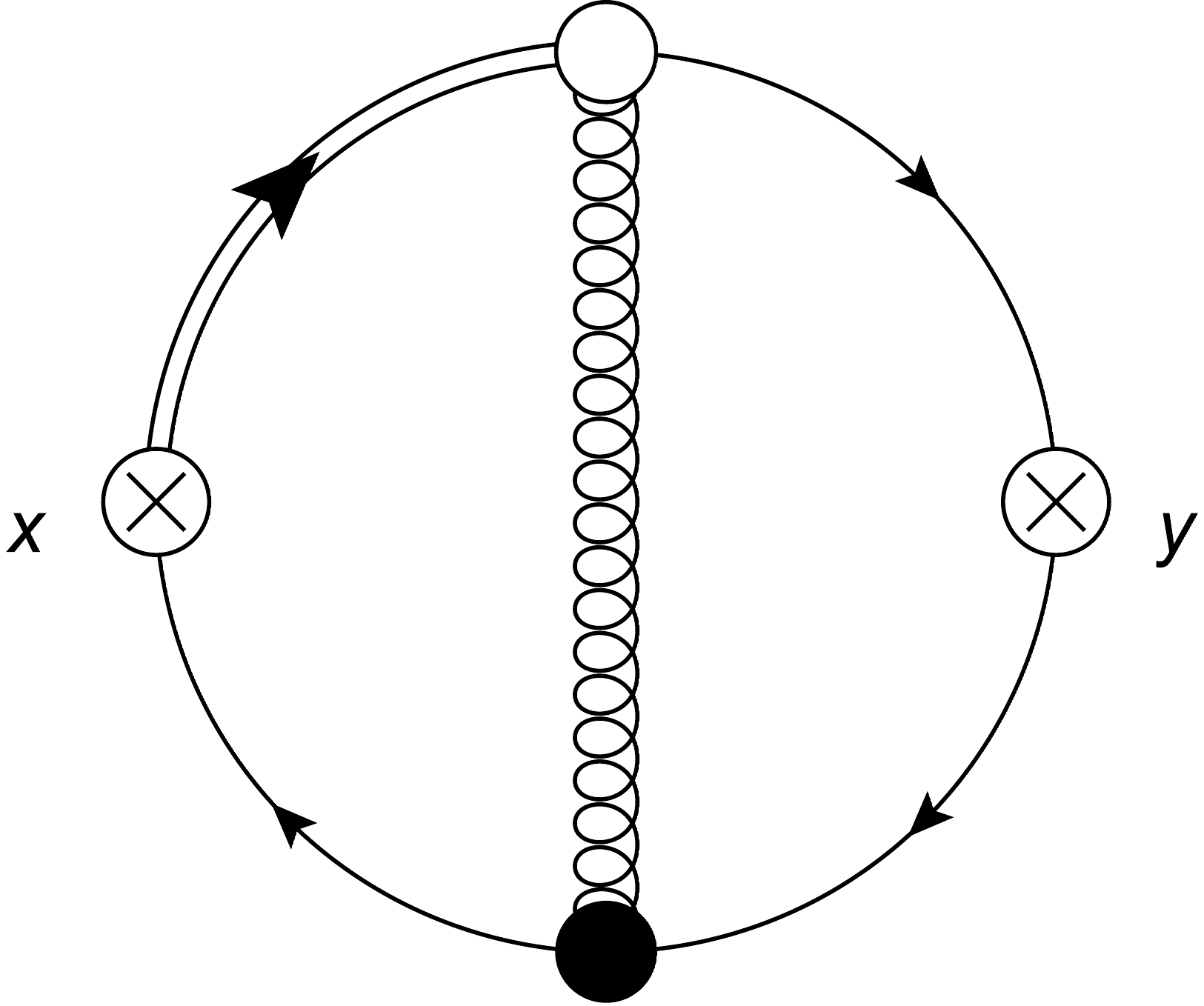}
		\caption*{(II.2)}
	\end{minipage}
\caption{Type II diagrams for the evolved 2-point vector correlator at $\order{g^2}$.}
\label{fig:classII}
\end{figure}
\begin{equation}\label{eq:II1}
\begin{split}
&\text{(II.1)}=4\text{tr}(T^aT^a)\int_0^tds\int_z\text{tr}\Big[\gmu S(\xbar_t-\ybar_t)\gnu S(\ybar_t-z)\\&\hspace{3cm}\times e^{(t-s)\Lap_x}\Big\{\delta^{(d)}(\xbar_s-z)D(\xbar_s-z)\Big\}\Big]+(x\leftrightarrow y;\mu\leftrightarrow\nu)
\end{split}
\end{equation}
and:
\begin{equation}
\begin{split}
&\text{(II.2)}=-4\text{tr}(T^aT^a)\int_0^tds\int_z\text{tr}\Big[\gmu S(\xbar_t-z)\grho S(z-\ybar_t)\gnu\\&\hspace{3cm}\times e^{(t-s)\Lap_x}\Big\{\partial_\rho^x S(\ybar_t-\xbar_s)D(\xbar_s-z)\Big\}\Big]+(x\leftrightarrow y;\mu\leftrightarrow\nu)
\end{split}
\end{equation}
Diagram II.1 is divergent, where the short-distance singularity arises when the black and white blobs in Fig.~\ref{fig:classII} coalesce. Diagram II.2 is instead finite, due to the presence of evolved propagators. Again, this is also manifest in the momentum integrals in App. \ref{app:momentumreps}, for the same reasons as I.2 in type I contribution. 
\subsubsection{Type III contribution}
\label{sec:NLOIII}
\begin{figure}[bt]
	\centering
	\begin{minipage}{.33\textwidth}
		\centering
		\includegraphics[width=.6\linewidth]{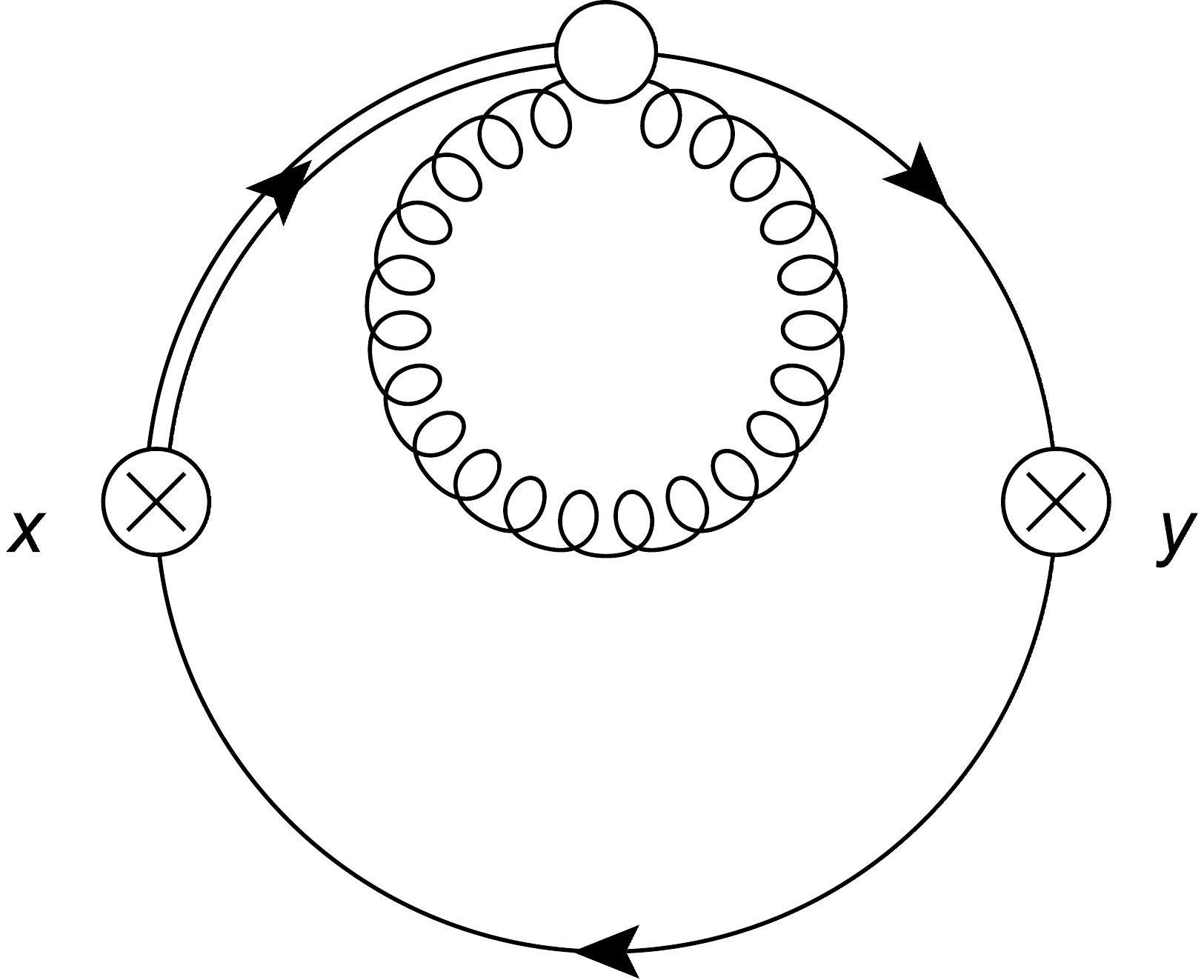}
		\caption*{(III.1)}
	\end{minipage}%
	\begin{minipage}{.33\textwidth}
		\centering
		\includegraphics[width=.6\linewidth]{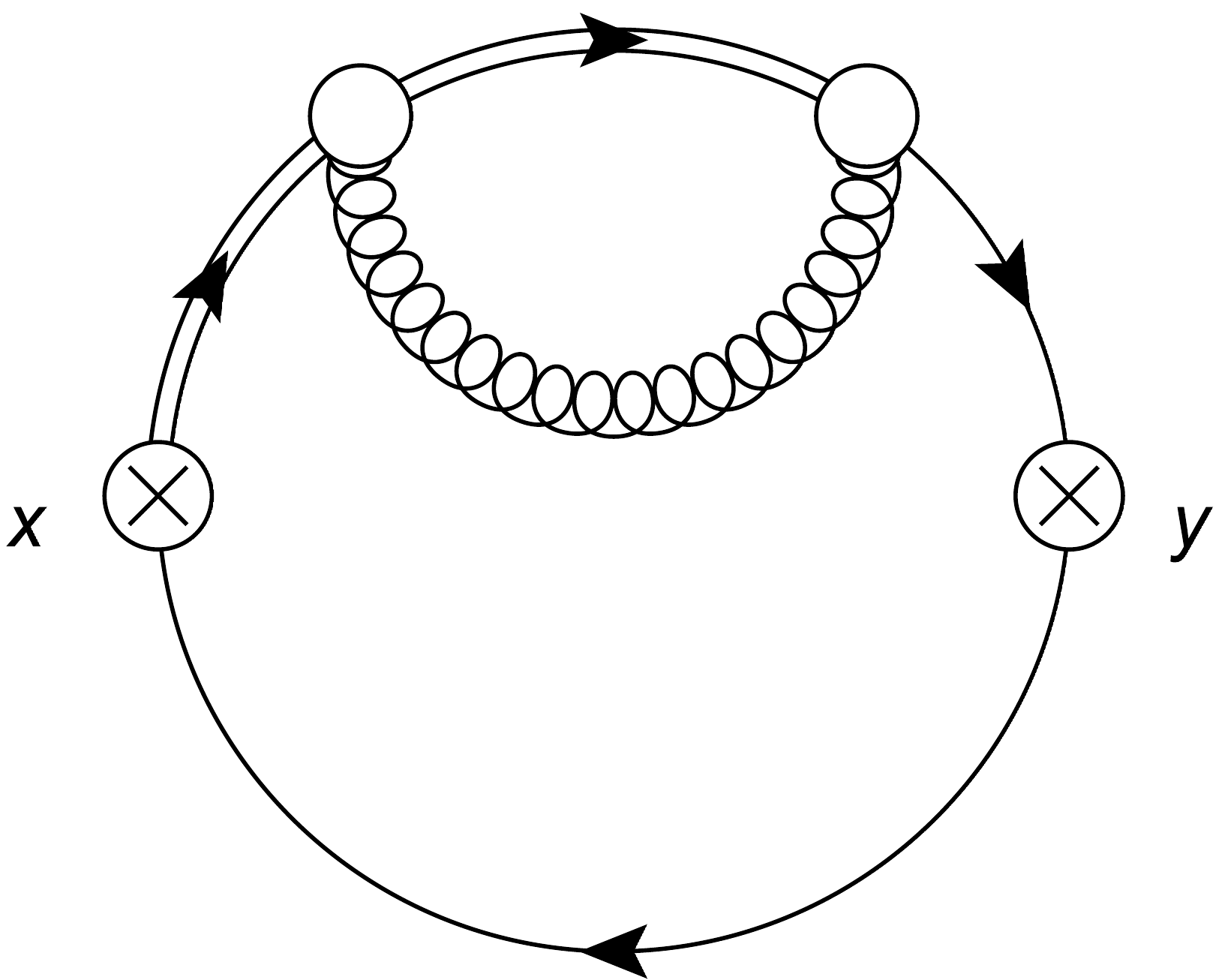}
		\caption*{(III.2)}
	\end{minipage}%
	\begin{minipage}{.33\textwidth}
		\centering
		\includegraphics[width=.6\linewidth]{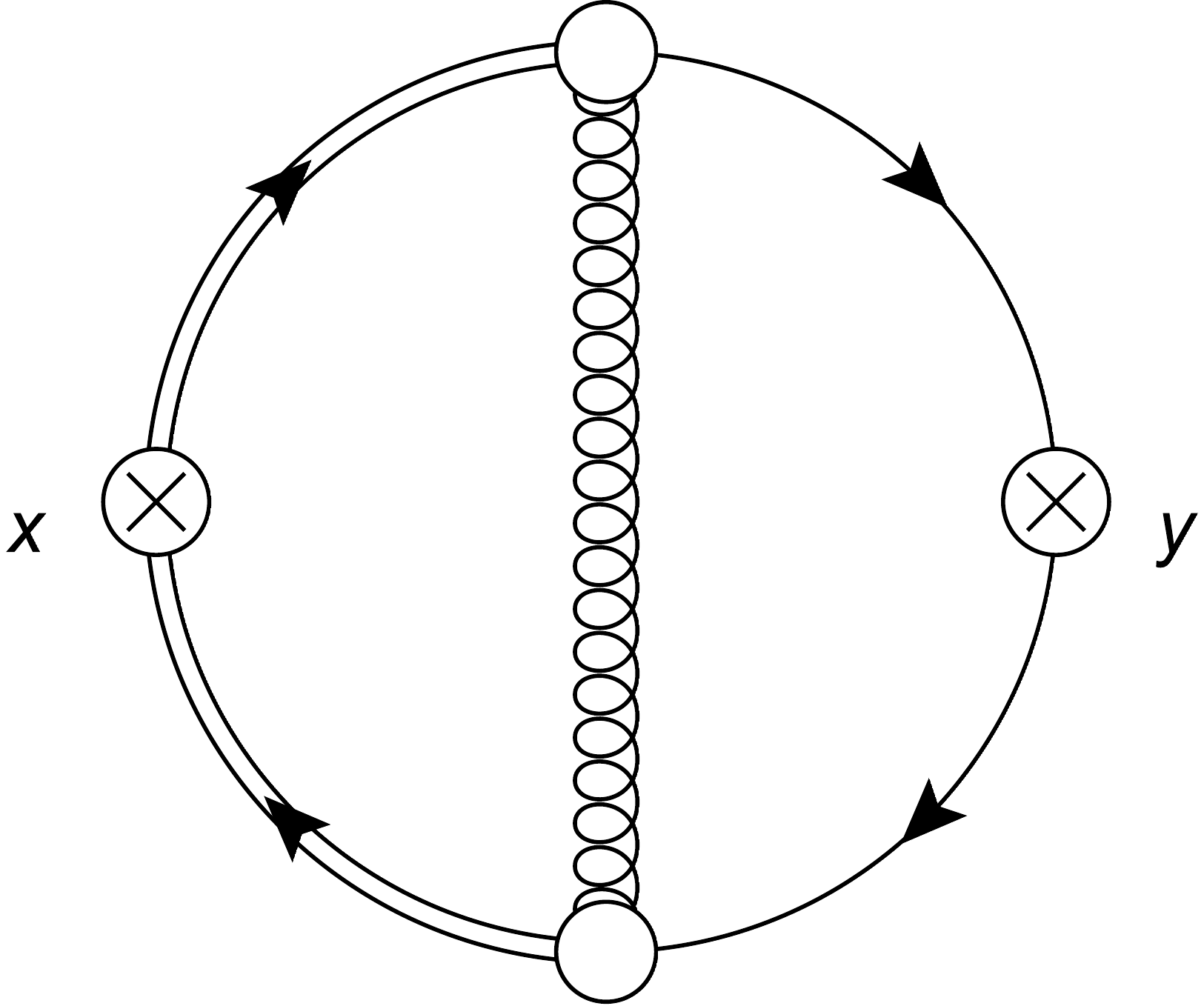}
		\caption*{(III.3)}
	\end{minipage}
\caption{Type III diagrams for the evolved 2-point vector correlator at $\order{g^2}$.}
\label{fig:classIII}
\end{figure}

The third contribution in Eq.~\eqref{eq:types} is associated to the diagrams in Fig.~\ref{fig:classIII}, and we employed:
\begin{equation}
\braket{(\chibar_2\gmu\chi_0)_x(\chibar_0\gnu\chi_0)_y}=
\braket{(\chibar_2\gmu\chi_0)_x(\chibar_0\gnu\chi_0)_y}^\dagger
=\braket{(\chibar_0\gmu\chi_2)_x(\chibar_0\gnu\chi_0)_y}
\end{equation}
when inserting $J_{\mu,2}^V$ of Eq.~\eqref{eq:cc} in the correlator. 
Diagrams III.1 and III.2 come from the term $\chibar_2\gmu\chi_0\,(\chibar_0\gmu\chi_2)$ in $J_{\mu,2}^V$. In fact, $\chibar_2\,(\chi_2)$ in Eq.~\eqref{eq:chiandBexpansions} has three terms, yet, the contribution coming from $B_{\mu,2}$ vanishes.
Diagram III.3 comes from the term $\chibar_1\gmu\chi_1$ in $J_{\mu,2}^V$.

It is convenient to rewrite the total contribution from III.1 and III.2 as follows, see App. \ref{app:manipulationsIII1III2} for the derivation:
\begin{equation}\label{eq:III1+III2}
\begin{split}
\text{(III.1)}+\text{(III.2)}
&=(\text{III}.1')+(\text{III}.2')-\text{(II.1)}
\end{split}
\end{equation}
where (II.1) is the type II contribution in Eq.~\eqref{eq:II1} and:
\begin{equation}\label{eq:III.1pr}
\begin{split}
(\text{III}.1')&=\frac{d-2}{d}(\text{III}.1)\\
&=-2(d-2)\text{tr}(T^aT^a)\int_0^tds\int_z\text{tr}\Big[\gmu S(\xbar_t-\ybar_t)\gnu\\
&\hspace{.3cm}\times e^{(t-s)\Lap_x}\Big\{S(\ybar_t-\xbar_s)\delta^{(d)}(x-z)D(\xbar_s-\zbar_s)\Big\}\Big]
+(x\leftrightarrow y;\mu\leftrightarrow\nu)
\end{split}
\end{equation}
\begin{equation}\label{eq:III.2pr}
\begin{split}
(\text{III}.2')&=8\text{tr}(T^aT^a)\int_0^tds\int_0^sdu\int_z\text{tr}\Big[\gmu S(\xbar_t-\ybar_t)\gnu\\
&\hspace{.3cm}\times e^{(t-s)\Lap_x}\Big\{\delta^{(d)}(x-z)e^{(s-u)\Lap_x}\big\{D(\xbar_u-\zbar_s)\dslash^x \delta^{(d)}(\ybar_t-\xbar_u)\big\}\Big\}\Big]\\
&\hspace{.3cm}+(x\leftrightarrow y;\mu\leftrightarrow\nu)
\end{split}
\end{equation}
Finally, III.3 is given by:
\begin{equation}
\begin{split}
\hspace{-0.2cm}\text{(III.3)}
&=4\text{tr}(T^aT^a)\int_0^tds_1\int_0^tds_2\int_z\text{tr}\Big[\gmu e^{(t-s_2)\Lap_x}\Big\{\delta^{(d)}(x-z)\del_\rho^x S(\xbar_{s_2}-\ybar_t)\Big\}\\
&\hspace{.3cm}\times\gnu e^{(t-s_1)\Lap_x}\Big\{D(\xbar_{s_1}-\zbar_{s_2})\del_\rho^x S(\ybar_t-\xbar_{s_1})\Big\}\Big]+(x\leftrightarrow y;\mu\leftrightarrow\nu)\\
\end{split}
\end{equation}
The contribution $\text{III}.1'$ is UV divergent and we further derive it in Sec. \ref{sec:isolatingdivs}. The remaining two contributions $\text{III}.2' $ and III.3 are both finite, though this is less straightforward to see in coordinate space expressions due to nested flow integrals and exponential-of-Laplacian actions. 
The finiteness of $\text{III}.2'$ is further established in App. \ref{app:finitenessIII2pr} using integration by parts in the momentum expression of App. 
\ref{app:momentumreps}. Finally, the finiteness of III.3 is manifest in its momentum expression in App. 
\ref{app:momentumreps}, analogously to I.2 and II.2.
\subsubsection{Type  IV contribution}
\label{sec:NLOIV}
The last contribution in Eq.~\eqref{eq:types} is associated to the diagrams in Fig.~\ref{fig:classIV} and it is given by:
\begin{figure}[tb]
	\centering
	\begin{minipage}{.35\textwidth}
		\centering
		\includegraphics[width=.6\linewidth]{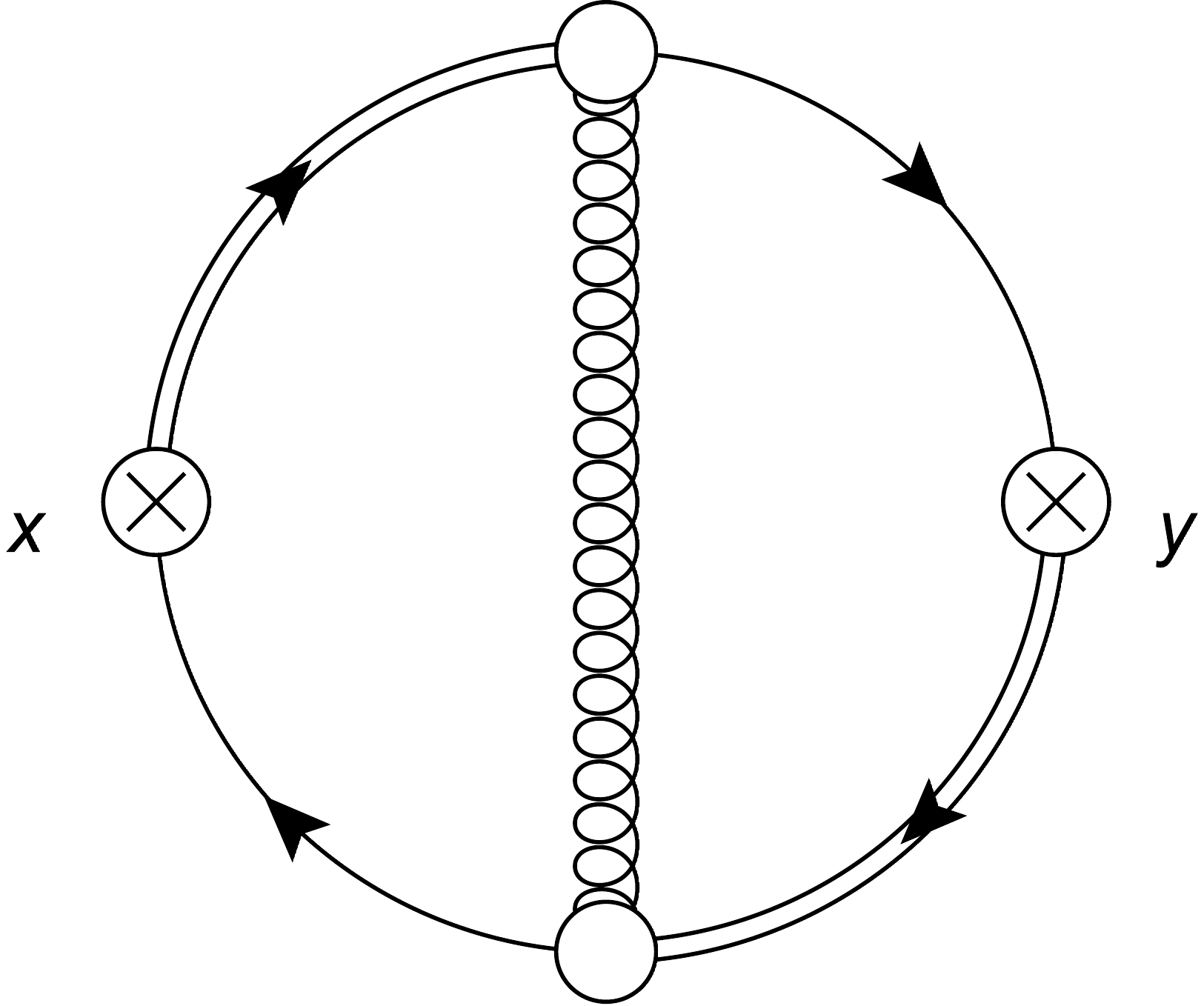}
		\caption*{(IV.1)}
	\end{minipage}%
	\begin{minipage}{.35\textwidth}
		\centering
		\includegraphics[width=.6\linewidth]{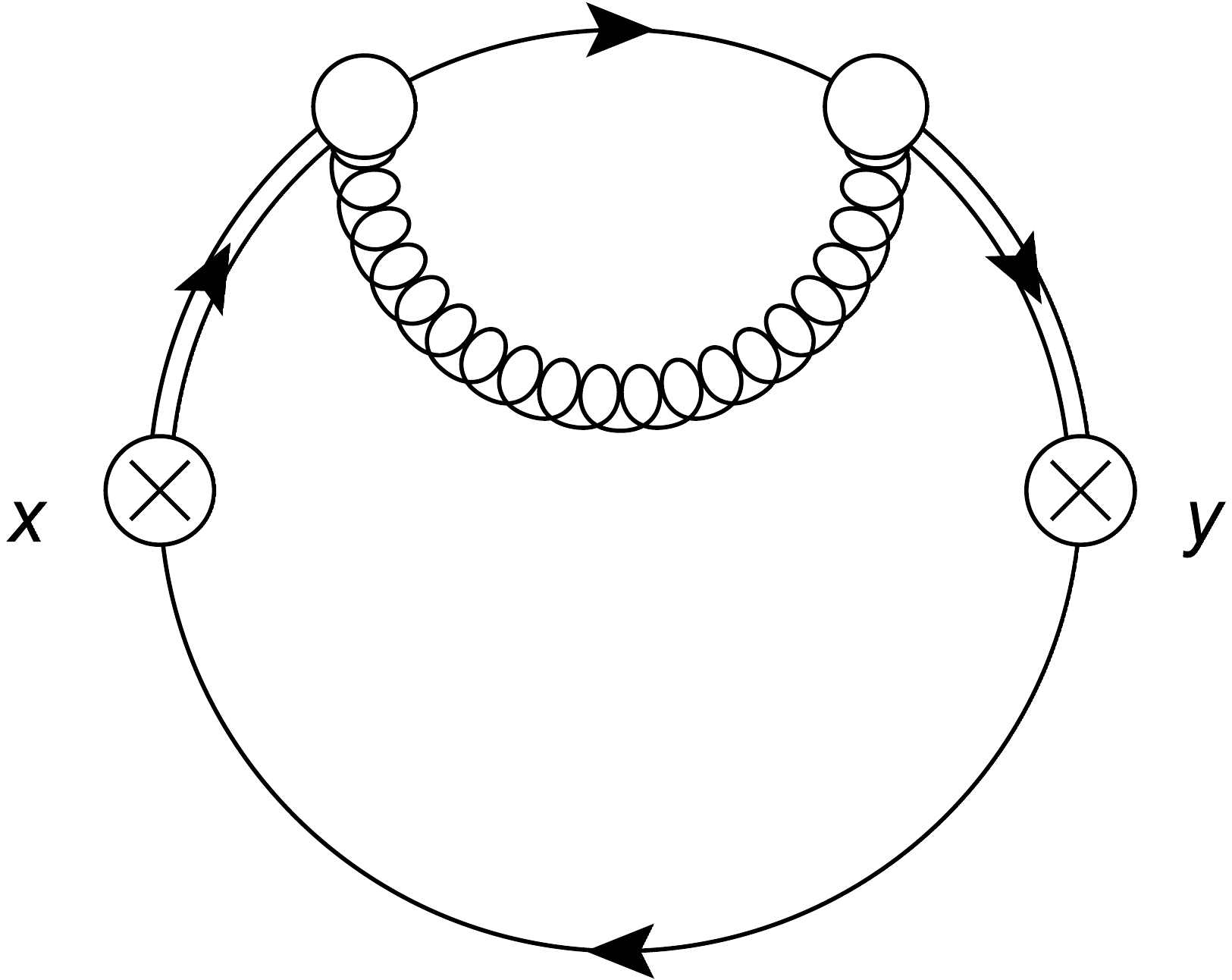}
		\caption*{(IV.2)}
	\end{minipage}
\caption{Type IV diagrams for the evolved 2-point vector correlator at $\order{g^2}$.}
\label{fig:classIV}
\end{figure}
\begin{equation}
\begin{split}
\text{(IV.1)}&=
-8\text{tr}(T^aT^a)\int_0^t ds_1\int_0^t ds_2\int_z\text{tr}\Big[\gmu e^{(t-s_2)\Lap_y}\Big\{\delta^{(d)}(y-z)\\
&\hspace{.3cm}\times\del_\rho^yS(\xbar_t-\ybar_{s_2})\Big\}\gnu e^{(t-s_1)\Lap_x}\Big\{D(\xbar_{s_1}-\zbar_{s_2})\del_\rho^x S(\ybar_t-\xbar_{s_1})\Big\}\Big]
\end{split}
\end{equation}
and:
\begin{equation}
\begin{split}
\text{(IV.2)}&=
8\text{tr}(T^aT^a)\int_0^tds_1\int_0^tds_2\,\text{tr}\Big[\gmu S(\xbar_t-\ybar_t)\gnu e^{(t-s_1)\Lap_x+(t-s_2)\Lap_y}\\
&\hspace{.3cm}\times \Big\{D(\xbar_{s_1}-\ybar_{s_2})\del_\rho^x\del_\rho^y S(\ybar_{s_2}-\xbar_{s_1})\Big\}\Big]
\end{split}
\end{equation}
Both IV.1 and IV.2  are finite. 
This is established by considering the momentum space expressions in App. \ref{app:momentumreps},  analogously to I.2, II.2 and III.3; the $k$ integrals for both contributions contain the factor $e^{-tk^2}$, thereby excluding the possibility of developing a divergence.
\subsection{Total UV divergence at $\order{g^2}$}\label{sec:isolatingdivs}
We have established in Sec. \ref{sec:NLO} that the contributions I.2, II.2, III.2$'$, III.3 and all type IV contributions to the evolved 2-point vector correlator are finite. We have also found that  
the contribution II.1, which is UV divergent, is cancelled by the type III contributions in Eq.~\eqref{eq:III1+III2}. Hence, we
 proceed to derive the UV divergent part of the remaining contributions\footnote{All contributions are IR finite. This is important, otherwise spurious UV divergences would be produced in dimensional regularization.} I.1 and III.1$'$ in dimensional regularization, with $d=4-2\eps$.

We have noticed in Sec. \ref{sec:NLOI} that the contribution I.1 in Eq.~\eqref{eq:I1}      
is UV divergent due to the one-loop fermion self-energy insertion, which is not modified by the flow.  
Appendix \ref{sec:eps-exp} shows the final result for I.1 in $d$ dimensions -- where the gradient-flow modifications occur outside the self-energy insertion -- and its $\eps$-expansion.  The latter yields:
\begin{equation}\label{eq:divI}
\text{(I.1)}=-\frac{2}{\eps}t^\eps\,C_2(R)\frac{1}{(4\pi)^2}\Pi_{\mn,0}^V(t,x-y) +\ldots
\end{equation}
where we employed $\text{tr}(T^aT^a)=-C_2(R)d(R)$,
$\Pi_{\mn,0}^V(t,x-y)$ is the leading order contribution in Eq.~\eqref{eq:evolved2pointLO}
and the dots stand for finite contributions that do not enter the renormalization of the correlator. 

The last divergent contribution is III$.1'$ in Eq.~\eqref{eq:III.1pr}. Its calculation is straightfoward upon noticing that the integral in $z$
in Eq.~\eqref{eq:III.1pr} yields:
\begin{equation}
\begin{split}
\int_z\delta^{(d)}(x-z)D(\xbar_s-\zbar_s)
&=\frac{1}{4\,\pi^{d/2}}\lim\limits_{z\rightarrow x}\Big\{\big((x-z)^2\big)^{1-d/2}\,\gamma\Big(\frac{d}{2}-1,\frac{(x-z)^2}{8s}\Big)\Big\}\\
&=\frac{2^{2-d/2}}{(d-2)(4\pi)^{d/2}}\,s^{1-d/2}
\end{split}
\end{equation}
where we employed the propagator in Eq.~\eqref{eq:propDGFevolved} and the expansion of the $\gamma$ function in Eq.~\eqref{eq:incomplgammaexpansion}, see App. \ref{sec:eps-exp} for the $\eps$-expansion.
Thus we obtain:
\begin{equation}
\begin{split}\label{eq:divIII}
(\text{III}.1')&=-\text{tr}(T^aT^a)\frac{2^{4-d/2}}{(4\pi)^{d/2}}\int_0^tds\,s^{1-d/2}\,\text{tr}\Big[\gmu S(\xbar_t-\ybar_t)\gnu S(\ybar_t-\xbar_t)\Big]\\
&=-\frac{2}{(4-d)}\text{tr}(T^aT^a)\frac{2^{4-d/2}}{(4\pi)^{d/2}}t^{2-d/2}\,\text{tr}\Big[\gmu S(\xbar_t-\ybar_t)\gnu S(\ybar_t-\xbar_t)\Big]\\
&=-\frac{4}{\eps}t^\eps\,\text{tr}(T^aT^a)\,\frac{1}{(4\pi)^2}\,\text{tr}\Big[\gmu S(\xbar_t-\ybar_t)\gnu S(\ybar_t-\xbar_t)\Big]+\ldots\\
&=-\frac{4}{\eps}t^\eps\,C_2(R)\frac{1}{(4\pi)^2}\Pi_{\mn,0}^V(t,x-y)+\ldots
\end{split}
\end{equation}
where $\Pi_{\mn,0}^V(t,x-y)$ is the leading order contribution in Eq.~\eqref{eq:evolved2pointLO} and  
the dots stand for finite contributions that do not enter the renormalization of the correlator.

We conclude that the bare 2-point vector correlator evolved by the gradient flow is no longer UV finite at the next-to-leading order in perturbation theory, i.e., $\order{g^2}$. The total UV divergence at $\order{g^2}$, from the sum of Eqs.~\eqref{eq:divI} and \eqref{eq:divIII} and after multiplying $\Pi_{\mn,2}^V(t,x-y)$ by the bare coupling $g^2$   reads:
\begin{equation}\label{eq:totdiv}
g^2\Pi_{\mn,2}^V(t,x-y)=-\frac{6}{\eps}C_2(R)\frac{g^2}{(4\pi)^2}\Pi_{\mn,0}^V(t,x-y) +\order{\eps^0}
\end{equation}
Thus the evolved correlator acquires a renormalization not present in the nonevolved case. 
\subsubsection{Including the gradient-flow renormalization factor, $Z_\chi$}
\label{sec:RG-GF}
The presence of a UV divergence and thus of a renormalization in the 2-point correlator of the evolved vector current may initially come as a surprise. 
However, this calculation with the result in Eq.~\eqref{eq:totdiv} provides an explicit verification 
of the fact that the renormalization factor $Z_\chi$ introduced in \cite{LuscherFerm} has indeed a universal nature, arising as a new renormalization of the fermion fields $\chi(t,x)$ and $\chibar(t,x)$ evolved by the gradient flow. In fact, by introducing the renormalized evolved fermion fields in the representation $R$:
\begin{equation}\label{eq:chiR}
\chi_R(t,x)=Z_\chi^{1/2}\chi(t,x),\hspace{.3cm}\chibar_R(t,x)=Z_\chi^{1/2}\chibar(t,x)
\end{equation}
with renormalization factor \cite{LuscherFerm}:
\begin{equation}\label{eq:Zchi}
Z_\chi^{1/2}(g(\mu),\eps)=1+\frac{g^2(\mu)}{(4\pi)^2}C_2(R)\frac{3}{2\eps}+\order{g^4}
\end{equation}
where $g(\mu)$ is the renormalized coupling, and employing the renormalization factor $Z_{J_t}=Z_\chi$ for the evolved current, 
we obtain the renormalized 2-point vector correlator at $\order{g^2(\mu)}$:
\begin{equation}\label{eq:2ptR}
\begin{split}
\Pi^V_{R,\mn}(t,x-y, \mu,g(\mu))&= Z_{J_t}^2(g(\mu),\eps) \,\Pi_{\mn}^V(t,x-y, \eps,g)  \\
&\hspace{-3.5cm}=\Big(1+\frac{g^2(\mu)}{(4\pi)^2}\frac{6}{\eps}C_2(R)\Big)\Pi^{V(\eps)}_{\mn,0}(t,x-y)+g^2(\mu)\mu^{2\eps}\Pi^{V(\eps)}_{\mn,2}(t,x-y)+\order{g^4}\\
&\hspace{-3.5cm}=
\Pi^V_{\mn,0}(t,x-y)
+\frac{g^2(\mu)}{(4\pi)^2}C_2(R)\Big(\frac{6}{\eps}-\frac{6}{\eps}-6\log{( t\mu^2)}\\
&\hspace{-3.0cm}+\text{finite terms}
\Big)\Pi^V_{\mn,0}(t,x-y)+\ldots \\
&\hspace{-3.5cm}=\Big( 1- \frac{g^2(\mu)}{(4\pi)^2}
6 C_2(R)\log{ (t\mu^2)} +\text{finite terms}
\Big)\Pi^V_{\mn,0}(t,x-y)+\ldots
\end{split}
\end{equation}
where $\Pi_{\mn}^V$ in the first equality is the bare correlator in dimensional regularization with bare coupling $g$, 
$\Pi^{V(\eps)}_{\mn,0}$ and $\Pi^{V(\eps)}_{\mn,2}$ in the second equality are derived in $d=4-2\eps$ dimensions in App. \ref{sec:eps-exp}, Eqs.~\eqref{eq:PiV0eps} and \eqref{eq:PiV2eps}, respectively, and $\Pi^V_{\mn,0}$ is the leading order contribution in Eq.~\eqref{eq:evolved2pointLO}. In the last two equalities, finite terms stand 
for finite non-logarithmic contributions at $\order{g^2}$ with the same Lorentz structure as $\Pi^V_{\mn,0}$, whereas the dots stand for possible finite non-logarithmic $\order{g^2}$ contributions with Lorentz structure different from $\Pi^V_{\mn,0}$\footnote{We refer to a Lorentz structure of the type $\delta_\mn/2 - a (x-y)_\mu(x-y)_\nu/(x-y)^2$ with $a\neq 1$.}  and $\order{g^4}$ contributions.

Equation \eqref{eq:2ptR} is written in terms of the renormalized vector current $J^V_{R,\mu}=Z_{J_t} J_\mu^V=\chibar_R\gmu\chi_R$ and the QCD renormalized coupling $g(\mu)$, which is related to the bare coupling $g$ as follows:
\be\label{eq:gR}
g^2=Z_g^2\mu^{2\eps}g^2(\mu)
\ee
with 
\be\label{eq:Zg2}
Z_g^2(g(\mu), \eps)=1-\f{\beta_0g^2(\mu)}{\eps}-\f{\beta_1g^4(\mu)}{2\eps}+\f{\beta_0^2g^4(\mu)}{\eps^2}+\order{g^6(\mu)}
\ee
Since the $1/\eps$ poles of the dimensionally regularized expression in Eq.~\eqref{eq:2ptR} exactly cancel, we conclude that the  renormalization of the evolved elementary fermion fields and that of the coupling is the only one required for the evolved 2-point vector correlator 
at $\order{g^2}$ in perturbation theory.  

The arguments presented in \cite{Luscher2, J, LuscherFerm} further suggest that this property extends  to all orders in perturbation theory. 
Moreover, the results in \cite{Luscher2, J, LuscherFerm} imply that all evolved fermion-bilinear currents acquire the same renormalization factor $Z_{J_t}$, and thus the same anomalous dimension. 
The latter can be obtained from its definition in Eq.~\eqref{eq:gammaV}, with $Z_J$ replaced by $Z_{J_t}$. In $\overline{MS}$-like schemes it reads:
\be\label{eq:gt}
\begin{split}
\gamma_{J_t}(g)&= -\f{d\log Z_{J_t}}{d\log\mu}=-\beta(g,\eps)\f{\partial\log Z_{J_t}}{\partial g}\\
&= - ( -\eps g+\beta(g))  \f{\partial\log Z_{J_t}}{\partial g}\\
&=6 C_2(R) \frac{g^2(\mu)}{(4\pi)^2}+\order{g^4}
\end{split}
\ee
with $\beta(g)$ in Eq.~\eqref{eq:gammaV} and $Z_{J_t}$ in Eq.~\eqref{eq:Zchi}\footnote{Note that in Eq.~\eqref{eq:gt} one should use the beta function in $d=4-2\eps$ dimensions.}. The $\order{g^4}$ contribution to $\gamma_{J_t}$ in Eq.~\eqref{eq:gt} is renormalization-scheme dependent. 
\subsection{OPE of the evolved currents from their 2-point correlators}
\label{sec:RGsolev}
In full analogy with the analysis in Sec. \ref{sec:CM} for the connected 2-point correlator of a generic nonevolved gauge-invariant and multiplicatively renormalizable current, the Callan-Symanzik equation for the connected 2-point correlator of an evolved gauge-invariant current of canonical dimension $D$, and that renormalizes as $J_R(t,x)=Z_{J_t} J(t,x)$ reads:
\be\label{eq:CS-GF}
\Big(\f{\partial}{\partial\log\mu}+\beta(g)\f{\partial}{\partial g} +2\gamma_{J_t} (g)  \Big){\Pi}_R(t, x-y, \mu, g(\mu))=0
\ee
with $\gamma_{J_t}$ in Eq.~\eqref{eq:gt}. Equation \eqref{eq:CS-GF} is implied by the renormalization-group invariance ($\mu$ independence) of the evolved bare correlator $\Pi$, related to $\Pi_R$ as:
\be\label{eq:RvsB-GF}
\Pi_R(t,x-y,\mu, g(\mu))= Z_{J_t}^2 \Big ( \f{\Lambda}{\mu}, g(\mu )\Big )\, \Pi (t, x-y, \Lambda, g(\Lambda))
\ee
analogously to the nonevolved case in Eq.~\eqref{eq:RvsB}, in a regularization with UV cutoff $\Lambda$. However, differently from the nonevolved case, the Callan-Symanzik Eq.~\eqref{eq:CS-GF} poses a two-scale problem, with scales $\sqrt{t}$ and $|x-y|$. We can solve the equation in the limit in which one of the two scales dominates in the UV. Specifically, we are interested in the limit $t\to 0$ at fixed $|x-y|$, hence $\sqrt{t}\ll |x-y|$, to establish an asymptotic relation between the evolved and nonevolved correlators. 

The validity of the OPE for composite operators evolved by the gradient flow in the small $t$ limit has been verified in \cite{Luscher2,LuscherFerm}. Thus, 
for a generic multiplicatively renormalizable evolved fermion-bilinear current in the small $t$ limit we write:
\be\label{eq:Jope}
J_R(t,x)=\braket{J(t,x)}_R+c(t)J_R(x)+\order{t}
\ee
where $\langle\ldots\rangle$ is its vacuum expectation value, $J_R(x)$ is the renormalized nonevolved local current and $c(t)$ is the leading coefficient of the OPE in the small $t$ limit. 

We can determine $c(t)$ from the small $t$ expansion of the corresponding evolved 2-point correlator $\Pi_R$. Such small $t$ expansion reads:
\be\label{eq:OPE}
\Pi_R(t,x-y,\mu, g(\mu))= C(\sqrt{t}\mu, g(\mu))\, \Pi_R(x-y,\mu, g(\mu)) +\Delta\Pi_R(t,x-y,\mu, g(\mu))
\ee
where in the right-hand side the dimensionless coefficient $C=c^2$ is the square of the OPE coefficient $c(t)$ in Eq.~\eqref{eq:Jope}, and $\Pi_R$ is the renormalized nonevolved correlator.  The latter is  given by Eq.~\eqref{eq:CSsol} for a multiplicatively renormalizable nonevolved current $J_R(x)=Z_JJ(x)$ in Eq.~\eqref{eq:Jope}. 

The factorization of the dependence on $\sqrt{t}$ and $|x-y|$ occurs in the first term of the right-hand side in Eq.~\eqref{eq:OPE}, which is of $\order{t^0}$ times logarithms of $\sqrt{t}\mu$\footnote{Note that in this expansion $\mu$ must be of order $1/\sqrt{t}$ to avoid large logarithms.}.
 
The term $\Delta\Pi_R$ refers to contributions of $\order{t}$ times logarithms of $\sqrt{t}\mu$ and $|x-y|\mu$. These higher order contributions have in general a nonfactorizable dependence on the two scales $\sqrt{t}$ and $|x-y|$. 

Therefore, for multiplicatively renormalizable $J_R(t,x)$ and $J_R(x)$ the renor\-mal\-iza\-tion-group equation for $C$ is now implied by Eq.~\eqref{eq:RvsB-GF} for the evolved correlator and Eq.~\eqref{eq:RvsB} for the nonevolved one, which yield:
\be\label{eq:CS-WilsonC}
\Big(\f{\partial}{\partial\log\mu}+\beta(g)\f{\partial}{\partial g} +2\gamma_{J_t} (g) -2\gamma_J(g) \Big)C(\sqrt{t}\mu, g(\mu))=0
\ee
Hence, $C$ has an anomalous dimension given by the difference of the anomalous dimensions of the evolved and the nonevolved $\Pi_R$. 
The solution can be written as:
\be\label{eq:CWilson}
C(\sqrt{t}\mu, g(\mu)) = F(g(\sqrt{t})) Z^2(g(\sqrt{t}), g(\mu))
\ee
where $F$ is a dimensionless RGI function of the RGI running coupling $g(\sqrt{t})$, with $F\to 1$ as $g(\sqrt{t})\to 0$ implied by perturbation theory, and: 
\be\label{eq:Zt}
\begin{split}
Z^2(g(\sqrt{t}),g(\mu))&=\exp \int_{g(\mu)}^{g(\sqrt{t})} 2\,\f{\gamma_{J_t}(g)-\gamma_J(g)  }{\beta(g)}dg 
\end{split}
\ee
Equation \eqref{eq:Zt} coincides with Eq.~\eqref{eq:Z} for the nonevolved case once $\gamma_J$ has been replaced with $\gamma_{J_t} -\gamma_J$ and $g(|x-y|)$ with $g(\sqrt{t})$. 
Then, by means of the latter substitutions, the UV asymptotic expression for $Z^2$ in Eq.~\eqref{eq:Zt}  is given by the second line of Eq.~\eqref{eq:Z}.

With $C$ given by Eq.~\eqref{eq:CWilson}, the small $t$ expansion of the evolved correlator thus reads:
\be\label{eq:OPEexp}
\begin{split}
\Pi_R(t,x-y,\mu, g(\mu))&= 
\Big(\f{ g^2(\sqrt{t}) }{g^2(\mu) }\Big)^{\f{\gamma_{J_t}^{(0)} - \gamma_{J}^{(0)}}{\beta_0}} \,e^{\order{g^2(\sqrt{t})-g^2(\mu)}}F(g(\sqrt{t})) \\
&\times \Pi_R(x-y,\mu, g(\mu))+\order{t}
\end{split}
\ee
showing that the evolved correlator has the same Lorentz structure of the nonevolved one at leading order in the small $t$ expansion, i.e., up to $\order{t}$ contributions. 

The leading universal UV asymptotics of Eq.~\eqref{eq:OPEexp} as $g(\sqrt{t})\to 0$ then reads:
\be\label{eq:UVasympPiR}
\Pi_R(t,x-y,\mu, g(\mu))\sim
\Big(\f{ g^2(\sqrt{t}) }{g^2(\mu) }\Big)^{\f{\gamma_{J_t}^{(0)} - \gamma_{J}^{(0)}}{\beta_0}} \,e^{\order{g^2(\mu)}}\, \Pi_R(x-y,\mu, g(\mu))
\ee
and the leading universal UV asymptotics of the first
 OPE coefficient: 
\be
c(\sqrt{t}\mu, g(\mu)) = F^{1/2}(g(\sqrt{t})) Z(g(\sqrt{t}), g(\mu))
\ee
for $J_R(t,x)$ in Eq.~\eqref{eq:Jope} thus follows:
\be\label{eq:UVct}
c(\sqrt{t}\mu, g(\mu)) \sim \Big(\f{ g(\sqrt{t}) }{g(\mu) }\Big)^{\f{\gamma_{J_t}^{(0)} - \gamma_{J}^{(0)}}{\beta_0}} \,e^{\order{g^2(\mu)}}
\ee
In the case of the evolved vector current one has $\gamma_{J}^{(0)}=0$. 
We have also noticed that in the nonevolved vector case, Eq.~\eqref{eq:CSsol} is replaced with Eq.~\eqref{eq:LS}, which takes into account a change of the Lorentz structure at higher orders in perturbation theory. 

Accordingly, by employing Eq.~\eqref{eq:LS} the small $t$ expansion 
for the evolved flavor singlet vector correlator has the explicit form:
\be\label{eq:OPEexpV}
\begin{split}
&\Pi^V_{R,\mn}(t,x-y,\mu, g(\mu))=
\Big(\f{ g^2(\sqrt{t}) }{g^2(\mu) }\Big)^{\f{\gamma_{J_t}^{(0)}}{\beta_0}} \,e^{\order{g^2(\sqrt{t})-g^2(\mu)}}F(g(\sqrt{t})) \\
&\times \f{1}{(x-y)^{2D}}\sum_{n=0}^\infty\, g^{2n}(|x-y|)\Big(A_{n}\f{\delta_\mn}{2}-B_n\f{(x-y)_\mu (x-y)_\nu}{(x-y)^2}\Big)     +\order{t}
\end{split}
\ee
with $A_n=B_n$ for $n=0,1$, and 
we employed the coefficient $C$ from Eqs.~\eqref{eq:CWilson} and \eqref{eq:Zt} with $\gamma_J=0$.
Equation \eqref{eq:OPEexpV} shows explicitly that the leading contribution to the small $t$ expansion of the evolved vector correlator in Eq.~\eqref{eq:OPEexpV}
inherits its Lorentz structure from the nonevolved correlator order by order in $g^{2}(|x-y|)$. 

Finally, we note that the leading $n=0$ term in Eq.~\eqref{eq:OPEexpV} reproduces the explicit $\order{g^2(\mu)}$ result in Eq.~\eqref{eq:2ptR}
by means of the perturbative expansion:
\be
 \Big(\f{ g^2(\sqrt{t}) }{g^2(\mu) }\Big)^{\f{\gamma_{J_t}^{(0)}}{\beta_0}} = 1+g^2(\mu)\gamma_{J_t}^{(0)}\log{(t\mu^2)}+\ldots
 \ee
 with  $\gamma_{J_t}^{(0)} =-6C_2(R)/(4\pi)^2$, and $F(g(\sqrt{t}))=1+\order{g^2(\sqrt{t})}$.

We conclude this section with a brief analysis of the finite contributions to the evolved $\Pi^V_{R,\mn}$ and their asymptotic behavior.

At $\order{g^2}$ the finite contributions are given by I.2, II.2, III.2$'$, III.3, IV.1 and IV.2, see Secs. \ref{sec:NLOI}-\ref{sec:NLOIV}. Only  I.2 is already present in the nonevolved case, where it is UV divergent at $d=4$ and cancels the UV divergence of I.1. Therefore, I.2 does not vanish as $t\to 0$ at fixed nonzero separation, and it contributes to the leading OPE term in Eq.~\eqref{eq:OPEexpV}.
All the other finite contributions manifestly vanish in the same limit and only contribute at higher orders to the OPE.   

In the opposite limit of vanishing separation at fixed $t>0$, all of the above finite contributions are expected to vanish as a consequence of the regulating effect of the gradient-flow smearing.

Finally, we mention that ``evolved contact terms'' resulting from the smearing of nonevolved contact terms, may be expected to contribute to evolved 2-point correlators.  In the vector case,  
dimensional analysis tells us that nonevolved contact terms are of the type $\delta_{\mu\nu}\Lap\delta^{(4)}(x-y)$ and 
$\partial_\mu \partial_\nu\delta^{(4)}(x-y)$. Hence, they may lead to evolved contact terms of the type
$\delta_{\mu\nu}\Lap K_t(x-y)$ and 
$\partial_\mu \partial_\nu K_t(x-y)$, respectively. These terms no longer vanish at nonzero separation at fixed $t>0$, and they can contribute starting at $\order{t}$ to the OPE in Eq.~\eqref{eq:OPEexpV}.
\section{Current conservation and renormalization}
\label{sec:nonrenorm}
\subsection{Nonevolved case: conservation implies nonrenormalization}
We briefly review  a simple argument for how conservation of a nonevolved gauge-invariant local current implies its nonrenormalization.
The conservation of the local and gauge-invariant vector current $J_\mu^V(x)$ is the consequence of an exact nonanomalous global $U(1)$ symmetry and it is encoded in the corresponding Ward identity once the theory is quantized.
 
The conservation of $J_\mu^V(x)$ then implies that the associated gauge-invariant and dimensionless charge $Q$ is also conserved. It follows that $Q$ cannot depend on any unphysical scale. Hence, it cannot acquire an anomalous dimension and the nonrenormalization of the vector current, $J^V_\mu(x)=J_{R,\mu}^V(x)$, thus follows. 

The nonrenormalization and conservation of $J^V_\mu(x)$  in turn imply the transversality of the corresponding 2-point correlator. 
\subsection{Evolved case: conservation does not imply nonrenormalization}
The situation is fundamentally different for the evolved vector current $J_\mu^V(t,x)$. We have seen that the latter acquires an anomalous dimension $\gamma_{J_t}$, which enters the leading term of the small $t$ expansion of the evolved 2-point vector correlator as shown in Eq.~\eqref{eq:OPEexpV}. Yet, the same leading term fulfils  transversality and current conservation, despite the presence of an anomalous dimension. Indeed, specifically:
\be\label{eq:CSsoln_der}
\begin{split}
\del_\mu^x\Pi^V_{R,\mn}(t, x-y, \mu, g(\mu))&= \\
&\hspace{-3.cm}= C(\sqrt{t}\mu, g(\mu))\, \del_\mu^x\Pi^V_{R,\mn}(x-y,\mu, g(\mu))
 +\del_\mu^x\Delta\Pi_{R,\mn}(t,x-y,\mu, g(\mu))\\
&\hspace{-3.cm}=\del_\mu^x\Delta\Pi_{R,\mn}(t,x-y,\mu, g(\mu))
\end{split}
\ee
and the first term in the right-hand side of the first equality has an anomalous dimension, but vanishes because the nonevolved correlator is transversal, i.e., $\del_\mu^x\Pi^V_{R,\mn}(x-y,\mu, g(\mu))=0$. 
It is then clear that a nonzero anomalous dimension  
is allowed because the evolved current depends on the additional (unphysical) gradient-flow scale $\sqrt{t}$, which parametrizes its nonlocality. 

On the other hand, the violation of transversality and the nonconservation of the evolved vector current do occur through the second term $\del_\mu^x\Delta\Pi_R $ in the right-hand side of Eq.~\eqref{eq:CSsoln_der}. Hence, they are a soft-breaking effect of $\order{t}$ induced by the smearing action of the gradient flow that vanishes as $t\to 0$. This agrees with the explicit results at $\order{g^0}$ in Eqs.~\eqref{eq:notrans-smallt} and \eqref{eq:notrans} and the classical leading-order consideration in App. \ref{sec:classic}.

As a side note, one could also relate the nonconservation of the evolved vector current  to the lack of a corresponding exact symmetry in a 
$d+1$-dimensional theory that includes the flow direction, along the lines of \cite{Luscher2,LuscherFerm,Shindler2013}. 
\section{Conclusions}
 \label{sec:conc}
 We have studied the renormalization properties of the gradient-flow evolved flavor singlet 2-point vector correlator in perturbatively massless QCD-like theories, and showed that, in contrast to the nonevolved case,  the correlator is renormalized and the evolved vector current acquires an anomalous dimension:
\be\label{eq:gconc}
\begin{split}
\gamma_{J_t}
&=6 C_2(R) \frac{g^2(\mu)}{(4\pi)^2}+\order{g^4}
\end{split}
\ee
Our result confirms that this anomalous dimension is induced by the renormalization of the evolved elementary fermion field first derived in \cite{LuscherFerm}, and thus applies to all evolved fermion-bilinear currents. Our result is also 
in agreement with results in the literature for 1-point correlators of evolved fermion bilinears \cite{LuscherFerm,MakinoSuzuki2014,Harlander3}.  

The Callan-Symanzik equation for the connected 2-point correlators of generic multiplicatively renormalizable evolved fermion-bilinear currents now involves two scales, i.e., the flow time $\sqrt{t}$ and the separation $|x-y|$.  
We made connection with the nonevolved case by deriving the solution of the Callan-Symanzik equation  in the limit of small gradient-flow time $\sqrt{t}$, at fixed separation $|x-y|$. Incidentally, the leading order contribution to this expansion also determines the leading OPE coefficient for the corresponding evolved current in the small $t$ limit. 

 We have also discussed how, interestingly, conservation of the evolved vector current and transversality of the corresponding 2-point correlator no longer imply nonrenormalization, at difference with the nonevolved case. 
In particular, the leading contribution to the small $t$ expansion of the evolved 2-point vector correlator --- which is $\order{t^0}$ times any power of logarithms -- fulfils transversality and the current is conserved despite the presence of the anomalous dimension $\gamma_{J_t}$. This is due to the presence of the additional gradient-flow scale $\sqrt{t}$, so that renormalization logarithms at $\order{t^0}$ only depend on the product $\sqrt{t}\mu$. Violation of transversality and nonconservation do occur at $\order{t}$ in the OPE as a soft breaking effect induced by the nonlocality of the evolved current. 

  \section*{Acknowledgments}
 We would like to thank Marco Bochicchio for the helpful comments. MB thanks Franz Herzog for a useful discussion on isolating divergences in $n$-loop Feynman diagrams. 
 
 \vspace{0.5cm}
\noindent{\bf Note added:} We are aware \footnote{M. Paupinutto, private communication.} that another group has been independently computing the gradient-flow evolved 2-point vector correlator at next-to-leading order with the same result. 
 
 \appendix

\section{Diagrammatic notation}\label{sec:diagramnotation}
In this appendix we introduce the diagrammatic notation employed in Fig. \ref{fig:D01-D06} and Figs. \ref{fig:2ptLO} - \ref{fig:classIV} for the gradient-flow evolved 1-point and 2-point correlators. The adopted notation follows
\cite{MakinoSuzuki2014}.
 \subsection*{Propagators}
The Feynman rule in Euclidean coordinate space for the gradient-flow evolved fermion propagator reads:
\\
\\
\begin{minipage}{.33\textwidth}
	\begin{figure}[H]
		\centering
		\includegraphics[width=0.85\linewidth]{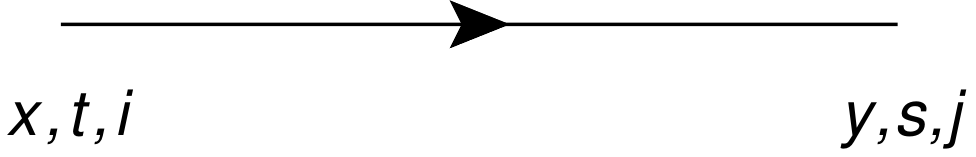}
	\end{figure}
\end{minipage}
\begin{minipage}{.66\textwidth}
	\begin{flalign}
	\hspace{1cm}=\hspace{2cm} \delta_{ij}S(\xbar_t-\ybar_s)&&
	\end{flalign}
\end{minipage}
\\
\\
\\
where $S(\xbar_t-\ybar_s)$ is given in Eq.~\eqref{eq:propf}, and we use the same line as for the nonevolved fermion propagator -- the latter is obtained for $t=s=0$.
Analogously, for the evolved gluon propagator one has:
\\
\\
\noindent
\begin{minipage}{.33\textwidth}
	\begin{figure}[H]
		\centering
		\includegraphics[width=0.85\linewidth]{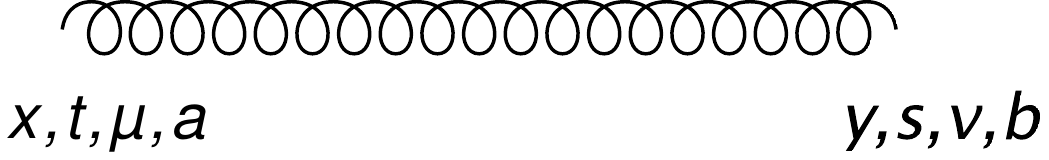}
	\end{figure}
\end{minipage}
\begin{minipage}{.66\textwidth}
	\begin{flalign}
	\hspace{1cm}=\hspace{2cm} \delta_{\mn}\delta^{ab}D(\xbar_t-\ybar_s)&&
	\end{flalign}
\end{minipage}
\\
\\
\\
\noindent
with $D(\xbar_t-\ybar_s)$  in Eq.~\eqref{eq:propDGFevolved}.
\subsection*{Flow-time integrals}
The flow-time integrals and associated kernels that are present in the
second term (interaction part) of the solution to the flow equation in Eq.~\eqref{eq:GFgaugesolutionnobox} for the gluon and Eq.~\eqref{eq:F} for the fermion are represented by a double line that always ends in a white blob representing a gradient-flow interaction vertex, i.e., a vertex induced by the flow. Explicitly, for the fermionic case in Eq.~\eqref{eq:F}:
\\ \noindent
\begin{minipage}{.33\textwidth}
	\begin{figure}[H]
		\centering
		\includegraphics[width=0.85\linewidth]{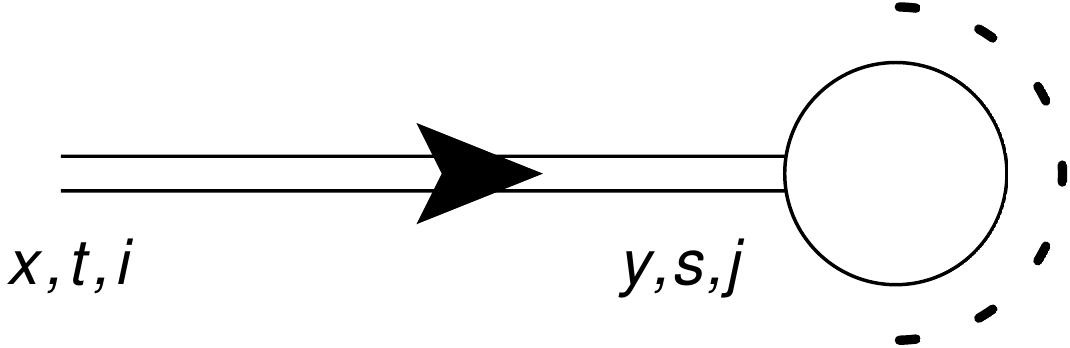}
	\end{figure}
\end{minipage}
\begin{minipage}{.66\textwidth}
	\begin{flalign}\label{eq:appflowtimeint}
	\hspace{1cm}=\hspace{2cm} \int_y\int_0^t ds\,K_{t-s}(x-y)\delta_{ij}\dots&&
	\end{flalign}
\end{minipage}
 \\
 \\
 \\ \noindent
where we only highlighted the flow-integral in the second term of Eq.~\eqref{eq:F} associated with the double line, and the dashes stand for a combination of lines emanating from the vertex. The structure of gradient-flow vertices is further explained around Eq.~\eqref{eq:FR}. 

The second term in the gauge field solution in Eq.~\eqref{eq:GFgaugesolutionnobox}, analogously represented by double gluon lines, does not occur in this work.  
 \subsection*{Vertices}
 The QCD vertex that enters our calculations is represented by a filled blob. The Feynman rule in Euclidean coordinate space and in the case of gradient-flow evolved  fields at the vertex reads:
 \\
 \noindent
 \begin{minipage}{.33\textwidth}
 	\begin{figure}[H]
 		\centering
 		\includegraphics[width=0.85\linewidth]{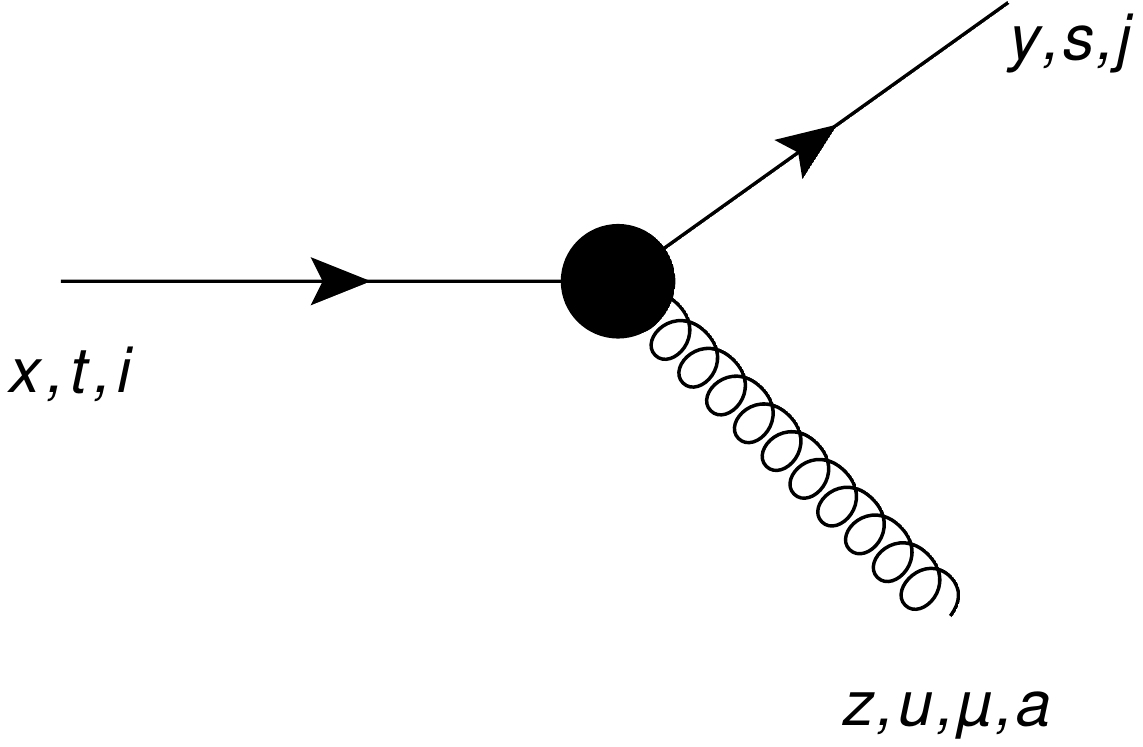}
 	\end{figure}
 \end{minipage}
 \begin{minipage}{.66\textwidth}
 	\begin{flalign}
 	\hspace{1cm}&=\hspace{1cm} -g T^a_{ij}\int_{x_1}S(\xbar_t-x_1)\gmu S(x_1-\ybar_s)&&\\ \nonumber &\hspace{3cm}\times D(x_1-\zbar_u)
 	\end{flalign}
 \end{minipage}
 \\
 \\
 \\ 
with obvious replacements in the case of nonevolved fields. 

We now discuss the structure of gradient-flow vertices that appear in this work, i.e., of the type in Eq.~\eqref{eq:appflowtimeint}.
Specifically, for the fermion field $\chi (t,x)$ 
they represent the second term in Eq.~\eqref{eq:F}, which can be written as an expansion in powers of $g$ starting at $\order{g}$. Thus at $\order{g^n}$ this vertex corresponds to $\chi_n(t,x)$ defined in Eq.~\eqref{eq:series}, with $n\geq1$. The explicit expressions for $\chi_{1,2}$ are in Eq.~\eqref{eq:chiandBexpansions}, and we reproduce them here together with the corresponding gradient-flow vertices.

The lowest order gradient-flow vertex in this work is given by:
\\
\\ \noindent
\begin{minipage}{1\textwidth}
\begin{equation}\label{eq:FR}
\chi_1(t,x)= 2\int_0^t ds\, e^{(t-s)\Lap_x} \left\{B_{\mu, 1}\partial_\mu\chi_0   \right\} 
\end{equation}
\end{minipage}
 \begin{minipage}{1\textwidth}
 	\begin{figure}[H]
 \hspace{5.8cm}\includegraphics[width=.28\linewidth]{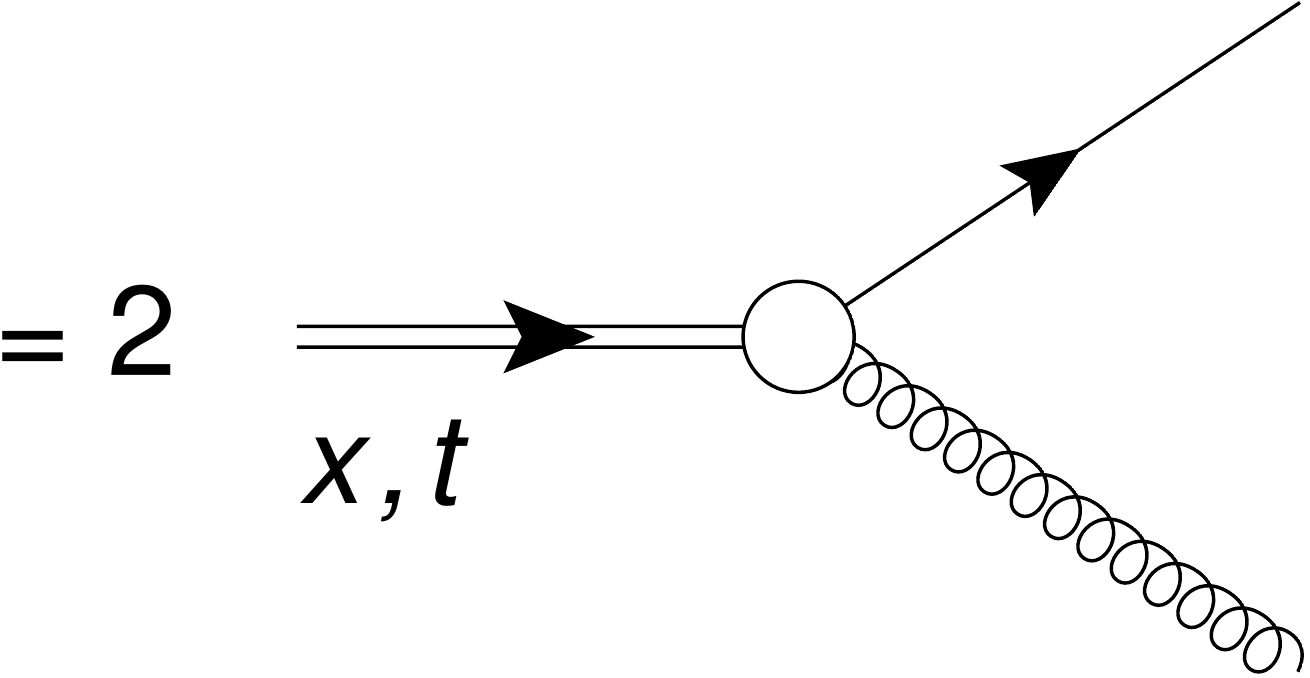}
 	\end{figure}
 \end{minipage}
\\ \\
and at the next-to-leading order one has:
\\
\\ \noindent
\begin{minipage}{1\textwidth}
\begin{equation}
\hspace{-0.8cm}\chi_2(t,x)=\int_0^tds\,e^{(t-s)\Lap_x}\left\{B_{\mu,1}B_{\mu,1}
\chi_0\hspace{.5cm}+2B_{\mu,2}\del_\mu\chi_0\hspace{.5cm}
+2B_{\mu,1}\del_\mu\chi_1\right\}
\end{equation}
\end{minipage}
\\ \noindent
\begin{minipage}{.8\textwidth}
\begin{figure}[H]
\hspace{2.cm}\includegraphics[width=1\linewidth]{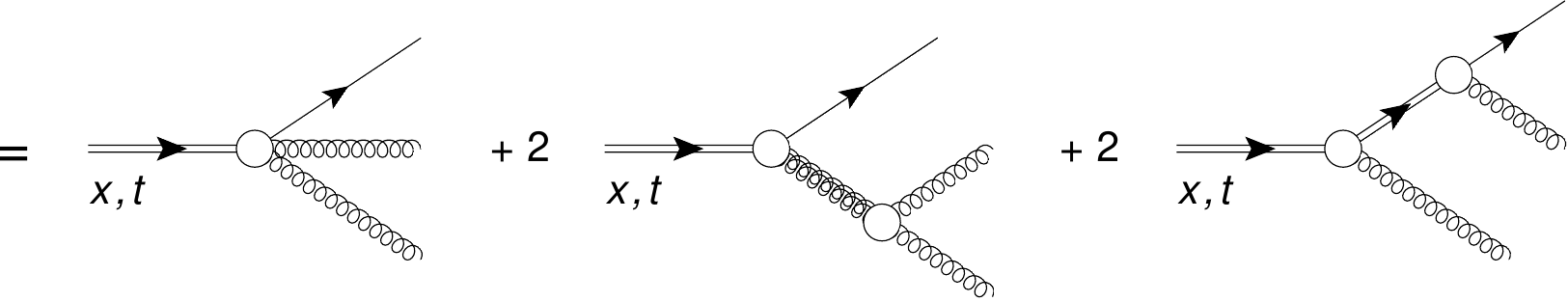}	
\end{figure}
\end{minipage}
\\
\\
\\
with $B_{\mu,2}$ in Eq.~\eqref{eq:chiandBexpansions}.
 \section{Derivation of Eq.~\eqref{eq:III1+III2} for 
(III.1)$+$(III.2)}\label{app:manipulationsIII1III2}
 In this appendix we derive Eq.~\eqref{eq:III1+III2}, which reads:
\begin{equation}\label{eq:III1+III2app}
\begin{split}
\text{(III.1)}+\text{(III.2)}
&=(\text{III}.1')+(\text{III}.2')-\text{(II.1)}
\end{split}
\end{equation}
We start with:
 \begin{equation}
 \begin{split}
\text{(III.1)}+\text{(III.2)}&=
 2\braket{(\chibar_2\gmu\chi_0)_x(\chibar_0\gnu\chi_0)_y}+(x\leftrightarrow y;\mu\leftrightarrow\nu)\\
& \hspace{-1.8cm}=2\int_0^tds\braket{(e^{(t-s)\Lap_x}\Big\{\chibar_0(s)B_{\rho,1}(s)B_{\rho,1}(s)\Big\}\gmu\chi_0)_x(\chibar_0\gnu\chi_0)_y}\\
 &\hspace{-1.5cm}+8\int_0^tds\int_0^sdu\langle
(e^{(t-s)\Lap_x}\Big\{\del^x_\sigma\big[e^{(s-u)\Lap_x}\big\{\big(\del_\rho^x\chibar_0(u)\big)B_{\rho,1}(u)\big\}\big]B_{\sigma,1}(s)\Big\} \\
 & \hspace{-1.5cm}  \times \gmu\chi_0)_x(\chibar_0\gnu\chi_0)_y\rangle
 +(x\leftrightarrow y;\mu\leftrightarrow\nu)
 \end{split}
 \end{equation}
associated to the diagrams (III.1) and (III.2) in Fig.~\ref{fig:classIII}. After performing the Wick contractions, we obtain:
 \begin{equation}\label{eq:classIIfirstandsecondterm}
 \begin{split}
\text{(III.1)}+\text{(III.2)}&=
 -2d\,\text{tr}(T^aT^a)\int_0^tds\int_z\text{tr}\Big[\gmu S(\xbar_t-\ybar_t)\gnu e^{(t-s)\Lap_x}\\
& \hspace{-1.5cm} \times \Big\{S(\ybar_t-\xbar_s) \delta^{(d)}(x-z)D(\xbar_s-\zbar_s)\Big\}\Big]\\
 & \hspace{-1.5cm}  -8\text{tr}(T^aT^a)\int_0^tds\int_0^sdu\int_z\text{tr}\Big[\gmu S(\xbar_t-\ybar_t)\gnu
 e^{(t-s)\Lap_x}\Big\{\delta^{(d)}(x-z)\\
& \hspace{-1.5cm}
\times e^{(s-u)\Lap_x}
\del^x_\rho\big\{D(\xbar_u-\zbar_s)\del_\rho^x S(\ybar_t-\xbar_u)\big\}\Big\}\Big]
 +(x\leftrightarrow y;\mu\leftrightarrow \nu)
 \end{split}
 \end{equation}
We then employ the relation:
 \begin{equation}
 \begin{split}
 &\int_0^sdu\frac{d}{du}\Big(e^{(s-u)\Lap_x}\Big\{D(\xbar_u-\zbar_s)S(\ybar_t-\xbar_u)\Big\}\Big)\\
 &=-\int_0^sdu\,e^{(s-u)\Lap_x}\Lap_x\Big\{D(\xbar_u-\zbar_s)S(\ybar_t-\xbar_u)\Big\}
+\int_0^sdu\,e^{(s-u)\Lap_x} \Big\{\Lap_xD(\xbar_u-\zbar_s)     \\
 &\hspace{.45cm}\times 
S(\ybar_t-\xbar_u)\Big\}+\int_0^sdu\,e^{(s-u)\Lap_x}\Big\{D(\xbar_u-\zbar_s)\Lap_xS(\ybar_t-\xbar_u)\Big\}\\
 &=-2\int_0^sdu\,e^{(s-u)\Lap_x}\Big\{\del_\rho^xD(\xbar_u-\zbar_s)
\del_\rho^xS(\ybar_t-\xbar_u)\Big\}
 \end{split}
 \end{equation}
to rewrite the second term in Eq.~\eqref{eq:classIIfirstandsecondterm} as:
 \begin{equation}\label{eq:classIIsecondtermmanip}
 \begin{split}
 -8&\text{tr}(T^aT^a)\int_0^tds\int_0^sdu\int_z\text{tr}\Big[\gmu S(\xbar_t-\ybar_t)\gnu\\
 &\hspace{1cm}\times e^{(t-s)\Lap_x}\Big\{\delta^{(d)}(x-z)e^{(s-u)\Lap_x}\del^x_\rho\big\{D(\xbar_u-\zbar_s)\del_\rho^x S(\ybar_t-\xbar_u)\big\}\Big\}\Big]\\
 =&-8\text{tr}(T^aT^a)\int_0^tds\int_0^sdu\int_z\text{tr}\Big[\gmu S(\xbar_t-\ybar_t)\gnu\\
 &\hspace{1cm}\times e^{(t-s)\Lap_x}\Big\{\delta^{(d)}(x-z)e^{(s-u)\Lap_x}\big\{D(\xbar_u-\zbar_s)\Lap_x S(\ybar_t-\xbar_u)\big\}\Big\}\Big]\\
 &+4\text{tr}(T^aT^a)\int_0^tds\int_z\text{tr}\Big[\gmu S(\xbar_t-\ybar_t)\gnu\\
 &\hspace{1cm}\times e^{(t-s)\Lap_x}\Big\{\delta^{(d)}(x-z)D(\xbar_s-\zbar_s)S(\ybar_t-\xbar_s)\Big\}\Big]\\
 &-4\text{tr}(T^aT^a)\int_0^tds\int_z\text{tr}\Big[\gmu S(\xbar_t-\ybar_t)\gnu\\
 &\hspace{1cm}\times e^{(t-s)\Lap_x}\Big\{\delta^{(d)}(x-z)e^{s\Lap_x}\big\{D(x-\zbar_s)S(\ybar_t-x)\big\}\Big\}\Big]
 \end{split}
 \end{equation}
The last term in Eq.~\eqref{eq:classIIsecondtermmanip} exactly cancels (II.1) in Eq.~\eqref{eq:II1}, after noting that:
 \begin{equation}
 \begin{split}
 &\int_z e^{(t-s)\Lap_x}\Big\{\delta^{(d)}(x-z)e^{s\Lap_x}\big\{D(x-\zbar_s)S(\ybar_t-x)\big\}\Big\}\\
 &=\int_z\int_{x'} e^{(t-s)\Lap_x}\Big\{\delta^{(d)}(x-z)e^{s\Lap_x}\big\{\delta^{(d)}(x-x')D(x'-\zbar_s)S(\ybar_t-x')\big\}\Big\}\\
&=\int_{x'} e^{(t-s)\Lap_x}\Big\{\int_z\delta^{(d)}(x-z)
\delta^{(d)}(\xbar_s-x')D(x'-\zbar_s)S(\ybar_t-x')\Big\}\\
 &=\int_{x'}S(\ybar_t-x')e^{(t-s)\Lap_x}\Big\{\delta^{(d)}(\xbar_s-x')D(x'-\xbar_s)\Big\}\\
 &=\int_z S(\ybar_t-z)e^{(t-s)\Lap_x}\Big\{\delta^{(d)}(\xbar_s-z)D(z-\xbar_s)\Big\}
 \end{split}
 \end{equation}
Then, the first term in Eq.~\eqref{eq:classIIsecondtermmanip} is III.2$'$ in Eq.~\eqref{eq:III.2pr}, after employing $\Lap_x S(\ybar_t-\xbar_u)=-\slashed{\partial}_x  \delta^{(d)}(\ybar_t-\xbar_u)$, and 
the second term in Eq.~\eqref{eq:classIIsecondtermmanip} combines 
with the first term in Eq.~\eqref{eq:classIIfirstandsecondterm} to give 
 III.1$'$ in Eq.~\eqref{eq:III.1pr}. Hence, the relation in Eq.~\eqref{eq:III1+III2app} follows. 
 \section{Momentum space expressions}\label{app:momentumreps}
We provide the Type I-IV contributions to the gradient-flow evolved 2-point vector correlator in terms of momentum-space integrals:
\begin{equation}
\begin{split}
\text{(I.1)}=-2\text{tr}(T^aT^a)&\int_{p,q,k}e^{i(p+q)(x-y)}e^{-2t(p^2+q^2)}\\&\frac{1}{(p^2)^2q^2k^2(p+k)^2}\,\text{tr}\big[\gamma_\mu\slashed{p}\gamma_\rho\slashed{k}\gamma_\rho\slashed{p}\gamma_\nu\slashed{q}\big]
\end{split}
\end{equation}
\begin{equation}
\begin{split}
\text{(I.2)}=\text{tr}(T^aT^a)&\int_{p,q,k}e^{i(p+q)(x-y)}e^{-t(p^2+q^2+k^2+(p+q+k)^2)}\\&\frac{1}{p^2q^2k^2(p+k)^2(p+q+k)^2}\,\text{tr}\big[\gamma_\mu(\pslash+\qslash+\kslash)\gamma_\rho\qslash\gamma_\nu\pslash\gamma_\rho\kslash\big]
\end{split}
\end{equation}
\begin{equation}
\begin{split}
\text{(II.1)}=8\text{tr}(T^aT^a)\int_0^tds&\int_{p,q,k}e^{i(p+q)(x-y)}e^{-2t(p^2+q^2)-2sk\cdot(p+k)}\\&\frac{1}{p^2q^2(p+k)^2}\text{tr}[\gamma_\mu\qslash\gamma_\nu\pslash]
\end{split}
\end{equation}
\begin{equation}
\begin{split}
\text{(II.2)}=4\text{tr}(T^aT^a)\int_0^tds&\int_{p,q,k}e^{i(p+q)(x-y)}e^{-t(p^2+q^2+k^2+(p+q+k)^2)+2sq\cdot(p+k)}\\&\frac{1}{p^2q^2k^2(p+k)^2}\,\text{tr}\big[\gamma_\mu\qslash\gamma_\nu(\pslash\qslash\kslash+\kslash\qslash\pslash)\big]
\end{split}
\end{equation}
\begin{equation}
\begin{split}
(\text{III}.1')=-4(d-2)\text{tr}(T^aT^a)\int_0^tds&\int_{p,q,k}e^{i(p+q)(x-y)}e^{-2t(p^2+q^2)-2sk^2}\\&\frac{1}{p^2q^2k^2}\text{tr}[\gamma_\mu\qslash\gamma_\nu\pslash]
\end{split}
\end{equation}
\begin{equation}
\begin{split}
(\text{III}.2')=16\text{tr}(T^aT^a)\int_0^tds\int_0^sdu&\int_{p,q,k}e^{i(p+q)(x-y)}e^{-2t(p^2+q^2)-2sk^2-2(s-u)p\cdot k}\\&\frac{1}{q^2k^2}\,\text{tr}\big[\gamma_\mu\qslash\gamma_\nu\pslash\big]
\end{split}
\end{equation}
\begin{equation}
\begin{split}
\text{(III.3)}=-8\text{tr}(T^aT^a)\int_0^tds_1\int_0^tds_2&\int_{p,q,k}e^{i(p+q)(x-y)}e^{-2t(p^2+q^2+k^2)-2(t-s_1)p\cdot k+2(t-s_2)q\cdot k}\\&\frac{q\cdot p}{p^2q^2k^2}\,\text{tr}\big[\gamma_\mu\qslash\gamma_\nu\pslash\big]
\end{split}
\end{equation}
\begin{equation}
\begin{split}
\text{(IV.1)}=-8\text{tr}(T^aT^a)&\int_0^tds_1\int_0^tds_2\int_{p,q,k}e^{i(p+q)(x-y)}\\&e^{-t(p^2+q^2+k^2+(p+q+k)^2)+2s_1q\cdot(p+k)-2s_2k\cdot(p+k)}\\&\frac{k\cdot q }{q^2k^2(p+k)^2}\,\text{tr}\big[\gamma_\mu\qslash\gamma_\nu\kslash\big]
\end{split}
\end{equation}
\begin{equation}
\begin{split}
\text{(IV.2)}=-8\text{tr}(T^aT^a)&\int_0^tds_1\int_0^tds_2\int_{p,q,k}e^{i(p+q)(x-y)}\\&e^{-2t(k^2+(p+q+k)^2)+2(s1+s2)q\cdot(p+k)}\\&\frac{1}{k^2(p+k)^2}\text{tr}[\gamma_\mu\qslash\gamma_\nu\kslash]
\end{split}
\end{equation}
  \section{Finiteness of (III.$2'$)}\label{app:finitenessIII2pr}
 It is convenient in this case to work directly with momentum-space integrals. The contribution (III.2$'$) then reads: 
 \be
\text{(III.2$'$)}=16\text{tr}(T^aT^a)\int_{p,q}
e^{i(p+q)(x-y)}e^{-2t(p^2+q^2)}
\frac{1}{q^2}\,\text{tr}\big[\gamma_\mu\qslash\gamma_\nu\pslash\big]\, \mathcal{I}(p)
\ee
with 
 \begin{equation}\label{eq:kintapp}
\begin{split}
 \mathcal{I}(p)&=\int_0^tds\int_0^sdu\int_k\frac{1}{k^2}e^{-2sk^2-2(s-u)p\cdot k}\\
&=\int_0^tds\int_0^sdu\int_k\frac{1}{k^2}e^{-2sk^2-2up\cdot k}
\end{split}
 \end{equation}
where in the last equality we let $u\rightarrow s-u$. We first perform the integral in $k$ in $\mathcal{I}(p)$, by inserting the identity:
\begin{equation}
1=\frac{1}{d}\frac{\del k_\mu}{\del k_\mu}
\end{equation}
then using integration by parts (IBP), while rewriting the resulting $p\cdot k$ appearing in the numerator as $-\frac{1}{2}\frac{d}{du}$, and using that in dimensional regularization:
\begin{equation}
\int_k \frac{\del}{\del k_\mu}f(k)=0
\end{equation}
Subsequently, IBP in $u$ yields:
\begin{equation}\label{eq:a.2equals}
\begin{split}
\mathcal{I}(p)&= \mathcal{I}_1(p)+\mathcal{I}_2(p)+\mathcal{I}_3(p)\\
&\mathcal{I}_1(p)\equiv\frac{3}{d}\int_0^tds\int_0^sdu\int_k\frac{1}{k^2}e^{-2sk^2-2up\cdot k}=\frac{3}{d}\mathcal{I}(p)\\
&\mathcal{I}_2(p)\equiv\frac{4}{d}\int_0^tds\int_0^sdu\int_k\,s\,e^{-2sk^2-2up\cdot k}\\
&\mathcal{I}_3(p)\equiv-\frac{1}{d}\int_0^tds\,s\int_k\frac{1}{k^2}e^{-2sk^2-2sp\cdot k}
\end{split}
\end{equation}
 Next we perform IBP in $k$ in $\mathcal{I}_3$, and we express the appearing $p\cdot k$ in the numerator as $-k^2-\frac{1}{2}\frac{d}{ds}$, and perform IBP in $s$ in the $\frac{d}{ds}$ term. This yields
 \begin{equation}
 \begin{split}
\mathcal{I}_3(p)&=\frac{1}{d(d-4)}t^2\int_k\frac{1}{k^2}e^{-2tk^2-2tp\cdot k}-\frac{2}{d(d-4)}\int_0^tds\,s^2\int_k e^{-2sk^2-2sp\cdot k}
 \end{split}
 \end{equation}
 Next we focus on $\mathcal{I}_2$. Using the same methods, we find
 \begin{equation}
 \begin{split}
\mathcal{I}_2(p)&=-\frac{2}{d}t\int_0^tdu\int_k\frac{1}{k^2}e^{-2tk^2-2up\cdot k}-2\,\mathcal{I}_3(p)+\frac{2}{d}\,\mathcal{I}(p)
\end{split}
 \end{equation}
 Plugging everything back in, we find:
 \begin{equation}\label{eq:intIlaststageappD}
 \begin{split}
\mathcal{I}(p)&=\frac{2}{(d-5)}\mathcal{I}_{2a}(p)+\frac{1}{(d-5)(d-4)}\left(\mathcal{I}_{3a}(p)+\mathcal{I}_{3b}(p)\right)\\
&\mathcal{I}_{2a}(p)\equiv - \int_0^tdu\int_k\frac{1}{k^2}e^{-2tk^2-2up\cdot k}\\
&\mathcal{I}_{3a}(p)\equiv - t^2\int_k\frac{1}{k^2}e^{-2tk^2-2tp\cdot k}\\
&\mathcal{I}_{3b}(p)\equiv 2\int_0^tds\,s^2\int_ke^{-2sk^2-2sp\cdot k}
 \end{split}
 \end{equation}
 The term containg $\mathcal{I}_{2a}$ is finite in $d=4$, due to the damping effect of  $e^{-2tk^2}$. Next we show that the terms with the $\frac{1}{(d-4)}$ prefactor cancel in $d=4$, rendering this term finite. For $\mathcal{I}_{3a}$ we use Schwinger parametrization:
\begin{equation}
\begin{split}
\mathcal{I}_{3a}(p)&=-t^2\int_0^\infty dv\int_k e^{-(2t+v)k^2-2tp\cdot k}\\
&=-(4\pi)^{-d/2}\,t^2\int_0^\infty dv\, (2t+v)^{-d/2}\,e^{\frac{t^2}{2t+v}p^2}\\
&=(-1)^{-d/2}(4\pi)^{-d/2}\,t^{4-d}(p^2)^{1-d/2}\,\gamma\left(\frac{d}{2}-1,-\frac{tp^2}{2}\right)\\
&\overset{d\rightarrow4}{=}\frac{1}{(4\pi)^2}\frac{1}{p^2}\,\gamma\left(1,-\frac{tp^2}{2}\right)
\end{split}
\end{equation}
 while for $\mathcal{I}_{3b}$ we have
\begin{equation}
\begin{split}
\mathcal{I}_{3b}(p)&=(4\pi)^{-d/2}\,2^{1-d/2}\int_0^t ds\, s^{2-d/2}\,e^{\half s p^2}\\
&=(-1)^{1-d/2}\,(4\pi)^{-d/2}\,2^{4-d}\,(p^2)^{d/2-3}\,\gamma\left(3-\frac{d}{2},-\frac{tp^2}{2}\right)\\
&\overset{d\rightarrow4}{=}-\frac{1}{(4\pi)^2}\frac{1}{p^2}\,\gamma\left(1,-\frac{tp^2}{2}\right)
\end{split}
\end{equation}
 and we see that these terms cancel when $d\rightarrow4$. Thus we conclude that $\mathcal{I}(p)$ is finite, which implies that (III.2$'$) is finite. See also \cite{Harlander3} for an automated implementation of IBP reduction for loop integrals involving flow-time integrations.
 \section{$\eps$-expansion of the evolved 2-point vector correlator at $\order{g^2}$}\label{sec:eps-exp}
Here we provide the explicit $\eps$-expansion of $\Pi^{V(\eps)}_{\mn,2}(t,x)$ in Eq.~\eqref{eq:2ptR}, which is given by (we temporarily replace $x-y\to x$ for brevity):
\begin{equation}\label{eq:Pi2eps}
\Pi^{V(\eps)}_{\mn,2}(t,x)=\text{(I.1)}^{(\eps)}+\text{(III.1}')^{(\eps)}+\text{finite contributions}
\end{equation}
where $\text{(I.1)}^{(\eps)}$ and $\text{(III.1}')^{(\eps)}$ are the contributions given in Eqs.~\eqref{eq:I1} and \eqref{eq:III.1pr}, evaluated in $d=4-2\eps$ dimensions. The remaining finite contributions consist of all terms other than $\text{(I.1)}$ and $\text{(III.1}')$ in Eq.~\eqref{eq:NLO_tot}.

The explicit expressions in $d$-dimensions are given by:
\begin{equation} \label{eq:I1ddim}
\begin{split}
 \text{(I.1)}^{(d)}&=\frac{\text{tr}(T^aT^a)}{(4\pi)^{\frac{3d}{2}}}\frac{16\,\Gamma\left(\frac{d}{2}\right)\Gamma\left(3-\frac{d}{2}\right) }{d(d-2)(4-d)}\,t^{2-d}(x^2)^{-\frac{d}{2}}\,\text{tr}[\gmu\slashed{x}\gnu\slashed{x}]\\
 &\hspace{.5cm}\times\gamma\left(\frac{d}{2},\frac{x^2}{8t}\right)\, {}_1 F_1\left(d-2,1+\frac{d}{2},-\frac{x^2}{8t}\right)
 \end{split}
\end{equation}
\begin{equation}\label{eq:III.1prddim}
\text{(III.1}')^{(d)}=\frac{\text{tr}(T^aT^a)}{(4\pi)^{\frac{3d}{2}}}\frac{2^{3+\frac{3d}{2}}}{(4-d)}\,t^{2-\frac{d}{2}}(x^2)^{-d}\,\text{tr}[\gmu\slashed{x}\gnu\slashed{x}]\,\gamma\left(\frac{d}{2},\frac{x^2}{8t}\right)^2
\end{equation}
while the leading order contribution in $d$-dimensions is given by:
\begin{equation}\label{eq:LOddim}
\Pi_{\mn,0}^{V(d)}(t,x)=\frac{d(R)}{(4\pi)^d}2^{2d-2}(x^2)^{-d}\,\text{tr}[\gmu\slashed{x}\gnu\slashed{x}]\,\gamma\left(\frac{d}{2},\frac{x^2}{8t}\right)^2
\end{equation}
where $d(R)$ is the dimension of the fermion representation $R$. We first provide the $\eps$-expansions of the lower incomplete gamma function and the confluent hypergeometric function ${}_1F_1$. For the former we obtain:
\begin{equation}\label{eq:epsexpincgamma}
\begin{split}
\gamma\left(\frac{d}{2},z\right)&=\gamma\left(2,z\right)+\frac{d}{d\eps}\gamma(2-\eps,z)\Big|_{\eps=0}\eps+\order{\eps^2}\\
&=\gamma(2,z)-\int^z_0 dt\, t\log(t)\,e^{-t}\,\eps+\order{\eps^2}\\
&=\gamma\left(2,z\right)+\Big[\gamma(1,z)-\text{Ein}(z)+\gamma(2,z)\log(z)\Big]\eps +\order{\eps^2}
\end{split}
\end{equation}
where we used the definition in Eq.~\eqref{eq:lower-gamma}, and $\text{Ein}(z)$ is the complementary exponential integral:
\begin{equation}\label{eq:Ein}
\text{Ein}(z)=\int_0^z dt\, \frac{1-e^{-t}}{t}
\end{equation}
For  the confluent hypergeometric function we use the following properties:
\begin{equation}\label{eq:conflhypsum}
{}_1F_1\left(a,b,z\right)=\sum_{n=0}^\infty\frac{(a)_n}{(b)_n n!}z^n
\end{equation}
where $(a)_n=\frac{\Gamma(a+n)}{\Gamma(a)}$ is the Pochhammer symbol, and the Kummer's transformation:
\begin{equation}\label{eq:kummerstrans}
{}_1F_1\left(a,b,-z\right)=e^{-z}\,{}_1F_1\left(b-a,b,z\right)
\end{equation}
 Further, the relation between the confluent hypergeometric function and the lower incomplete gamma function is given by
\begin{equation}\label{eq:conflhypandincomplgamma}
{}_1F_1\left(1,1+s,z\right)=s\, z^{-s}\, e^z \,\gamma\left(s,z\right)
\end{equation}
and we will use the following expressions for the derivatives of the confluent hypergeometric function with respect to its parameters $a,b$
\cite{Ancarani2008}:
\begin{subequations}\label{eq:conflhypder}
\begin{equation}
\frac{d}{da}\,{}_1F_1(a,b,z)=\frac{a}{b}z\sum_{m_1=0}^{\infty}\sum_{m_2=0}^\infty \frac{1}{a+m_1}\frac{(a+1)_{m_1+m_2}}{(2)_{m_1+m_2}(b+1)_{m_1+m_2}}z^{m_1+m_2}
\end{equation}
\begin{equation}
\frac{d}{db}\,{}_1F_1(a,b,z)=-\frac{a}{b}z\sum_{m_1=0}^{\infty}\sum_{m_2=0}^\infty \frac{1}{b+m_1}\frac{(a+1)_{m_1+m_2}}{(2)_{m_1+m_2}(b+1)_{m_1+m_2}}z^{m_1+m_2}
\end{equation}
\end{subequations}
Using these properties we derive
\begin{equation}\label{eq:epsexpconflhyp}
\begin{split}
{}_1 F_1\left(d-2,1+\frac{d}{2},-z\right)
&=e^{-z}\,{}_1 F_1\left(1+\eps,3-\eps,z\right)\\
&=\frac{2}{z^2}\gamma\left(2,z\right)+e^{-z}\frac{d}{d\eps}\,{}_1F_1\left(1+\eps,3-\eps,z\right)\bigg|_{\eps=0}\eps\\
&=\frac{2}{z^2}\gamma\left(2,z\right)+\frac{4}{z^2}\sum_{n=0}^\infty\frac{\gamma(n+3,z)}{\Gamma(n+4)}\frac{n+2}{n+1}\,\eps+\order{\eps^2}
\end{split}
\end{equation}
where in the first equality we use Eq.~\eqref{eq:kummerstrans}, and to obtain the third equality we first employ Eq.~\eqref{eq:conflhypder}, then combine Eqs.~\eqref{eq:conflhypsum} and \eqref{eq:conflhypandincomplgamma} to rewrite the sum in $m_2$.

Plugging the expansions into Eqs.~\eqref{eq:I1ddim} and \eqref{eq:III.1prddim}, we obtain:
\begin{equation}\label{eq:I1eps}
\begin{split}
\text{(I.1)}^{(\eps)}&=-C_2(R)\frac{2}{(4\pi)^2}\Pi_{\mn,0}^{V(\eps)}(t,x)\bigg\{\frac{1}{\eps}+\log(t)+\gamma\left(2,\xt\right)^{-1}\bigg[\gamma\left(1,\xt\right)\\
&\hspace{.5cm}-\text{Ein}\left(\xt\right)+2\sum_{n=0}^\infty\frac{\gamma\left(n+3,\xt\right)}{\Gamma(n+4)}\frac{n+2}{n+1}\bigg]+\text{finite parts}\bigg\}
\end{split}
\end{equation}
\begin{equation}\label{eq:III1preps}
\text{(III.1}')^{(\eps)}=-C_2(R)\frac{4}{(4\pi)^2}\Pi_{\mn,0}^{V(\eps)}(t,x)\left\{\frac{1}{\eps}+\log(t)+\text{finite parts}\right\}
\end{equation}
where we employed $\text{tr}(T^aT^a)=-C_2(R)d(R)$, and adding them we have for Eq.~\eqref{eq:Pi2eps}:
\begin{equation}\label{eq:PiV2eps} 
\begin{split}
\Pi^{V(\eps)}_{\mn,2}(t,x)&=-C_2(R)\frac{2}{(4\pi)^2}\Pi_{\mn,0}^{V(\eps)}(t,x)\bigg\{\frac{3}{\eps}+3\log(t)+\gamma\left(2,\xt\right)^{-1}\\
&\hspace{-1.5cm}\bigg[\gamma\left(1,\xt\right)-\text{Ein}\left(\xt\right)+2\sum_{n=0}^\infty\frac{\gamma\left(n+3,\xt\right)}{\Gamma(n+4)}\frac{n+2}{n+1}\bigg]+\text{finite parts}\bigg\}\\
&\hspace{-1.5cm}+\text{finite contributions}
\end{split}
\end{equation}
For completeness we also explicitly present $\Pi_{\mn,0}^{V(\eps)}$, though the $\order{\eps}$ term will not contribute in the final expression for $\Pi_{R,\mn}^V$ at $\order{g^2}$ in Eq.~\eqref{eq:2ptR}:
\begin{equation}\label{eq:PiV0eps} 
\begin{split}
&\Pi_{\mn,0}^{V(\eps)}(t,x)=\Pi_{\mn,0}^V(t,x)\bigg(1+2\bigg\{\log(x^2)+\log\left(\frac{x^2}{8t}\right)\\
&\hspace{1.7cm}+\gamma\left(\xt\right)^{-1}\left[\gamma\left(1,\xt\right)-\text{Ein}\left(\xt\right)\right]+\text{finite parts}\bigg\}\eps+\order{\eps^2}\bigg)
\end{split}
\end{equation}
with $\Pi_{\mn,0}^V$ given in Eq.~\eqref{eq:evolved2pointLO}.

To summarize, we have for Eq.~\eqref{eq:2ptR}:
\begin{equation}
\begin{split}
&\Pi_{R,\mn}^V(t,x,\mu,g(\mu))\\
&\hspace{.3cm}=\bigg(1-\frac{g^2(\mu)}{(4\pi)^2}C_2(R)\bigg\{6\log(t\mu^2)+2\gamma\left(2,\xt\right)^{-1}\bigg[\gamma\left(1,\xt\right)-\text{Ein}\left(\xt\right)\\
&\hspace{.8cm}+2\sum_{n=0}^\infty\frac{\gamma\left(n+3,\xt\right)}{\Gamma(n+4)}\frac{n+2}{n+1}\bigg]+\text{finite terms}\bigg\}
\bigg)\Pi_{\mn,0}^V(t,x)+\ldots
\end{split}
\end{equation}
where here the finite terms include the finite parts from the $\eps$-expansions in Eqs.~\eqref{eq:I1eps} and \eqref{eq:III1preps}, 
while the dots stand for $\order{g^4}$ contributions and additional $\order{g^2}$ finite terms. 

We add that at $t>0$ the short-distance limit of the $\order{g^2}$  contributions computed here is nonsingular and vanishes.
This is shown using:
\begin{equation}
\frac{d}{dz}\gamma(a,z)=z^{a-1}e^{-z},\hspace{.5cm}\frac{d}{dz}\text{Ein}(z)=\frac{1-e^{-z}}{z}
\end{equation}
which both follow from the definitions in Eqs.~\eqref{eq:lower-gamma}  and \eqref{eq:Ein}, and employing l'Hopital's rule to find
\begin{equation}
\lim_{z\rightarrow 0}\, \gamma(2,z)^{-1}\left(\gamma(1,z)-\text{Ein}(z)+2\sum_{n=0}^\infty\frac{\gamma(n+3,z)}{\Gamma(n+4)}\frac{n+2}{n+1}\right)=-\half
\end{equation}
which then combines with the vanishing limit of $\Pi_{\mn,0}^V(t,x)$.

\section{Leading order nonconservation of the evolved vector current}
\label{sec:classic}
We explicitly derive the classical nonconservation of the vector current evolved by the gradient flow: 
\begin{equation}
\del_\mu J^V_{\mu}(t,x)\neq0
\end{equation}
at leading order in the coupling $g$. At this order, the evolved current reads (see Eq.~\eqref{eq:cc}):
\begin{equation}\label{eq:LOGFJ}
J^V_{\mu,0}(t,x)=\chibar_0(t,x)\gmu\chi_0(t,x)=e^{t\Lap_x}\psibar(x)\gmu e^{t\Lap_x}\psi(x)
\end{equation}
with the exponentials-of-Laplacian
acting only on what is directly to the right of them. Taking into account the commutativity of the derivative and the exponential-of-Laplacian, the derivative of $J^V_{\mu,0}$ in Eq.~\eqref{eq:LOGFJ} yields (see also Eq.~\eqref{eq:chiandBexpansions}):
\begin{equation}
\begin{split}
\del_\mu J^V_{\mu,0}(t,x)&=e^{t\Lap}\left\{\del_\mu\psibar(x)\right\}\gmu\chi_0(t,x)+\chibar_0(t,x)\gmu e^{t\Lap}\left\{\del_\mu\psi(x)\right\}\\
&=e^{t\Lap}\left\{\psibar(x)\overleftarrow{\dslash}\right\}\chi_0(t,x)+\chibar_0(t,x)e^{t\Lap}\left\{\dslash \psi(x)\right\}
\end{split}
\end{equation}
Next we employ the classical equations of motion for $\psibar$ and $\psi$:
\begin{equation}
\left(\dslash+\aslash\right)\psi=0,\hspace{1cm}\psibar\left(\overleftarrow{\dslash}-\aslash\right)=0
\end{equation}
to arrive at
\begin{equation}
\del_\mu J^V_{\mu,0}(t,x)=e^{t\Lap}\left\{\psibar(x)\aslash(x)\right\}e^{t\Lap}\psi(x)-e^{t\Lap}\psibar(x)e^{t\Lap}\left\{\aslash(x)\psi(x)\right\}
\end{equation}
As a final step, we employ the Fourier transforms for $\psibar$, $\psi$ and $A_\mu$:
\begin{equation}
\psibar(x)=\int_p e^{ipx}\,\tilde{\psibar}(p),\hspace{.5cm} A_\mu(x)=\int_k e^{ikx} \,\tilde{A}_\mu(k),\hspace{.5cm} \psi(x)=\int_q e^{iqx}\,\tilde{\psi}(q)
\end{equation}
yielding
\begin{equation}
\begin{split}
\del_\mu J^V_{\mu,0}(t,x)&=\int_{p,q,k}e^{i(p+q+k)x}\left(e^{-t(p+k)^2-tq^2}-e^{-tp^2-t(k+q)^2}\right)\tilde{\psibar}(p)\tilde{\aslash}(k)\tilde{\psi}(q)\\
&=\int_{p,q,k}e^{i(p+q+k)x}e^{-t(p+k)^2-tq^2}\left(\tilde{\psibar}(p)\tilde{\aslash}(k)\tilde{\psi}(q)-\tilde{\psibar}(q)\tilde{\aslash}(k)\tilde{\psi}(p)\right)
\end{split}
\end{equation}
which does not vanish at $t>0$, while  we recover the conservation of the nonevolved current at  $t=0$.

\end{document}